\documentclass[pra,aps,psfig,epsf,eqsecnum,twocolumn,amsmath,showpacs,amssymb,groupedaddress]{revtex4}
\usepackage{bm}

\newcommand{\eps}{\varepsilon}
\newcommand{\as}{\tilde{a}}
\newcommand{\oh}{{\frac{1}{2}}}
\newcommand{\g}{\tilde{g}}

\newcommand{\kh}{{\hat{k}}}

\newcommand{\rv}{{\bf r}}
\newcommand{\qv}{{\bf q}}
\newcommand{\kv}{{\bf k}}

\newcommand{\be}{\begin{equation}}
\newcommand{\ee}{\end{equation}}
\newcommand{\bea}{\begin{eqnarray}}
\newcommand{\eea}{\end{eqnarray}}

\newcommand{\bse}{\begin{subequations}}
\newcommand{\ese}{\end{subequations}}

\def\rf#1{(\ref{#1})}
\def\rfs#1{Eq.~\rf{#1}}
\usepackage[pdftex]{graphicx}
\begin{document}

\title{Post-quench dynamics and pre-thermalization in a resonant Bose
  gas} 

\author{Xiao Yin}
\email[]{xiao.yin@colorado.edu}

\author{Leo Radzihovsky}
\email[]{radzihov@colorado.edu}
\affiliation{Department of Physics, University of Colorado,
   Boulder, CO 80309}
\date{\today}
\begin{abstract}
  We explore the dynamics of a resonant Bose gas following its quench
  to a strongly interacting regime near a Feshbach resonance. For such
  deep quenches, we utilize a self-consistent dynamic field
  approximation and find that after an initial regime of many-body
  Rabi-like oscillations between the condensate and finite-momentum
  quasiparticle pairs, at long times, the gas reaches a pre-thermalized
  nonequilibrium steady state.  We explore the resulting state through
  its broad stationary momentum distribution function, that exhibits a
  power-law high momentum tail. We study the dynamics and steady-state
  form of the associated enhanced depletion, quench-rate dependent
  excitation energy, Tan's contact, structure function and radio
  frequency spectroscopy. We find these predictions to be in a
  qualitative agreement with recent experiments.
\end{abstract}
\pacs{67.85.De, 67.85.Jk}
\maketitle
\section{Introduction}

\subsection{Background and motivation}
Degenerate atomic gases have radically expanded the scope of quantum
many-body physics beyond the traditional solid-state counter part,
offering opportunity to study highly coherent, strongly interacting,
and well-characterized, defects-free systems. Atomic field-tuned
Feshbach resonances (FRs)
\cite{FRrmp,BlochRMP,ZweirleinReview,GurarieRadzihovskyAOP} have
become a powerful experimental tool that has been extensively utilized
to explore strong resonant interactions in these systems. Feshbach
resonances have thus led to a seminal realization of paired $s$-wave
fermionic superfluidity, with the associated BCS-to-Bose-Einstein
condensate (BEC) crossover~\cite{GrimmBCS-BEC, JinBCS-BEC,
  ZweirleinReview,GurarieRadzihovskyAOP} through a universal unitary
regime \cite{JasonHo,VeillettePRA,PowellSachdevPRA}, and phase
transitions driven by species imbalance~\cite{PartridgePolarized,
  RSpolarized} and by Mott-insulating physics in an optical lattice
\cite{GreinerMI,Doniach,JackishZoller,refsFermionsFR_opticallattice,
  ZhaochuanPRL}.  Numerous other promising many-body states and phase
transitions, such a $p$-wave fermionic superfluidity
\cite{pwaveGR,pwaveYip,pwaveJin} and Stoner ferromagnetism
\cite{magnetismKetterle} have been proposed and
continue to be explored.

Unmatched by their extreme coherence and high tunability of system
parameters, such as FR interactions and single-particle (trap and
lattice) potentials, atomic gases have also enabled numerous
experimental realizations of highly {\em nonequilibrium},
strongly-interacting many-body states and associated phase
transitions \cite{GreinerMI,JinBCS-BEC,BlochRMP}.

This has motivated extensive theoretical
studies \cite{PolkovnikovRMP2011,CazalillaRMP2011,Schmiedmayer}, with a particular
focus on nonequilibrium dynamics following a quench of Hamiltonian
parameters, $\hat H_i\rightarrow \hat H_f$. In addition to studies of specific
physical systems, experiments on these closed and highly coherent
systems have driven theory to address fundamental questions in quantum
statistical mechanics. These include the conditions for and nature of
thermalization under unitary time evolution $|\hat \psi(t)\rangle=e^{i
 \hat H_f t}|\hat \psi_i(0)\rangle$ of a closed quantum system vis-\'a-vis
eigenstate thermalization hypothesis \cite{Srednicki,Rigola}, role of
conservation laws and obstruction to full equilibration of integrable
models argued to instead be characterized by a generalized Gibb's
ensemble (GGE), emergence of statistical mechanics under unitary time
evolution for equilibrated and nonequilibrium stationary
states \cite{KinoshitaWengerWeissNature2006,RigolOlshanniNaturePRL07}.
These questions of post-quench dynamics have been extensively explored
in a large number of
systems \cite{AlmanVishwanath06,Barankov06,AGR06,Yuzbashan07,
  MitraGiamarchi13,GurarieIsing13,CalabreseCardy07,
  SotiriadisCardy10,Sondhi13,NatuMueller13,ChinGurarie13,YinLR14,
  BacsiBalazs13,MitraSpinChain14,Essler14,NessiCazalilla14}

Early studies of a Feshbach-resonant Fermi gas predicted persistent
coherent post-quench oscillations \cite{Barankov,AGR06} and, more
recently found topological nonequilibrium steady states and phase
transitions \cite{Yuzbashan,Foster}.

Resonant Bose gas quenched dynamics studies date back to seminal
experiments on $^{85}$Rb \cite{Donley,Claussen}, that demonstrated
coherent Rabi-like oscillations between atomic and molecular
condensates \cite{HollandKokkelmans}, enabling a measurement of the
molecular binding energy. More recently, oscillations in the dynamic
structure function have also been observed in quasi-2D bosonic
$^{133}$Cs \cite{ChinGurarie13} and studied theoretically \cite{NatuMueller13,Ranon13} for shallow quenches between weakly-repulsive interactions (small gas parameter $n a^3_s \ll 1$ where $a_s$ is the s-wave scattering length).

Such resonant bosonic gases were also predicted to exhibit distinct
atomic and molecular superfluid phases, separated by a quantum Ising
phase transition (rather than just a fermionic smooth BCS-BEC
crossover) and other rich
phenomenology \cite{RPWmolecularBECprl04,StoofMolecularBEC04,
  RPWmolecularBECpra08,ZhouPRA08,BonnesWessePRB12}, thereby providing
additional motivation for their studies.

Important recent developments are experiments by Makotyn, et
al, \cite{Makotyn}, that explored dynamics of $^{85}$Rb following a
{\em deep} quench to the vacinity of the unitary point on the
molecular (positive scattering length, $a_s > 0$) side of the Feshbach
resonance. It was discovered that even near the unitary point, where a
Bose gas is expected to be unstable \cite{PetrovShlypnikov}, the
three-body decay rate $\gamma_3$ (on the order of an inverse
milli-second) appears to be more than an order of magnitude slower
than the two-body equilibration rate $\gamma_2$ (both measured to be
proportional to Fermi energy, as
expected \cite{FeiZhou,LRunpublished}. This thereby opened a window of
time scales from a microsecond (a scale of the quench) to a
milli-second for observation of a metastable strongly-interacting
nonequilibrium dynamics.

Stimulated by these fascinating experimental developments and
motivated by the aforementioned earlier work, in a recent brief
publication \cite{YinLR14} we reported on results for the upper-branch
repulsive dynamics of a resonant Bose gas following a deep-detuning
quench close to the unitary point on the molecular side ($a_s > 0$) of
the FR \cite{Makotyn}. Taking the aforementioned slowness of
$\gamma_3\ll\gamma_2$ as an empirical fact, consistent with
experimental observations we predicted a fast evolution to a
pre-thermalized strongly-interacting stationary state, characterized
by a broad, power-law steady-state momentum distribution function,
$n_k^{ss}$, with a time scale $\tau_k = \hbar/E_k$ for the
pre-thermalization of momenta $k$ set by the inverse of the excitation
spectrum, $E_k$. The associated condensate depletion was found to
exhibit a monotonic growth to a nonequilibrium value exceeding that of
the corresponding ground state. In the current manuscript we present the
details of the analyses that led to these results as well as a large
number of other predictions.

\subsection{Outline}
The rest of the paper is organized as follows.  We conclude the
Introduction with a summary of our key results. In Section II,
starting with a one-channel model of a Feshbach-resonant Bose gas, we
develop its approximate Bogoluibov and self-consistent dynamic field
forms. In Section III, as a warmup we analyze the equilibrium
self-consistent model for the strongly interacting case and compare
its predictions to that of the Bogoluibov approximation.  In Section
IV we utilize the Bogoluibov model to study the {\em nonequilibrium}
dynamics following a shallow-quench, computing the momentum
distribution function $n_k(t)$ probed in the time-of-flight, the
radio-frequency (RF) spectroscopy signal, $I(\omega,t)$, and the
structure function $S_k(t)$ probed via Bragg spectroscopy. Then in
Section V we generalize the quench to a more experimentally realistic case of a finite-rate ramp and study the effect of ramp rate. In Section VI we employ the self-consistent dynamic field theory to study these and a number of other observables for deep quenches in a strongly interacting regime relevant to JILA
experiments \cite{Makotyn}.  In Section VII we study excitation energy, an important measure of long time nonequilibrium stationary state, for both sudden quench and finite ramp-rate cases, and discuss its dependence on quench depth and ramp rate. We generalize Tan's Contact to nonequilibrium process and study its long time behavior in Section VIII. Finally in Section IX we conclude with a discussion of our predictions for experiments and of the future directions for this work. We relegate the details of most calculations to Appendices.

\subsection{Summary of results}
Before turning to the derivation and analysis, we briefly summarize
the key predictions of our work. Working within the upper-branch of a
single-channel model of a resonantly interacting Bose gas we studied
an array of nonequilibrium observables following its Feshbach
resonance quench toward the unitary point. One central quantity
extensively studied in recent time of flight measurements
\cite{ChinGurarie13,Makotyn} is the momentum distribution function,
$n_k(t)=\langle gs_i|a^\dagger_k(t)a_k(t)|gs_i\rangle$ at time $t$
after a quench from a ground state $|gs_i\rangle$ of an initial
Hamiltonian $\hat H_i$ to a final Hamiltonian $\hat H_f$.  Motivated by
experiments we take $|gs_i\rangle$ to be a superfluid BEC ground state
in the upper branch of the repulsive Bose gas \cite{footnote}. For a shallow quench in
the scattering length $a_i\rightarrow a_f$, away from the immediate
vicinity of the unitary point, the calculation is controlled by an
expansion in a small interaction parameter, $n a_{i,f}^3\ll 1$. Within
the lowest, Bogoluibov approximation the momentum distribution
function is given by (choosing units where $\hbar = 1$ and $k_B=1$ throughout)
\cite{NatuMueller13}
\begin{equation}
\begin{split}
n_{\hat{k}}(\hat{t})=\frac{\hat{k}^2+\sigma
  +\frac{2(1-\sigma)}{\hat{k}^2+2}\sin^2(\hat{t}
\sqrt{\hat{k}^2(\hat{k}^2+2)})}{2\sqrt{\hat{k}^2(\hat{k}^2+2\sigma)}}
-\frac{1}{2},\\
\end{split}
\label{nkBogoliubovResults}
\end{equation}
where $\sigma\equiv a_i/a_f$ characterizes the ``depth'' of the
quench, and we have rescaled the momentum $k$ and time $t$ with the coherence length $\xi\equiv1/\sqrt{2 m n g_f}$ and pre-thermalization timescale $t_0=1/ng_f$, as $\hat{k}=k\xi$ and $\hat{t}=t/t_0$, respectively. We start the system in a weakly
interacting state, characterized by a short positive scattering length
$a_i$ and quench it to $a_f > a_i$ ($\sigma\leq 1$). Following
coherent oscillations, the gas then exhibits pre-thermalization
dynamics, where after a dephasing time $\tau_k$, set by the inverse of
the excitation spectrum $1/E_k=1/\sqrt{\hat{k}^2(\hat{k}^2+2)}$
consistent with experiments \cite{Makotyn}, the initial narrow
Bogoluibov momentum distribution evolves to a stationary state,
characterized by a broadened distribution function 
\bse
\begin{eqnarray}
n^{ss}_{\hat{k}}&=&\oh\left[\frac{(\hat{k}^2+\sigma)(\hat{k}^2+2)+ 1
    -\sigma}{(\hat{k}^2+2)\sqrt{\hat{k}^2(\hat{k}^2+2\sigma)}}-1\right],\\
&\sim&\left\{\begin{array}{ll}
C^{ss}/k^4,&\mbox{for $k\xi\gg 1$},\\
1/k^2,&\mbox{for $\sigma\ll k\xi\ll 1$},\\
1/k,&\mbox{for $k\xi\ll \sigma$},
\end{array}\right.
\label{nkssBogoliubovResults}
\end{eqnarray}
\ese
where we defined $C^{ss}$ as the nonequilibrium analog of Tan's
contact for the nonequilibrium steady state, given by
\begin{eqnarray}
C^{ss}=(4\pi a_f n)^2[1+(1-\sigma)^2].
\end{eqnarray}
Within above approximation the quasi-particles do not scatter,
precluding full thermalization, and the above final state remains
nonequilibrium, completely determined by the depth-quench parameter
$\sigma$, with the associated diagonal density matrix ensemble.

The associated condensate depletion $n_d(t) =\frac{1}{N}\sum_{\kv\neq 0}
n_k(t)$ is then straightforwardly computed and monotonically
pre-thermalizes to
\begin{equation}
\label{ndssIntro}
\begin{split}
  n_d^{ss}(\sigma)=\frac{8}{3\sqrt{\pi}}\left({n a_f^3}\right)^{1/2}
  \left[\sigma^{3/2}+\frac{3}{2}\sqrt{1-\sigma}
    \arccos(\sqrt{\sigma})\right],
\end{split}
\end{equation}
a value exceeding that for the ground state of the final scattering
length $a_f$ and greater than the initial ground state depletion
$n_{d}^i=n_d^{ss}(\sigma=1)=\frac{8}{3\sqrt{\pi}}\left({na_i^3}\right)^{1/2}$
at scattering length $a_i$.

\begin{figure}[htb]
 \centering
   \includegraphics[width=80mm]{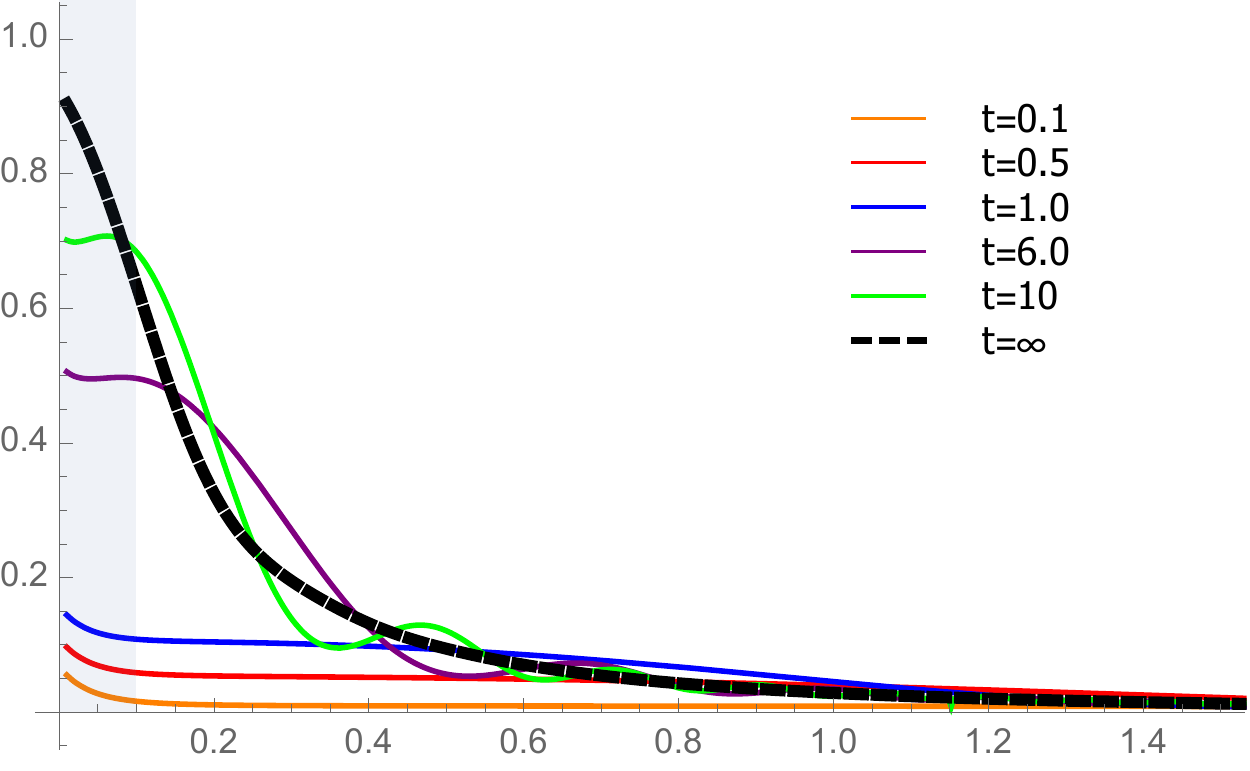}
\caption{(Color online) Time evolution of the (column-density)
  momentum distribution function, $\tilde{n}_{\kv_\perp}(t)\equiv
 \int dk_z n_\kv(t)$ following a deep scattering length
  quench $k_na_i=0.01\rightarrow k_na_f=1$ in a resonant Bose gas (where $k_n\equiv n^{1/3}$), computed within a self-consistent dynamic field approximation. Here momentum is rescaled by the coherence length $\xi$ as $\hat{k}=k\xi\equiv k/\sqrt{2 m n g_f}$. Lowest curve corresponds to earlier time at $\hat{t}\equiv t/t_0=0.1$ in units of pre-thermalization timescale $t_0= 1/ng_f=m/(4\pi a_f n )$ while the dashed-thick black one represents the asymptotic steady-state distribution. The figure illustrates the initial narrow
  momentum distribution (lowest curve) evolving to a much broader  momentum distribution (highest curve), corresponding to a  pre-thermalized steady state. The grey region indicates a range of momenta not resolved in JILA experiments, due to initial inhomogeneous real space density profile and finite trap size.} 
\label{nktplot}
\end{figure}

\begin{figure}[htb]
\centering
\includegraphics[width=0.45\textwidth]{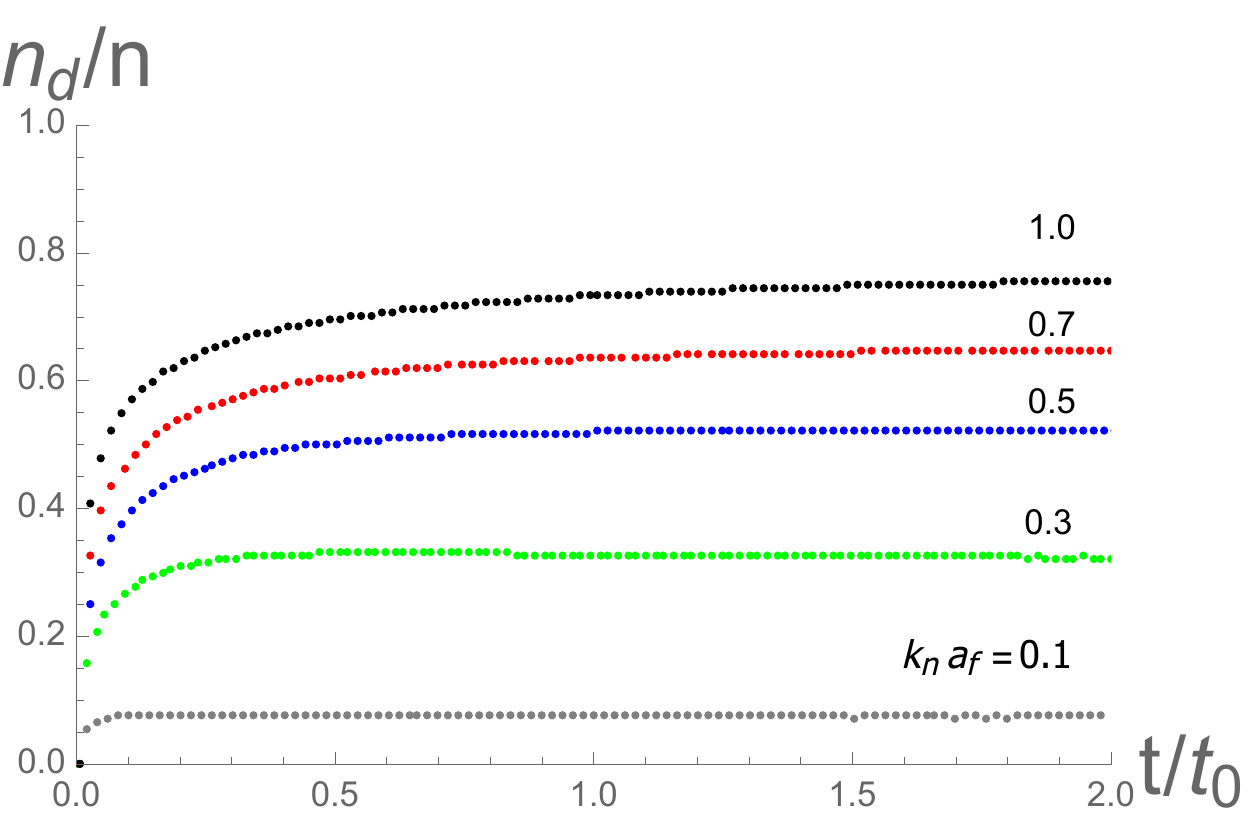}
\caption{(Color online) Time evolution of the condensate depletion
  fraction $n_d(t)/n$ (treated within a self-consistent dynamic field
  analysis), following a scattering length quench from $k_na_i=0.01$
  to various $k_na_f$ in a resonant Bose gas. Here we normalize the
  time with the pre-thermalization timescale $t_0= 1/ng_f=m/(4\pi a_f
  n)$ associated with $k_na_f=1$ (where $k_n\equiv n^{1/3}$).}
\label{ndtplot}
\end{figure}

With the goal of understanding deep quenches of a strongly interacting
Bose gas \cite{Makotyn,YinLR14,SykesPRA14} near a Feshbach resonance,
we developed a self-consistent dynamic field theory of coupled
Gross-Petaevskii equation for the condensate $n_c(t)$ and a Heisenberg
equation for atoms $\hat a_{\kv\neq 0}(t)$ excited out of the condensate. It
accounts for strong time-dependent depletion of the condensate, with
feedback on dynamics of excitations. Within this nonpertubative (but
uncontrolled) approximation this amounts to solving for a Heisenberg
evolution of $\hat a_{\kv}(t)$ with a time-dependent Bogoluibov-like
Hamiltonian, parameterized by a condensate density $n_c(t)$. The
latter is self-consistently determined by the atom-number constraint
equation, $n_c(t)=n-\sum_\kv
n_k(t,[n_c(t)])$ \cite{AGR06,YinLR14}. Our treatment here is closely
related to the analysis of post-quench quantum coarsenning dynamics of
the $O(N)$ \cite{Sondhi13} and Ising \cite{SotiriadisCardy10} models. The
resulting momentum distribution function, $\tilde{n}_{\kv_\perp}(t)$ (projected
column density measured in experiments \cite{Makotyn}) and the corresponding depletion $n_d(t)$ are illustrated in Figs.~\ref{nktplot},\ref{ndtplot}.

\begin{figure}[htb]
\centering
\includegraphics[width=0.45\textwidth]{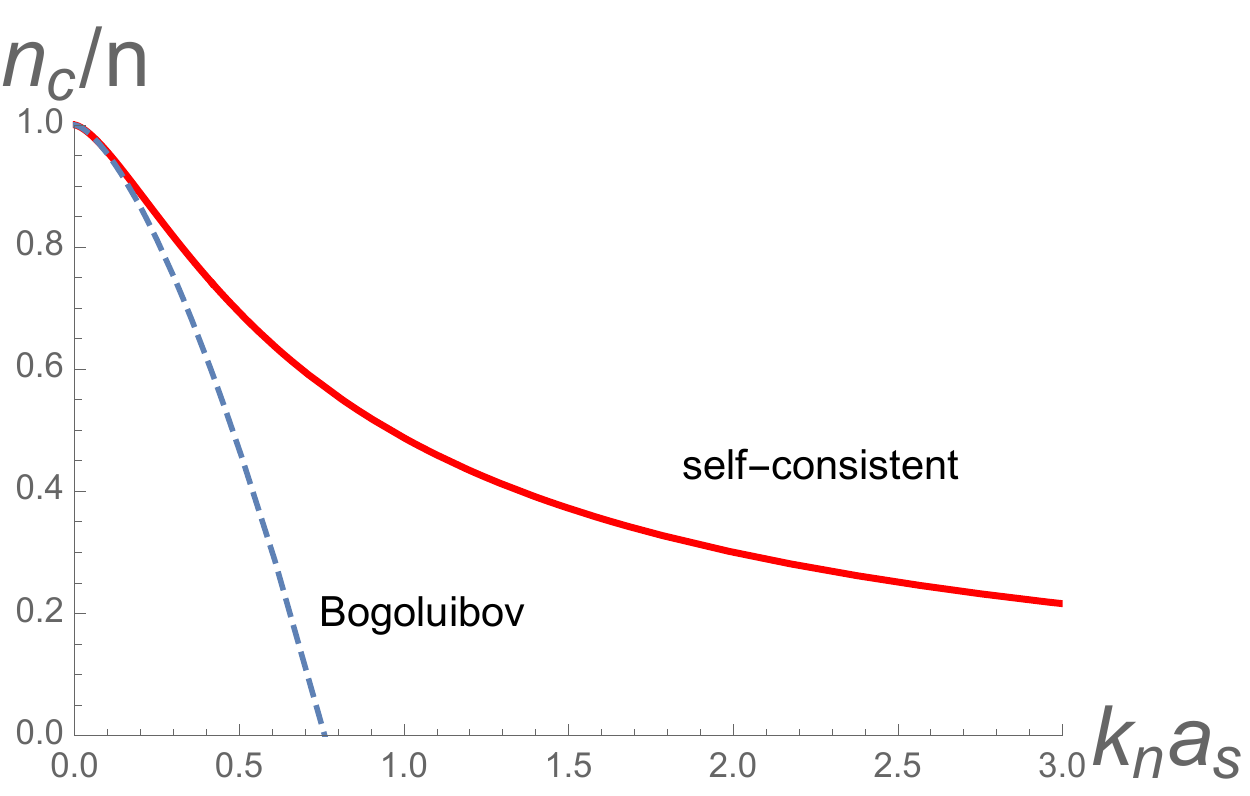}
\caption{(Color online) Ground state condensate fraction as a function of a
  dimensionless measure of atom density and interaction, $k_n a_s$(with $k_n\equiv n^{1/3}$),
  computed within a self-consistent dynamic field
   approximation (solid red curve), as compared to Bogoluibov approximation result (dashed blue curve).}
\label{static_SC}
\end{figure}

\begin{figure}[htb]
 \centering
  \includegraphics[width=80mm]{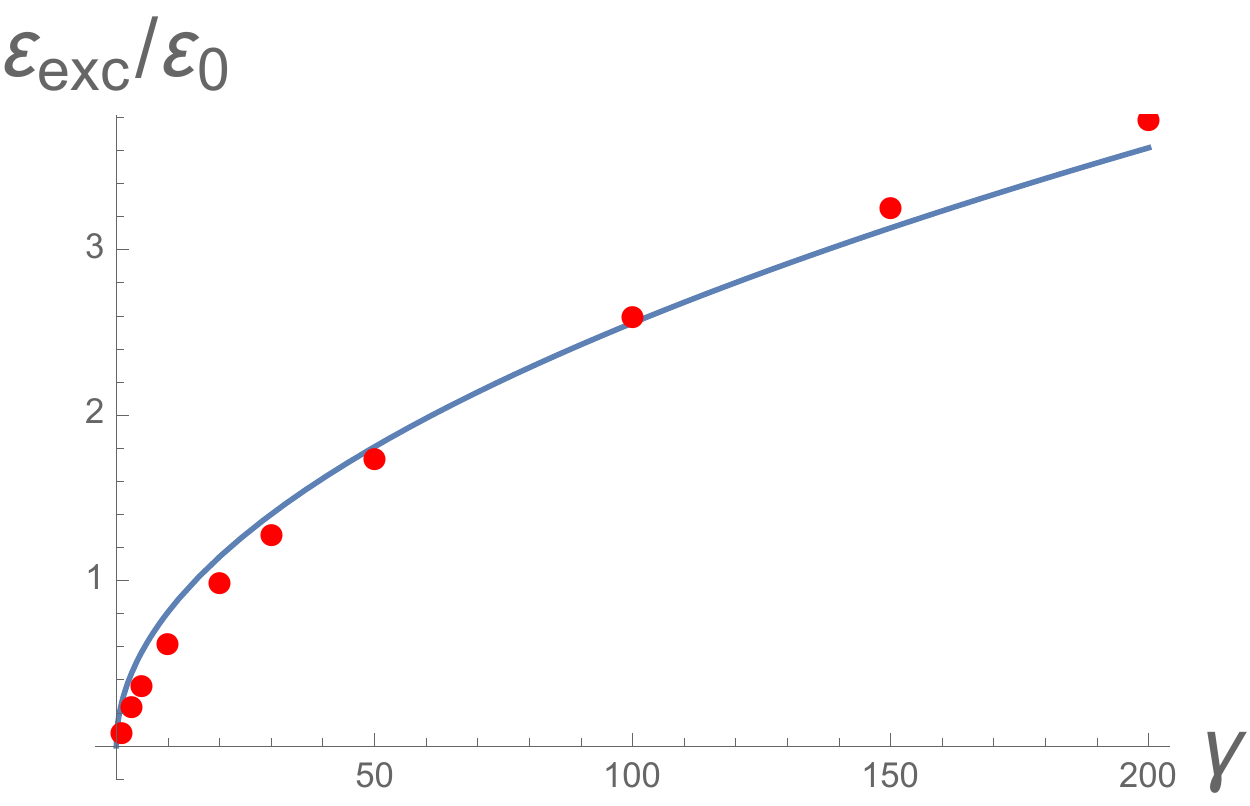}
 \caption{(Color online) Excitation energy (scaled by LHY correction to the ground state energy) following a scattering
   length ramp as a function of ramp rate $\gamma$ (as a ``zoom-in'' for Fig.~\ref{excratedep}, see Sec.~\ref{appendix:excitationenergy}). The red data points are obtained for each chosen $\gamma$ at
   $a_i/a_f={1}/{2}$, with scaled dimensionless momentum cutoff $\hat{\Lambda}=\Lambda\xi=100$ ($\xi\equiv1/\sqrt{2 m n g_f}$ is the coherence length); the blue curve
   represents the fitting function $y=0.26 \sqrt{x}$. }
\label{energyrampplot}
\end{figure}

We also studied the excitation energy after a constant ramp rate
$\gamma$ between $a_i$ and $a_f$ scattering lengths. As illustrated in
Fig.~\ref{energyrampplot}, we found that it displays a $\sqrt{\gamma}$
form
\bse
\begin{eqnarray}
 \frac{ E_{exc}(\gamma)}{V}&=&\frac{4(\sigma - 1)^2  n^2 a_f}{m}a_f\Lambda f(\gamma/E_\Lambda),\\
  &\propto&\mbox{$(1-\sigma)^{3/2}
\sqrt{\gamma}$, \;\; for $\gamma\ll E_{\Lambda}$}\nonumber,\\
  &\propto&\mbox{$ (1-\sigma)^2 a_f\Lambda~$,\;\;\;\;\;for $\gamma\gg E_{\Lambda}$},\nonumber\\
\end{eqnarray}
\ese
for a ramp-rate below the microscopic energy cutoff $E_\Lambda=\Lambda^2/2m$.

To further characterize the post-quench evolution and the resulting
pre-thermalized steady-state we have also computed a time dependent
structure function $S(q,t) =\langle
gs_i|n(-\qv,t)n(\qv,t)|gs_i\rangle$, a Fourier transform of the
density-density connected correlation function. For the weakly
interacting, shallow-quench regime, at temperature $1/\beta$ it
is given by
\begin{equation}
\begin{split}
  S(q,t)=\frac{n_0\epsilon_q}{E^2_{qf}} \coth(\beta
  E_{qi}/2)\left[1+\frac{E^2_{qi}-E^2_{qf}}{E^2_{qf}}\sin^2 (E_{qf} t)\right],
\end{split}
\end{equation}
first computed and measured in Ref. \cite{ChinGurarie13}, and after
pre-thermalization reduces to a time-independent form \cite{YinLR14},
\begin{equation}
\begin{split}
S^{ss}_q=
\frac{n_0\epsilon_q}{2E^2_{qf}}\coth(\beta
  E_{qi}/2)\left(1+\frac{E^2_{qi}}{E^2_{qf}}\right).
\end{split}
\end{equation}

\begin{figure}[htb]
 \centering
  \includegraphics[width=80mm]{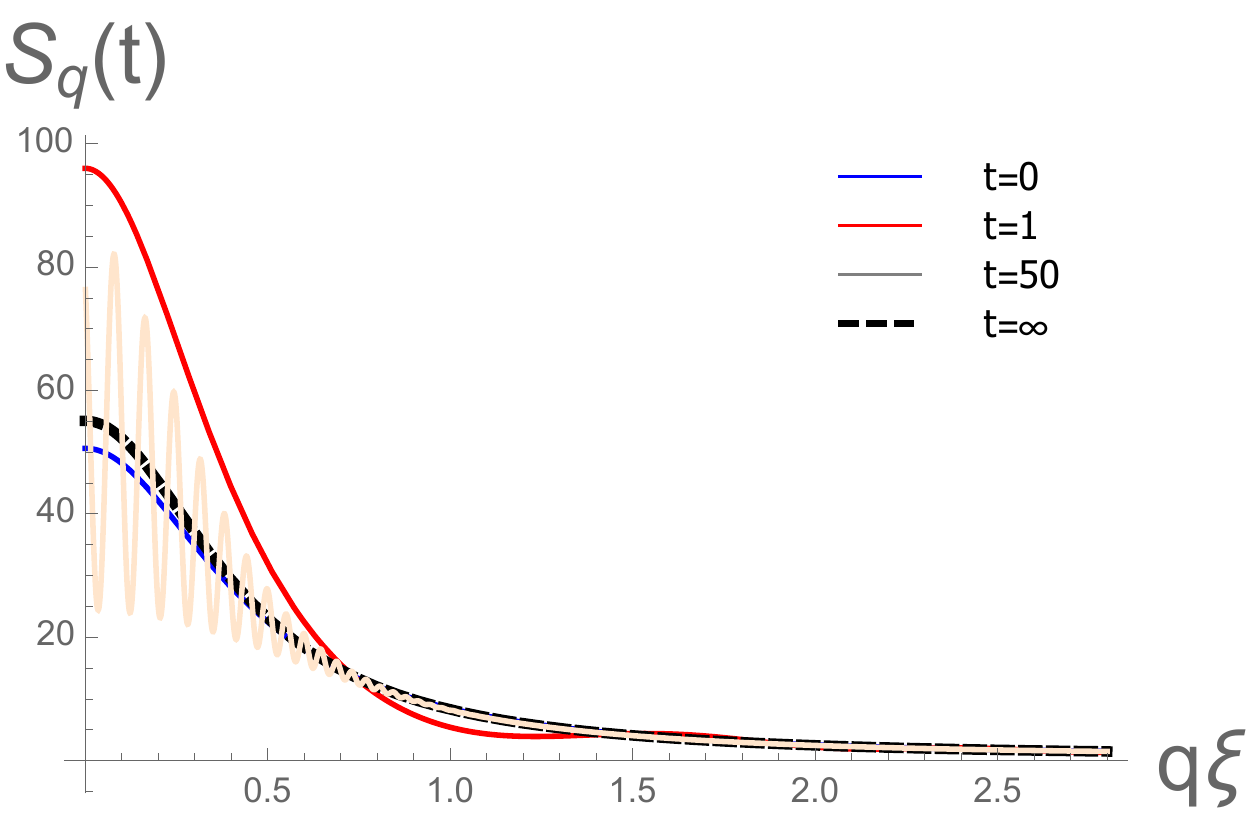}
 \caption{(Color online) Time evolution of the structure function
   $S_\qv(t)$ defined in the text following a scattering length quench
   from $0.1a_f\to a_f$ with $k_na_f=0.7$ (where $k_n\equiv n^{1/3}$), referring to
   Eq.~\eqref{quasiscf2} using quasi-adiabatic self-consistent
   approximation (see Sec.~\ref{ndSCqs}). It illustrates the initial
   ground state structure function (blue curve), that following the
   quench develops oscillations and after a pre-thermalization time
   approaches a steady-state distribution (dashed black curve), which
   within-quasi-adiabatic approximation almost collapses with the
   initial ground state curve. Here momentum and time are rescaled
   with $\xi\equiv 1/\sqrt{2 m n g_f}$ and $t_0\equiv 1/(n g_f)$,
   respectively.}
\label{SqDMFT}
\end{figure}

Utilizing our self-consistent dynamic field theory we extended above
calculation of $S(q,t)$ to deep quenches of strongly interacting
resonant condensates. The resulting time-dependent structure function
and its steady-state form are illustrated in Fig.~\ref{SqDMFT}.

We also computed the RF spectroscopy signal $I(\omega_{RF})$
\cite{JinRF,Wild}, that measures the transition rate of atoms from two
resonantly interacting hyperfine states into a third weakly
interacting hyperfine state, for the quench process. Within the Bogoluibov approximation the
response is given by
\begin{eqnarray}
  I(\omega_{RF})=\frac{\sqrt{2}\tau VI^2_0}{4\sqrt{\pi m}}\frac{(4\pi n a_f )^2}{\omega_{RF}^{3/2}},
\end{eqnarray}
as measured experimentally, with the amplitude proportional to Tan's
contact, that in the simplest Bogoluibov approximation is given by $C
=(4\pi n a_f )^2 $.

We next turn to a single-channel Feshbach resonant model, followed by
its detailed analysis that led to above and other results.

\section{Model of a resonant superfluid}
\label{model}
A resonant gas of bosonic atoms can be modeled by a single-channel
grand-canonical Hamiltonian, 
(defining $\int_\rv\equiv\int d^3r$)
\begin{equation}
\begin{split}
  \hat H = \int_\rv [\hat\psi^{\dagger}(\hat\epsilon-\mu)\hat\psi
  +\frac{g}{2}\hat\psi^{\dagger}\hat\psi^{\dagger}\hat\psi\hat\psi],
\end{split}
\label{Hmodel}
\end{equation}
where $\hat\psi(\rv)$ is a bosonic atom field operator, $\hat
\epsilon=-\frac{\nabla^2}{2m}$ is a single-particle
Hamiltonian, $\mu$ is the chemical potential, and the pseudo-potential
$g$ characterizes the atomic two-body interaction on the scale longer
than its microscopic range $r_0 = 1/\Lambda$, typically on the order
of ten angstroms. For simplicy, we have set $\hbar=1$.

\begin{figure}[htb]
 \centering
  \includegraphics[width=80mm]{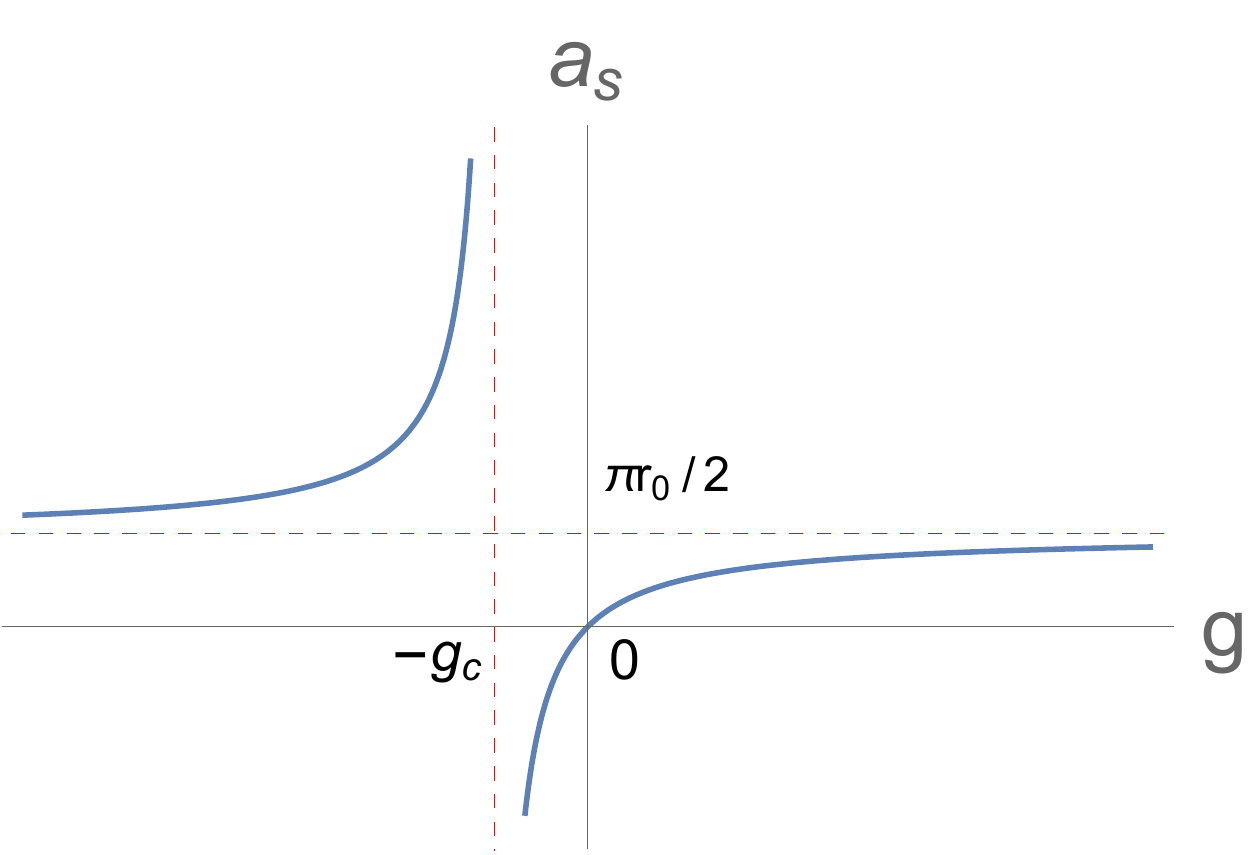}
 \caption{(Color online) A plot of the s-wave scattering length $a_s$
   (renormalized coupling $\tilde g$) as a function of bare coupling
   $g$ in a Feshbach resonance. Here $g_c=2\pi^2 r_0 /m$ is the
   critical coupling strength at which $a_s$ diverges.}
\label{fig:frdiagram}
\end{figure}

As discussed in detail in Ref.~\cite{GurarieRadzihovskyAOP} and
references therein, near a Feshbach resonance the magnetic
field-dependent coupling $g(B)$ controls the s-wave scattering length
$a_s$ through the renormalized coupling ($T$-matrix) $\tilde{g}^{-1} =
g^{-1}+ \int_\kv\frac{1}{2\epsilon_k} = g^{-1}+
m\Lambda/(2\pi^2 )$, 
\begin{eqnarray}
\tilde{g}=\frac{g}{1 + g/g_c},
\label{as}
\end{eqnarray}
related to the scattering length via $\tilde{g}=4\pi a_s/m$. As
illustrated in Fig.~\ref{fig:frdiagram}, for a sufficiently strong
attractive interaction, in a vacuum, the two-atom scattering length
diverges at $g_c = 2\pi^2 /(m\Lambda)=2\pi^2 r_0/m$, as the two-body bound state
forms for $g < -g_c$ and $a_s$ turns positive on the so-called ``BEC''
side of the Feshbach resonance. $r_0$ is the range of the potential and $\Lambda$ is the corresponding momentum cutoff.  It is this scattering-length
tunability that enables studies of phase transitions in resonant Bose
\cite{RPWmolecularBECprl04,StoofMolecularBEC04,
  RPWmolecularBECpra08,ZhouPRA08,BonnesWessePRB12} (and BCS-BEC
crossover in Fermi \cite{GrimmBCS-BEC, JinBCS-BEC,
  ZweirleinReview,FRrmp,BlochRMP,ZweirleinReview,GurarieRadzihovskyAOP})
gases and quenched dynamics
\cite{ChinGurarie13,Makotyn,YinLR14,SykesPRA14} that is our focus
here.

To allow for dynamics within a Bose-condensed state explored
experimentally \cite{ChinGurarie13,Makotyn}, we decompose the atomic field
operator $\hat\psi(\rv) = \frac{1}{\sqrt{V}}\sum_\kv \hat a_\kv e^{i
  \kv\cdot\rv}$, into a c-field condensate $\Psi_0$ and a fluctuation
field $\hat{a}(\rv)$,
\begin{equation}
\begin{split}
\hat{\psi}=\Psi_0+\hat{a}.
\end{split}
\label{decomp}
\end{equation}
Expressing the Hamiltonian, \rf{Hmodel} in terms of the operator $\hat
a$, it decomposes into
\begin{equation}
\begin{split}
\hat{H}=\hat{H}_0+\hat{H}_1+\hat{H}_2+\hat{H}_3+\hat{H}_4,
\end{split}
\label{Htot}
\end{equation}
where
\begin{equation}
\begin{split}
\hat{H}_0=\int_\rv[\Psi_0^*(\hat\epsilon-\mu)\Psi_0+\frac{g}{2}|\Psi_0|^4],
\end{split}
\label{H0}
\end{equation}
is the lowest order mean-field ground-state energy, and
\bse
\begin{equation}
\begin{split}
  \hat{H}_1=\int_\rv[\hat 
a^{\dagger}(\hat\epsilon+g|\Psi_0|^2-\mu)\Psi_0]+h.c.,\\
\end{split}
\label{H1}
\end{equation}
\begin{equation}
\begin{split}
\hat{H}_2=\int_\rv\left[\hat
a^{\dagger}\left(\hat\epsilon+2g|\Psi_0|^2-\mu\right)\hat a
+\frac{g}{2}\left({\Psi_0^*}^2\hat a\hat a+\Psi_0^2 \hat a^{\dagger}\hat a^{\dagger}\right)\right],
\end{split}
\label{H2}
\end{equation}
\begin{equation}
\begin{split}
\hat{H}_3=g \int_\rv[\Psi_0 \hat a^{\dagger}\hat a^{\dagger}\hat a+\Psi_0^*\hat a^{\dagger}\hat a\hat a],
\end{split}
\end{equation}
\begin{equation}
\begin{split}
\hat{H}_4=\frac{g}{2}\int_\rv \hat a^{\dagger}\hat a^{\dagger}\hat
a\hat a.
\end{split}
\end{equation}
\ese
are the operator components organized by respective orders in
the excitation $\hat a$.

\subsection{Bogoluibov approximation for weakly interacting bosons}
\label{sec:BdG}
We set the stage for the study of dynamics following a shallow quench \cite{ChinGurarie13} and of a self-consistent dynamic field
treatment \cite{YinLR14} of a deep quench \cite{Makotyn} by first
briefly summarizing the results for the ground state and excitations
in the Bogoluibov approximation \cite{Bogoliubov,Fetter}.

In the weakly interacting limit the atomic gas is characterized by a
small gas parameter $n a_s^3\ll 1$, well-approximated by the
Bogoluibov quadratic Hamiltonian, neglecting the nonlinear $\hat
H_{3,4}$ components of $\hat H$.  Focusing on the uniform (bulk)
condensate and eliminating the chemical potential in favor the
condensate density by requiring the vanishing of the $\hat H_1$
component (equivalent to a minimization of $\hat H_0$ over $\Psi_0$),
$\mu = g|\Psi_0|^2 \approx g n$, neglecting the difference between the
condensate density, $|\Psi_0|^2\equiv n_c$ and total atom density,
$n$, the grand-canonical Hamiltonian reduces to $\hat H \approx -\oh V
g n^2 + \hat H_B$,
\begin{eqnarray}
\hat{H}_{B}
&=&-\oh\sum_{\kv\neq 0}\varepsilon_k
+\oh\sum_{\kv\neq 0} 
\begin{pmatrix}
 \hat a_\kv^\dagger & \hat a_{-\kv}\\
\end{pmatrix}
\begin{pmatrix}
  \varepsilon_k & g n_c\\
  g n_c & \varepsilon_k\\
\end{pmatrix}
\begin{pmatrix}
\hat   a_\kv\\
\hat   a_{-\kv}^\dagger\\
\end{pmatrix},\nonumber\\
&=& -\oh\sum_{\kv\neq 0}\varepsilon_k
+\oh\sum_{\kv\neq 0} 
\hat \Phi^\dagger_{\kv,i} h_{\kv,ij}\hat \Phi_{\kv,j},\nonumber\\
&=& -\oh\sum_{\kv\neq 0}\varepsilon_k
+\oh\sum_{\kv\neq 0}E_k\hat \Psi^\dagger_{\kv,s}\hat \Psi_{\kv,s},\nonumber\\
&=&-\oh\sum_{\kv\neq 0}(\varepsilon_k
- E_k) + \sum_{\kv\neq 0} E_k\hat \alpha_\kv^\dagger\hat  \alpha_\kv,
\label{Hbog}
\end{eqnarray}
where the quadratic Hamiltonian was straightforwardly diagonalized in
terms of the Bogoluibov quasi-particles  $\hat \Psi_\kv =
(\hat \alpha_\kv,\hat \alpha_{-\kv}^\dagger)$, related to the atomic Nambu
spinor $\hat \Phi_\kv = (\hat a_\kv,\hat a_{-\kv}^\dagger)$ by a pseudo-unitary
transformation, $U_\kv$
\bse
\begin{eqnarray}
\begin{pmatrix}
\hat a_\kv\\ 
\hat a_{-\kv}^\dagger\\
\end{pmatrix}
&=&
\begin{pmatrix}
u_\kv & v_\kv\\
v^*_\kv & u^*_\kv\\
\end{pmatrix}
\begin{pmatrix}
\hat \alpha_\kv,\\ 
\hat \alpha_{-\kv}^\dagger,\\
\end{pmatrix}\\
\hat \Phi_\kv &=& U_\kv\hat  \Psi_\kv.
\end{eqnarray}
\ese 
$U_\kv$ satisfies a pseudo eigenvalue equation $h_\kv U_\kv =
E_\kv\sigma^z U_\kv$ and preserves the canonical commutation relation,
$[a_\kv,a_{\kv'}^\dagger]=\delta_{\kv,\kv'}$, corresponding to
$[\hat \Phi_{i\kv},\hat \Phi^\dagger_{j\kv'}]=\sigma^z_{ij}\delta_{\kv,\kv'}$,
defined by
\begin{eqnarray}
  U\sigma_z U^\dagger = \sigma_z,
\end{eqnarray}
with $|u_\kv|^2 - |v_\kv|^2 = 1$ and $\sigma^z$ the third Pauli
matrix.  

With $\eps_k= k^2/2m + g n$ in \rf{Hbog}, the Bogoluibov
spectrum is given by a well-known gapless form,
\begin{eqnarray}
E_k &=& \sqrt{\varepsilon_k^2 - g^2 n^2} = \sqrt{\epsilon_k^2 + 2g
  n\epsilon_k}= c   k\sqrt{1+\xi^2 k^2/2},\nonumber\\
\label{Ebog}
\end{eqnarray}
that interpolates between the low-momentum zeroth-sound with velocity
$c=\sqrt{g n/m}$ (a Goldstone mode of the $U(1)$ symmetry breaking)
and the high-momentum quadratic dispersion, with crossover scale set
by the correlation length $\xi = 1/\sqrt{2m g n}$. The
corresponding coherence factors defining $U_\kv$ are given by
\begin{eqnarray}
u_k^2 &=& \oh\left(\frac{\varepsilon_k}{E_k} + 1\right),\ \ 
v_k^2 = \oh\left(\frac{\varepsilon_k}{E_k} - 1\right).
\end{eqnarray}

The ground state is a vacuum of Bogoluibov quasi-particles,
$\hat\alpha_\kv|gs\rangle=0$, with the energy density ${\cal
  E}_{gs}=V^{-1}\langle gs|\hat H|gs \rangle$ given by
\bse
\begin{eqnarray}
  {\cal E}_{gs}&=&\oh g n^2 - \frac{1}{V}\sum_{\kv\neq0}E_k n_k,\\
  &=&\frac{2\pi  a_s}{m} n^2\left[1 +\frac{128}{15\sqrt{\pi}}(n
    a^3_s)^{1/2}\right],
    \label{EbogLHY}
\end{eqnarray}
\ese
where the $T=0$ momentum distribution function 
\begin{eqnarray}
\label{nkt0}
n_k&=&\langle gs|\hat a_k^\dagger \hat a_k|gs\rangle
= |v_k|^2\approx_{k\rightarrow\infty}C/k^4,
\end{eqnarray}
with Tan's contact $C = \partial{\cal E}_{gs}/\partial a^{-1}_s =
16\pi^2 n^2 a^2_s[(1 + \frac{64}{3\sqrt{\pi}}(na^3_s)^{1/2}]$ and
\begin{eqnarray}
  \mu &=& \frac{4\pi  a n}{m}\left[1 +\frac{32}{3\sqrt{\pi}}(n a^3_s)^{1/2}\right].
\label{eps_gsLHY}
\end{eqnarray}

The interaction-driven condensate depletion, $n_d\equiv n-n_c$ is
given by
\begin{eqnarray}
\label{depletionbg}
n_d&=&\frac{1}{V}\sum_{\boldsymbol{k}\neq0}
n_k\approx\frac{8}{3\sqrt{\pi}}\left(n a^3_s\right)^{1/2}n,
\end{eqnarray}
and provides an important measure of the validity of the Bogoluibov
approximation that neglects the difference between $n$ and $n_c$.

Clearly, for a large gas parameter, $n a^3_s\gg 1$ the depletion is
substantial and must be accounted for. Although there is no currently
available systematic analysis in this nonperturbative limit, as we
will show in subsequent sections, an uncontrolled self-consistent
method, akin to a spherical, large-$N$
model \cite{ZinnJustin,ChaikinLubensky,SotiriadisCardy10,Sondhi13,LRunpublished}
captures important qualitative physics in this resonantly interacting
regime.

\subsection{Generalization for large scattering length}
To extend the analysis to a large $na^3_s$ we need to account (even if
approximately) for the nonlinear components of the Hamiltonian, $\hat
H_{3,4}$ neglected in the Bogoluibov model. To this end, in the spirit
of variational theory or a spherical model \cite{ChaikinLubensky}, we
replace these nonlinear operators by their ``best'' approximation in
terms of operators up to a quadratic order in fluctuation field $\hat
a$. Using Wick's theorem, we have
\bse
\begin{eqnarray}
  \hat a^{\dagger}\hat a^{\dagger}\hat a\hat a&\rightarrow& 4 \langle
  \hat a^{\dagger}\hat a\rangle \hat a^{\dagger}\hat a+\langle \hat
  a^{\dagger}\hat a^{\dagger}
  \rangle \hat a\hat a
  +\langle \hat a\hat a\rangle \hat a^{\dagger}\hat
  a^{\dagger}\nonumber\\
&&-2\langle \hat a^{\dagger}\hat a\rangle \langle \hat
  a^{\dagger}\hat a\rangle
-\langle \hat a^\dagger \hat a^\dagger\rangle \langle \hat a\hat
a\rangle,\nonumber\\
&\approx& 4 n_d\hat a^{\dagger}\hat a -2n_d^2,\\
\label{a4_a2}
\hat a^{\dagger}\hat a \hat a&\rightarrow& 2\langle \hat a^{\dagger}\hat
a\rangle \hat a \approx 2 n_d \hat a,\\
\hat a^{\dagger}\hat a^{\dagger}\hat a&\rightarrow& 2\hat
a^{\dagger}\langle \hat a^{\dagger}\hat a\rangle
\approx 2 n_d \hat a^\dagger,
\label{a3_a1}
\end{eqnarray}
\ese
where we kept the depletion density $n_d=\langle \hat a^{\dagger}\hat
a\rangle$ and neglected ``anomalous'' averages (e.g., $\langle\hat
a\hat a\rangle=0$) and high order correlators (e.g., $\langle
\hat a^{\dagger}\hat a\hat a\rangle=0$) that we expect to be subdominant.

With these we approximate $\hat H_3$ and $\hat H_4$ by a linear and
quadratic forms
\begin{eqnarray}
\hat{H}_3&=&g\int_\rv[\Psi_0 \hat a^{\dagger}\hat a^{\dagger}\hat{a}
+\Psi_0^*\hat a^{\dagger}\hat a\hat a]\to \delta \hat H_1,
\end{eqnarray}
where
\begin{eqnarray}
\delta\hat H_1&=&g\int_\rv(2\Psi_0 n_d\hat a^{\dagger}+h.c.),
\label{deltaH1}
\end{eqnarray}
and
\begin{eqnarray}
\hat{H}_4&=&\frac{g}{2}\int_\rv\hat a^{\dagger}\hat a^{\dagger}\hat
a\hat a
\to \delta\hat H_0+\delta\hat H_2,
\end{eqnarray}
where
\bse
\begin{eqnarray}
\delta\hat H_0&=&-g V n_d^2,\\
\delta\hat H_2&=&2g\int_\rv n_d \hat a^{\dagger}\hat a.
\label{H4_H01}
\end{eqnarray}
\ese 
The grand-canonical Hamiltonians now take the following forms:
$\hat{H}\approx \hat{H}'_0+ \hat{H}'_1+ \hat{H}'_2$, where
\bse
\begin{equation}
\begin{split}
\hat{H}'_0&=\hat H_0+\delta \hat H_0,\\
&=\int_\rv\left[\Psi_0^*(\hat\epsilon-\mu)\Psi_0+\frac{g}{2}n^2_c-g n^2_d\right],
\end{split}
\label{kprime0}
\end{equation}
\begin{equation}
\begin{split}
\hat{H}'_1&=\hat H_1+\delta \hat H_1,\\
&=\int_\rv\left[\hat a^{\dagger}(\hat\epsilon+gn_c+2gn_d-\mu)\Psi_0\right]+h.c.,
\end{split}
\label{kprime1}
\end{equation}
\begin{equation}
\begin{split}
\hat{H}'_2&=\hat H_2+\delta\hat  H_2,\\
&=\int_\rv \left[\hat a^{\dagger}(\hat\epsilon+2g(n_c+n_d)-\mu)\hat a+\frac{gn_c}{2}(\hat a\hat  a+\hat  a^{\dagger}\hat a^{\dagger})\right].
\end{split}
\label{kprime2}
\end{equation}
\ese 
Above, for simplicity, we have defined $n_c\equiv|\Psi_0|^2$ and
$n_d\equiv \langle \hat a^{\dagger}\hat a\rangle$ and in
Eqs.~\eqref{kprime0},\eqref{kprime1},\eqref{kprime2} have discarded the
"anomalous average" term $\tilde{m}\equiv\langle \hat a\hat a\rangle$ to satisfy
the constraint of Goldstone theorem, which requires a gapless
excitation spectrum. This amounts to the widely used Popov
approximation \cite{HFB_Popov}.

Following what was done in the last subsection, we fix the chemical
potential $\mu$ by requiring $ \hat{H}'_1=0$
\begin{equation}
\begin{split}
(\hat\epsilon+g|\Psi_0|^2+2gn_d)\Psi_0=\mu\Psi_0.
\end{split}
\end{equation}
For a uniform system, this gives
\begin{equation}
\begin{split}
\mu=gn_c+2gn_d.
\end{split}
\end{equation}
Thus we obtain the grand-canonical Hamiltonian
\begin{equation}
\begin{split}
  \hat{H}&=\int_\rv \left[\hat a^{\dagger}(\hat\epsilon+gn_c)\hat a
    +\frac{g}{2}n_c(\hat a\hat a+\hat a^{\dagger}\hat a^{\dagger})\right]-E_0\\
  &=\sum_{\boldsymbol{k}\neq0}\left[(\epsilon_k+gn_c)\hat a^{\dagger}_\kv \hat a_\kv
    +\frac{1}{2}gn_c(\hat a^{\dagger}_\kv \hat a^{\dagger}_{-\kv}+\hat a_\kv \hat a_{-\kv})\right]-E_0,
\end{split}
\label{Htotal}
\end{equation}
where $E_0/V = \frac{g}{2}n^2_c + 2gn_cn_d + g n_d^2$. It exhibits the
standard Bogoluibov form with gapless spectrum, but also approximately
accounts for a potentially strong depletion through the condensate
density $n_c$ replacing the full density $n$ as the self-consistently
determined parameter of the Hamiltonian.

\section{self-consistent analysis for strongly interacting ground
  state}
  \label{sec:ndSCgs}
Before turning to our main focus of nonequilibrium post-quench
dynamics, we examine the ground state properties of a strongly
interacting resonant Bose gas, characterized by a large scattering
length and gas parameter $n a^3_s\gg 1$. This regime lies beyond the
scope of the standard Bogoluibov theory. Nevertheless we expect to be
able to treat it qualitatively correctly (even if quantitatively
uncontrolled) by taking into account the large depletion $n - n_c > 0$
through the Hamiltonian \eqref{Htotal} and the self-consistency
condition through the atom number conservation 
\bse
\begin{eqnarray}
\label{SCcondition}
n&=&n_c+\frac{1}{V}\sum_{\boldsymbol{k}\neq0}\langle \hat a^{\dagger}_\kv\hat a_\kv\rangle,\\
&=&n_c+\frac{8}{3\sqrt{\pi}}\left({n_ca^3_s}\right)^{1/2}n_c,
\label{ncSC}
\end{eqnarray}
\ese
where in the second line we calculated the depletion by diagonalizing
\eqref{Htotal} as in Sec.~\ref{sec:BdG} of the conventional Bogoluibov
theory, but with $n_c$ replacing $n$. Such treatment is quite close in
spirit to the self-consistent Hartree-Fock approximations, and the BCS
and other mean-field gap equations.

In the dimensionless form for $\hat{n}_c=n_c/n$, the self-consistency
equation reduces to
\begin{equation}
\label{screduced}
\begin{split}
1-\hat{n}_c-\lambda {\hat{n}}^{3/2}_c=0,
\end{split}
\end{equation}
where $\lambda=8(na^3_s)^{1/2}/(3\sqrt{\pi})\equiv
8(k_na_s)^{3/2}/(3\sqrt{\pi})$, with $k_n\equiv n^{1/3}$ the mometum
scale set by the boson density $n$.

The solution to Eq.~\eqref{screduced} is illustrated in
Fig.~\ref{static_SC}. We find that the self-consistency constraint suppresses condensate depletion, leading to a higher condensate
fraction $n_c$ than the Bogoluibov approximation for the same strength
of the interaction parameter $k_n a_s$. We also observe that, as expected
the correction to Bogoluibov theory from the self-consistency
condition grows (from zero) with increasing gas parameter $k_na_s$,
thereby avoiding the spurious transition to a vanishing condensate
state appearing in the Bogoluibov theory.

\section{Dynamics for shallow quench}
\label{sec:ShallowQuench}
We now turn to nonequilibrium dynamics following a change in the
scattering length $a_s$ from its initial value $a_i$ to the final
value $a_f$, as can be realized experimentally in a Feshbach resonant
Bose gas by a change in the external magnetic
field \cite{Makotyn}. Here we assume the change is
instantaneous (sudden quench), allowing analytical analysis. In this
section, we focus on shallow quenches characterized by both
$na^3_i\ll1$ and $na^3_f\ll1$, so that the Bogoluibov approximation
remains rigorously valid.

For shallow quenches, the system is well approximated by
Hamiltonian \eqref{Hbog} with $g_i$ and $g_f$ for the initial and final
Hamiltonians, respectively, with corresponding Bogoluibov
quasi-particle bases $(\hat \alpha_\kv,\hat \alpha^{\dagger}_\kv)$ prior to the
quench and $(\hat \beta_\kv,\hat \beta^{\dagger}_\kv)$ post the quench.
Focussing on a sudden quench, the two sets of bases are related to the
atomic basis $(\hat a_\kv,\hat a^{\dagger}_\kv)$ via a pseudo-unitary
transformations
\bse
\begin{eqnarray}
\begin{pmatrix}
\hat a_\kv\\\hat a^{\dagger}_{-\kv}\end{pmatrix}
&=&
\begin{pmatrix}u^{\prime}_k&v^{\prime}_k\\v^{\prime}_k&u^{\prime}_k
\end{pmatrix}
\begin{pmatrix}\hat \alpha_\kv\\
\hat \alpha^{\dagger}_{-\kv}
\end{pmatrix},\\
\hat \Phi_\kv(0) &=&U_k(0^-)\hat \Psi_{\kv}(0^-),
\end{eqnarray}
\ese
and
\bse
\begin{eqnarray}
\label{pseudounitarytrans}
\begin{pmatrix}
\hat a_\kv\\\hat a^{\dagger}_{-\kv}
\end{pmatrix}&=&
\begin{pmatrix}u_k&v_k\\v_k&u_k\end{pmatrix}
\begin{pmatrix}\hat \beta_\kv\\\hat \beta^{\dagger}_{-\kv}\end{pmatrix},\\
\hat \Phi_\kv(0) &=&U_k(0^+)\hat \Psi_{\kv}(0^+),
\end{eqnarray}
\ese
where
\bse
\begin{eqnarray}
u^{\prime}_k
&=&\sqrt{\frac{1}{2}\left(\frac{\epsilon_k+ng_i}{E_{ki}}+1\right)},\;\;
v^{\prime}_k=-\sqrt{\frac{1}{2}\left(\frac{\epsilon_k+ng_i}{E_{ki}}-1\right)},
\label{uvprime}\nonumber\\
&&\\
u_k&=&\sqrt{\frac{1}{2}\left(\frac{\epsilon_k+ng_f}{E_{kf}}+1\right)},\;\;
v_k=-\sqrt{\frac{1}{2}\left(\frac{\epsilon_k+ng_f}{E_{kf}}-1\right)},
\label{uv}\nonumber\\
\end{eqnarray}
\ese
define Bogoluibov transformations for Hamiltonians $ \hat{H}_i$ (with
interaction $g_i\equiv g(0^-)$) before and $ \hat{H}_f$ (with interaction
$g_f\equiv g(0^+)$) after the quench, respectively. The corresponding
excitation spectra are
\begin{equation}
\label{Ek}
E_{ki}=\sqrt{{\epsilon_k}^2+2ng_i\epsilon_k},\;\;
E_{kf}=\sqrt{{\epsilon_k}^2+2ng_f\epsilon_k},
\end{equation}
and the two quasi-particle bases are related by
\begin{eqnarray}
\label{betaalpha}
\begin{pmatrix}
\hat \beta_\kv\\\hat \beta^{\dagger}_{-\kv}\end{pmatrix}&=&U_k^{-1}(0^{+})U_k(0^{-})
\begin{pmatrix}\hat \alpha_\kv\\ \hat \alpha^{\dagger}_{-\kv}\end{pmatrix},\nonumber\\
&=&\begin{pmatrix}\cosh\Delta\theta_k&\sinh\Delta\theta_k\\
\sinh\Delta\theta_k&\cosh\Delta\theta_k\end{pmatrix}
\begin{pmatrix}\hat \alpha_\kv\\\hat \alpha^{\dagger}_{-\kv}\end{pmatrix},
\end{eqnarray}
with
\begin{equation}
\label{eq65}
\Delta\theta_k
=\frac{1}{2}\cosh^{-1}\left[\oh\left(\frac{E_{kf}}{E_{ki}}
+\frac{E_{ki}}{E_{kf}}\right)\right].
\end{equation}

We take the initial state $|0^-\rangle$ to be the ground state of the
pre-quenched Hamiltonian $\hat H_i$ \cite{footnote}, and thus a vacuum of
$ \hat\alpha$ quasi-particles, $\hat\alpha_\kv|0^-\rangle = 0$. At finite
temperature this generalizes to Bose-Einstein distribution of
$\hat \alpha$ quasi-particle occupation,
\begin{eqnarray}
\langle\hat{\alpha}^{\dagger}_{\kv}\hat{\alpha}_\kv\rangle_{0^-}
=\frac{1}{e^{E_{k i}/ T}-1}.
\end{eqnarray}

Because experiments probe physical observables expressed in terms of
atomic operators, we need to compute time evolution of $ \hat\Phi_\kv(t) =
( \hat a_\kv(t), \hat a^{\dagger}_\kv(t))$. Using free post-quench evolution of
$ \hat\beta$ quasi-particles
\begin{equation}
\begin{pmatrix}
\hat \beta_\kv(t)\\
\hat \beta^{\dagger}_{-\kv}(t)\end{pmatrix}
=\begin{pmatrix}e^{-iE_{kf}t}&0\\0&e^{iE_{kf}t}
\end{pmatrix}\begin{pmatrix}\hat \beta_\kv(0)\\\hat \beta^{\dagger}_{-\kv}(0)\end{pmatrix}
\equiv T_k(t)\hat \Psi_\kv(0^+),
\label{quenchevo}
\end{equation}
the relation \rf{betaalpha}, together with the simplicity of matrix
elements of $\hat \alpha$ quasi-particles in the pre-quench ground state
(vacuum of $\hat\alpha_\kv$), we find
\bse
\label{transfm}
\begin{eqnarray}
\begin{pmatrix}\hat a_\kv(t)\\\hat a^{\dagger}_{-\kv}(t)\end{pmatrix}
&=&U_k(0^+)
\begin{pmatrix}\hat \beta_\kv(t)\\\hat \beta^{\dagger}_{-\kv}(t)\end{pmatrix},\\
&=&U_k(0^+)T_k(t)
\begin{pmatrix}\hat \beta_\kv(0)\\\hat \beta^{\dagger}_{-\kv}(0)\end{pmatrix},\\
&=&U_k(0^+)T_k(t)U_k^{-1}(0^{+})U_k(0^{-})
\begin{pmatrix}\hat \alpha_\kv(0)\\\hat \alpha^{\dagger}_{-\kv}(0)\end{pmatrix},
\nonumber\\
&&\\
\hat \Phi_\kv(t) &=&U_k(t)\hat \Psi_{\kv}(0^-)=R_k(t)U_k(0^-)\hat \Psi_{\kv}(0^-),\nonumber\\
\end{eqnarray}
\ese
where the matrix
\bse
\begin{eqnarray}
\hspace{-0.4cm}
R_{ij}(t)&=&U_{il}T_{lm}(t)U_{mn}^{-1},\\
&=& (\cos E_{kf} t) I_{ij} + i\frac{\sin E_{kf} t}{E_{kf}}
\begin{pmatrix}
\epsilon_k + g_f n & g_f n\\
-g_f n & -\epsilon_k - g_f n \\
\end{pmatrix},\nonumber\\
\label{Revolv}
\end{eqnarray}
\ese
evolves the initial Bogoluibov spinor
$\left(u_k(0^-),v_k(0^-)\right)\rightarrow \left(u_k(t),v_k(t)\right)$,
and
\bse
\label{Uevolv}
\begin{eqnarray}
U_k(t) &=& U_k(0^+)T_k(t)U_k^{-1}(0^{+})U_k(0^{-}),\\
&=&\begin{pmatrix}u_ke^{-i E_{kf} t}&v_k e^{i E_{kf} t}\\
v_k e^{-i E_{kf} t}&u_k e^{i E_{kf} t}\end{pmatrix}
\begin{pmatrix}\cosh\Delta\theta_k&\sinh\Delta\theta_k\\
\sinh\Delta\theta_k&\cosh\Delta\theta_k\end{pmatrix}.\nonumber\\
\end{eqnarray}
\ese
%

Having derived the evolution of the atomic fields $\hat \Phi_\kv(t) =
(\hat a_\kv(t), \hat a^{\dagger}_\kv(t))$, we can now compute the basic atomic
bilinear correlator (supressing the momentum $\kv$ argument on the
right hand-side):
\begin{eqnarray}
\label{Cij}
C_\kv^{ij}(t,t')&=&\langle \hat\Phi_i^\dagger(t) \hat\Phi_j(t')\rangle,\nonumber\\
 &=&\langle \hat\Psi^\dagger_{m}(0^-)
U^\dagger_{m i}(t)U_{j n}(t') \hat\Psi_{n}(0^-)\rangle,\nonumber\\
&=&U^\dagger_{m i}(t) N_{m n} U_{j n}(t'),
\end{eqnarray}
in terms of the pre-quench ($t = 0^-$) quasi-particle occupation
matrix
\bse
\label{Nmn}
\begin{eqnarray}
N_{m n}&=&\langle \hat\Psi^\dagger_{m}(0^-) \hat\Psi_{n}(0^-)\rangle,\\
&=&\begin{pmatrix}
\langle\hat \alpha^\dagger_\kv\hat \alpha_\kv\rangle 
&\langle\hat \alpha^\dagger_\kv\hat \alpha^\dagger_{-\kv}\rangle\\
\langle\hat \alpha_{-\kv}\hat \alpha_\kv\rangle & 
\langle\hat \alpha_{-\kv}\hat \alpha^\dagger_{-\kv}\rangle\\
\end{pmatrix}_{mn},\\
&=&\begin{pmatrix}
n_k(0^-)& 0 \\
0& n_{-k}(0^-) + 1\\
\end{pmatrix}_{mn},\\
&=&\begin{pmatrix}
0 & 0 \\
0&  1\\
\end{pmatrix}_{mn},\ \ \text{for $T=0$},
\end{eqnarray}
\ese
from which physical observables, such as the momentum distribution
function, structure function, RF spectroscopy signal, and many others
can be obtained. We turn to their computation in the following
subsections.

\subsection{Time of flight: momentum distribution function}
Time of flight measurements, where a gas is released from its trap and
its density profile is measured at long times, is one of the central
experimental probes dating back to the realization of BEC in dilute
alkali gases \cite{CornellWiemannBEC1995,KetterleBEC1995}. A straightforward
analysis demonstrates \cite{RSpolarized}, that at long times the density
profile is proportional to the momentum distribution function.
At $T=0$, that is our main focus here, we obtain
\begin{eqnarray}
  n_k(t)&=&\langle 0^-| \hat a^\dagger_\kv(t) \hat a_\kv(t)|0^-\rangle=C_k^{11}(t,t),\nonumber\\
  &=&|(u_ke^{-i E_{kf} t}\sinh\Delta\theta_k+v_ke^{i E_{kf} t}\cosh\Delta\theta_k)|^2,\nonumber\\
  &=&\frac{\epsilon_k+g_i n
    +\frac{2g_f(g_f-g_i)n^2}{\epsilon_k+2g_fn}\sin^2(E_{kf} t)}{2\sqrt{\epsilon_k(\epsilon_k+2g_i n)}}-\frac{1}{2},\quad\quad\quad
\label{nkbogoliubov}
\end{eqnarray}
at $t=0$ reducing to the ground-state momentum distribution
Eq.~\eqref{nkt0}, as expected by continuity of evolution.  Rescaling
momentum as $\hat{k}=k \xi\equiv k/\sqrt{2m n g_f}$ and time as
$\hat{t}=t/t_0\equiv t n g_f$, we obtain the momentum distribution in
terms of dimensionless variables as
\begin{equation}
\begin{split}
n_{\hat{k}}(\hat{t})=\frac{[\hat{k}^2+\sigma
+\frac{2(1-\sigma)}{\hat{k}^2+2}\sin^2(\hat{t}\sqrt{\hat{k}^2(\hat{k}^2+2)})]}{2\sqrt{\hat{k}^2(\hat{k}^2+2\sigma)}}-\frac{1}{2},\\
\end{split}
\label{nkbogoliubov1}
\end{equation}
where the initial-to-final scattering length ratio, $\sigma\equiv
a_i/a_f$ characterizes the ``depth'' of the quench.

\begin{figure}[htb]
\centering
\includegraphics[width=0.45\textwidth]{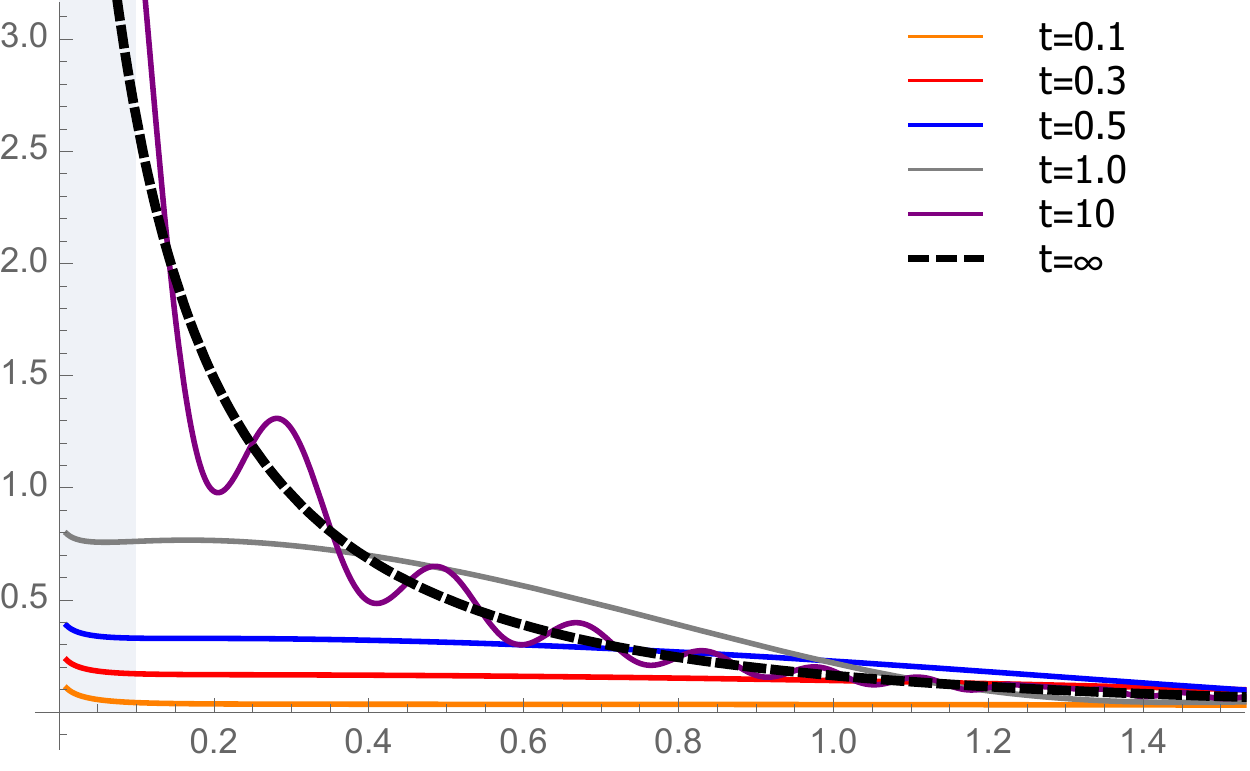}
\caption{(Color online) Time evolution of the column momentum
  distribution defined in the text following a scattering length
  quench from $0.1a_f\to a_f$, referring to Eq.~\eqref{nkbogoliubov1} using Bogoluibov approximation. It illustrates the initial narrow
  momentum distribution (lowest curve) evolving to a much broader
  momentum distribution (highest curve), corresponding to a
  pre-thermalized steady state. Here momentum and time are rescaled with $\xi\equiv 1/\sqrt{2 m n g_f}$ and $t_0\equiv 1/(n g_f)$, respectively. The grey region indicates a range of momenta not resolved in JILA experiments, due to initial inhomogeneous real space density
profile and finite trap size.}
\label{nkt_bog}
\end{figure}

The column momentum distribution $\tilde{n}_{\hat{k}}(\hat{t})\equiv
\int d\hat{k}_z n_{\hat{k}}(\hat{t})$ is a more
experimentally relevant quantity that we plot at different times in
Fig.~\ref{nkt_bog}.
We observe that starting with a narrow BEC peak, the column momentum
distribution function quickly broadens and develops a large momentum
tail. The momentum distribution approaches a pre-thermalized
steady-state $\tilde{n}^{ss}_k$ from high momenta, with momenta $k >
k_{pth}(t)$ taking time $t_{pth}(k)\approx 1/E_{kf}$ to
pre-thermalize \cite{footnote}.  Thus we obtain
\begin{equation}
\begin{split}
t_{pth}(\hat{k})=1/\sqrt{\hat k^2(\hat k^2+2)},
\end{split}
\end{equation}
consistent with experiments \cite{Makotyn} scaling as $1/k$ and $1/k^2$
at small and large momenta, respectively.

\begin{figure}[htb]
\centering
\includegraphics[width=0.45\textwidth]{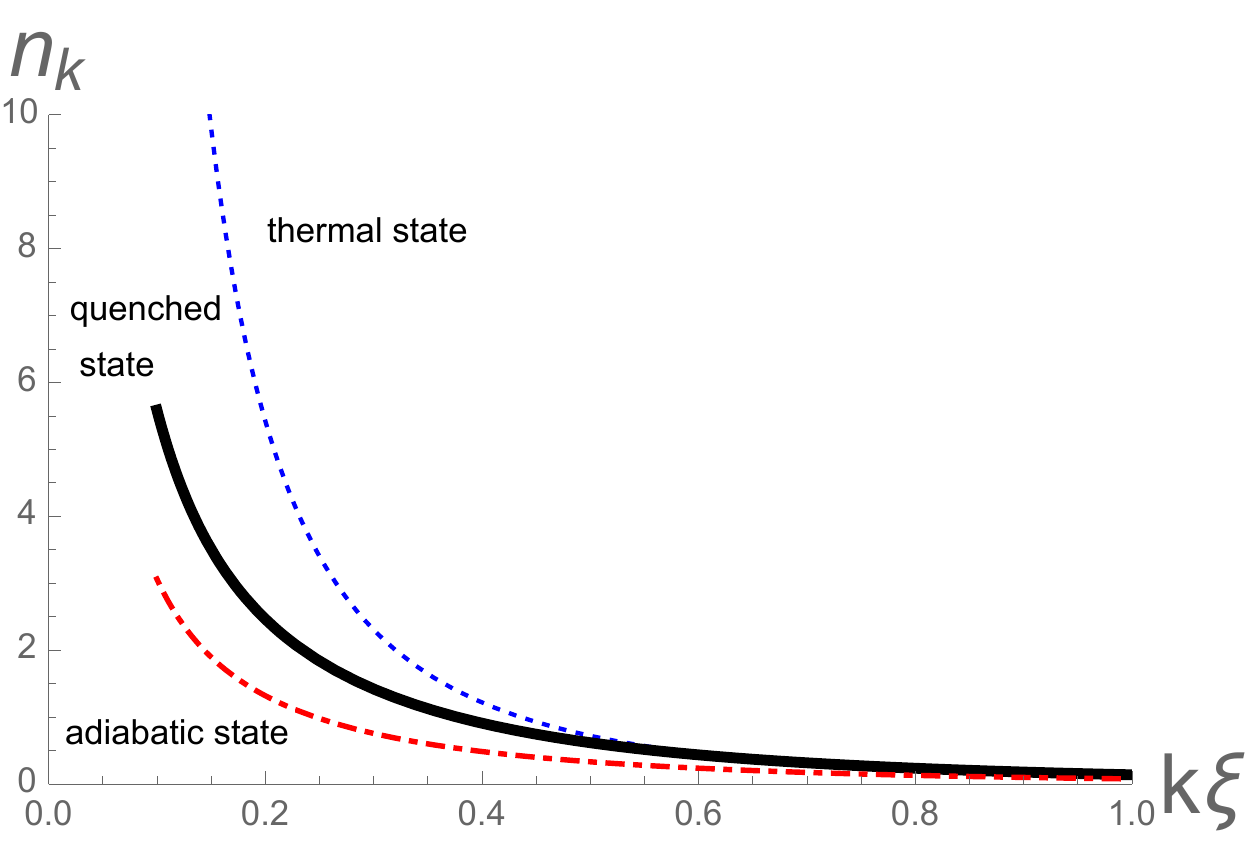}
\caption{(Color online) Quenched steady-state momentum distribution
  function $n^{ss}_k$ following a scattering length quench
  $a_i=0.1a_f\to a_f$ (thick black curve), as compared to the ground
  state momentum distribution at $a_f$ (dash-dotted red) and the corresponding
  Bogoluibov thermalized distribution (dotted blue) at temperature $T =
  0.45ng_f$.}
\label{nk_ssFig}
\end{figure}

The steady-state momentum distribution, $n^{ss}_k$ for a $a_i = 0.1
a_f\rightarrow a_f$ is plotted in Fig.~\ref{nk_ssFig} and compared to the ground state $n_k$ for the same $a_f$ as well as thermal state $n_k$ at finite temperature. We observe that this
steady-state momentum distribution lies above the ground state one,
indicating that even in the long time limit the post-quench system
remains in the excited states, as required by energy
conservation. However, it also differs significantly from the
corresponding finite-temperature thermal-equilibrium distribution,
$n^T_k =(u^2_k+v^2_k)\langle\hat{\alpha}^{\dagger}_{\kv}\hat{\alpha}_\kv\rangle_{0^-}+v^2_k
=1/(e^{E_{kf}/T}-1)+v_k^2\coth(E_{kf}/2T)$, demonstrating that even in the long time, stationary state limit the system is only {\em pre-thermalized}. This is expected because of the
quadratic, fully integrable form of the Bogoluibov Hamiltonian.  The
latter guarantees the absence of scattering of the Bogoluibov
quasi-particles $\hat\beta_\kv$, with a conserved momentum distribution
function, that is directly related to the initial distribution by
\rf{betaalpha}.

\begin{figure}[htb]
\centering
\includegraphics[width=0.45\textwidth]{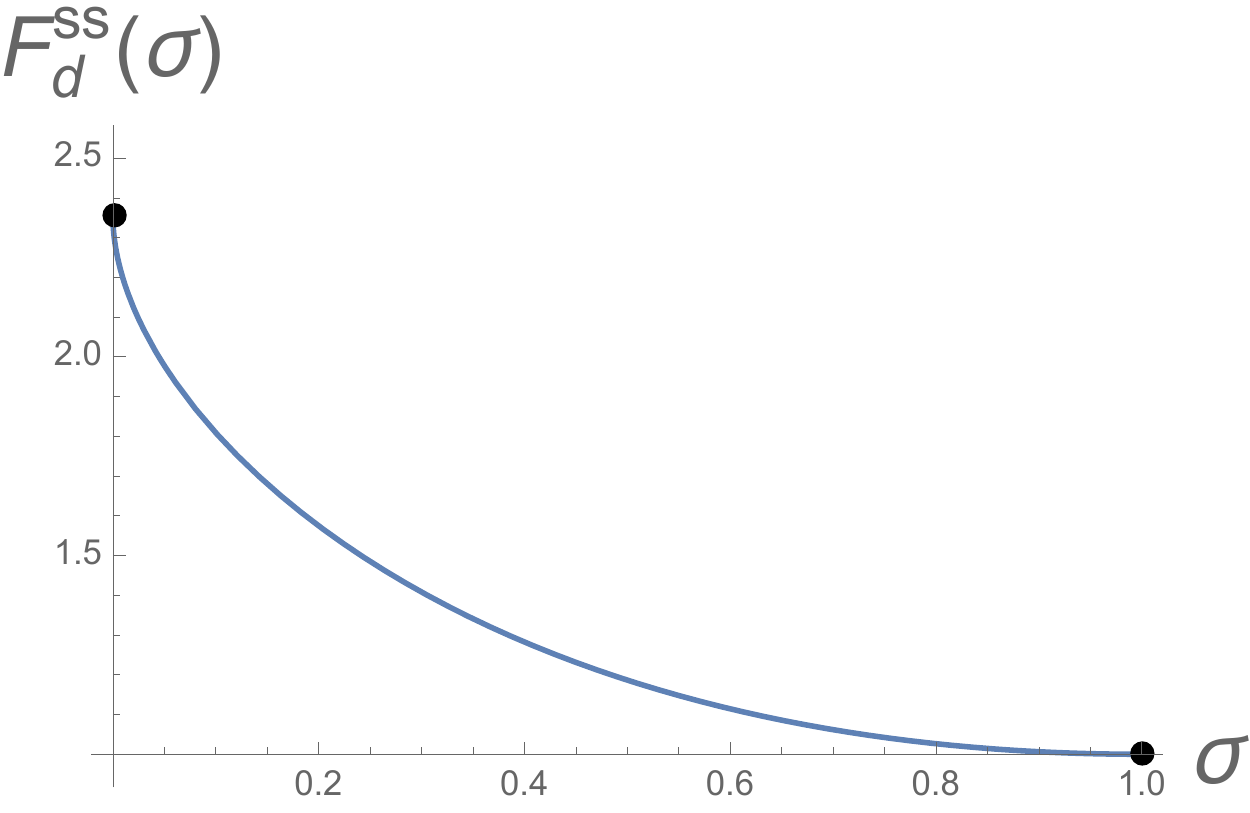}
\caption{(Color online) Quenched steady-state depletion enhancement factor $F^{ss}_d(\sigma)$ above the corresponding ground state value as a
  function of $\sigma=a_i/a_f$ following a quench from $a_i\to a_f$. The two dots coprrespond to the
  maximum enhancement $F^{ss}_d(0)=3\pi/4$ (quenching a non-interacting gas
  or quenching to unitarity) and minimum enhancement $F^{ss}_d(1)=1$ (no
  quench), respectively. }
\label{ndFactor}
\end{figure}

A simpler measure of the post-quench dynamics is the evolution of the
condensate depletion, obtained from the momentum distribution
function, $n_k(t)$, \eqref{nkbogoliubov},
\begin{eqnarray}
\label{depquenchbg}
n_d(t)&=&\sum_\kv n_k(t) = V\int\frac{d^3k}{(2\pi)^3}n_k(t),\nonumber\\
 &=& n^0_d F_d(\sigma,t),
\end{eqnarray}
where $n^0_d=8/(3\sqrt{\pi})(n a_f^3)^{1/2}$ is the ground-state
depletion for $a_s = a_f$.
\begin{eqnarray}
\label{depquenchbg1}
F_d(\sigma,t)&=&(\sigma)^{3/2}+\frac{3}{2}\sqrt{1-\sigma}\mathrm{Arccos}(\sqrt{\sigma})
\nonumber\\
&-&\frac{3\sqrt{2}}{2}\int y
dy\frac{(1-\sigma)\cos(2ty\sqrt{y^2+2})}{(y^2+2)(y^2+2\sigma)^{1/2}}
\ \ \ \ \ \ \ \
\end{eqnarray}
is the nonequilibrium depletion enhancement factor above the corresponding ground state, that interpolates
between $\sigma^{3/2}$ (giving the initial depletion at $t=0^-$ for
$a_s = a_i$) and the asymptotic depletion
\begin{equation}
\label{Ffactor}
\begin{split}
F^{ss}_d(\sigma)\equiv F_d(\sigma,t\to\infty)=\sigma^{3/2}+\frac{3}{2}\sqrt{1-\sigma}\mathrm{Arccos}(\sqrt{\sigma})
\end{split}
\end{equation}
of the pre-thermalized state, plotted in Fig.~\ref{ndFactor}.

\begin{figure}[htb]
\centering
\includegraphics[width=0.45\textwidth]{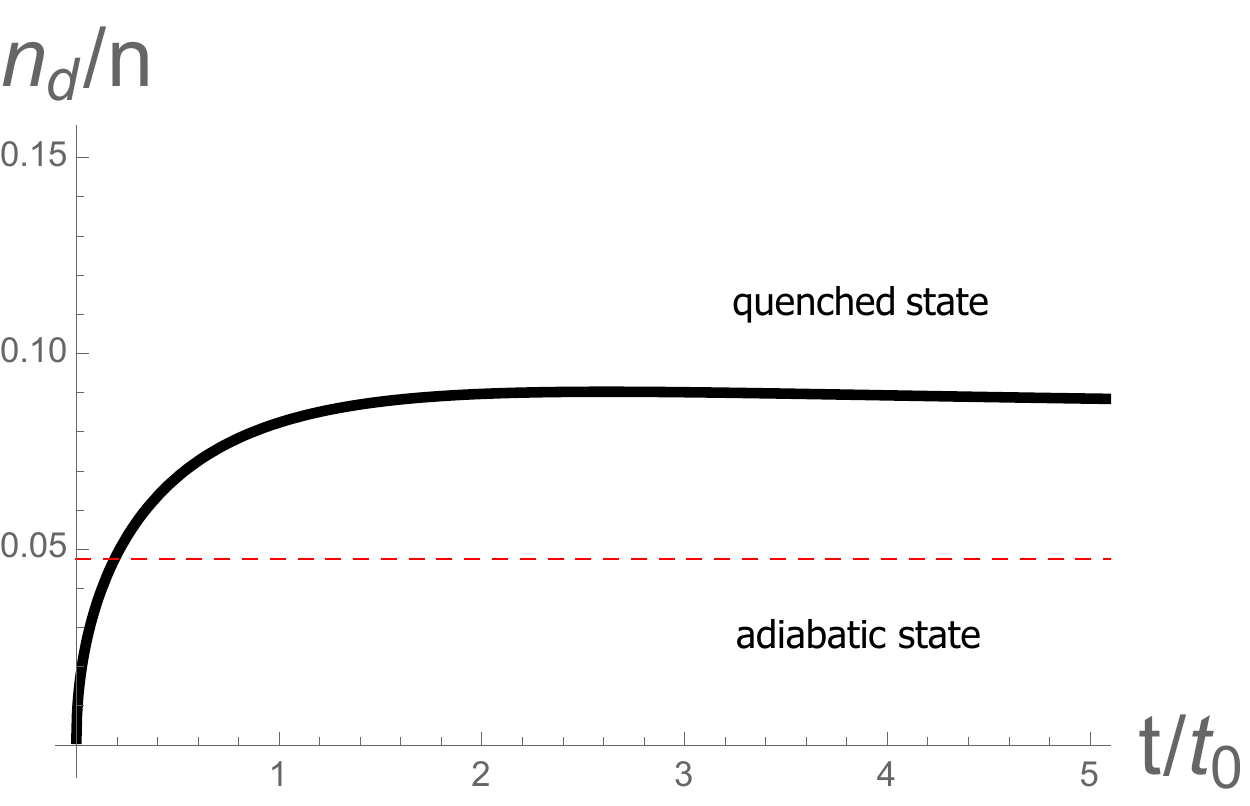}
\caption{(Color online) Post quench dynamics of the condensate depletion fraction as a function of
  rescaled time in units of pre-thermalization timescale $t_{0}= \hbar/(n g_f)=m/(4\pi a_f n\hbar )$ (solid black curve), following a scattering length quench from $0.1a_f\to a_f$ with $k_na_f=0.1$ (where $k_n\equiv n^{1/3}$), as compared to the ground state depletion at $k_na_f$ (dashed red line). For a typical $^{85}$Rb
  experiment with $n=5\times10^{12} cm^{-3}$, $a_f=1100a_0$ (here $a_0=5.29\times10^{-11}m$ is the Bohr radius), $t_{0}\approx360\mu
  s$.}
\label{ndt_bog}
\end{figure}

As is clear from the asymptotics of $F_d(\sigma,t)$ defined by
\rf{Ffactor} and illustrated in Fig.~\ref{ndt_bog}, the depletion
fraction monotonically increases as $\sqrt{t}$ over a characteristic
time
\begin{eqnarray}
t_{pth}\approx \frac{1}{n g_f},
\end{eqnarray}
approaching its asymptotic pre-thermalized value, that is always higher
than that of the ground state with the same scattering length $a_s=
a_f$.
The quenched steady-state depletion enhancement,
$F^{ss}_d(\sigma)$ monotonically increasing with
decreasing $\sigma$ (deeper quench), reaching a minimum at $\sigma=1$
(no quench), and exhibiting a maximum at $\sigma = 0$, corresponding
to initially noninteracting gas or a quench deep into unitary regime,
where $a_f\to\infty$. We note, however, that the latter
strongly-interacting resonant regime, clearly lies outside of the
perturbative Bogoluibov theory.
We will treat this $k_n a_f\gg 1$ nonperturbative regime in a subsequent
section, using an approximate self-consistent treatment.

\subsection{Bragg spectroscopy: structure function}
A two-time structure function $S_\qv(t,t')=\langle\delta \hat
n(-\qv,t)\delta \hat n(\qv,t')\rangle$ is another central probe of the
nonequilibrium dynamics of degenerate atomic gases. It can be measured
via Bragg spectroscopy through a stimulated two-photon
transitions \cite{Papp}, and via a correlation function of a measured
density excitation, $\delta\hat n(\qv,t)$ at momentum $\qv$ and time
$t$ \cite{ChinGurarie13}.  The former thus allowed measurements of the
excitation spectrum of a strongly interacting ($^{85}$Rb) BEC, near
unitarity ($n a^3_s\gg 1$), demonstrating a large deviation from the
Bogoluibov and Lee-Huang-Yang (LHY) prediction of Sec. \rf{sec:BdG}. The latter
technique was used to characterize dynamics of a Feshbach-resonant
Cesium gas, following a shallow quench in its scattering
length \cite{ChinGurarie13}.

With current experiments in mind, for simplicity we focus on the
equal-time $t=t'$ structure function (nontrivial for nonequilibrium
dynamics),
\begin{eqnarray}
  S_\qv(t) &=& \langle\delta \hat n(-\qv,t)\delta \hat n(\qv,t)\rangle,\nonumber\\
  &=&\frac{1}{V}\int_{\rv,\rv'}e^{i\qv\cdot(\rv-\rv')}
  \langle\hat\psi^\dagger(\rv,t)\hat\psi(\rv,t)
  \hat\psi^\dagger(\rv',t)\hat\psi(\rv',t)\rangle_c
,\nonumber\\
&\approx&S^0_\qv(t) + \delta S^B_\qv(t),
\label{Sqt}
\end{eqnarray}
where
\bse
\begin{eqnarray}
S^0_\qv(t)
&=&n_c\left[\langle \hat a_\qv^\dagger(t) \hat a_\qv(t)\rangle + \langle
\hat a_{-\qv}(t) \hat a_{-\qv}^\dagger(t)\rangle\right.\nonumber\\
&&\left.+ \langle\hat a_{-\qv}(t) \hat a_\qv(t)\rangle 
+ \langle \hat a_\qv^\dagger(t) \hat a_{-\qv}^\dagger(t)\rangle\right],\\
&=&n_c\left[C_q^{11}(t) + C_q^{22}(t) + C_q^{21}(t) +
  C_q^{12}(t)\right],\nonumber\\
\label{S0qt}
\end{eqnarray}
\ese
and 
\begin{eqnarray}
\delta S^B_\qv(t) 
&=&\frac{1}{V}\sum_{\kv\neq 0}\left[\langle\hat a_\kv^\dagger(t) \hat a_\kv(t)\rangle 
\langle \hat a_{\kv-\qv}(t) \hat
a_{\kv-\qv}^\dagger(t)\rangle\right.\nonumber\\
&&\left.
+\langle\hat a_{\kv}^\dagger(t) \hat a_{-\kv}^\dagger(t)\rangle
\langle \hat a_{\kv-\qv}(t)\hat a_{-\kv+\qv}(t)\rangle\right],
\nonumber\\
&=&\frac{1}{V}\sum_{\kv\neq 0}\left[C_k^{11}(t)C_{-k+q}^{22}(t) 
+ C_k^{(12)}(t)C_{-k+q}^{(21)}(t)\right],\nonumber\\
\label{deltaSqt}
\end{eqnarray}
are, respectively the quadratic and quartic contribution to
$S_\qv(t)$, both computed within the Bogoluibov approximation.

Utilizing the Bogoluibov analysis of the nonequilibrium quenched
dynamics from the previous subsection, (Eqs.\rf{transfm}, \rf{Revolv},
\rf{Uevolv}, \rf{Cij}, \rf{Nmn}) the leading quadratic contribution
to $S^B_\qv(t)$ is given by \cite{ChinGurarie13}
\begin{eqnarray}
S^0_\qv(t)&=&
S^0_q\left[1 + \frac{E_{qi}^2-E_{qf}^2}{E_{qf}^2}\sin^2(E_{qf}t)\right],
\end{eqnarray}
where as a check, at initial time $S_\qv^0(t=0)$ and/or for no-quench
$g_i=g_f$ above expression reduces to the pre-quench $t = 0^-$
structure function,
\begin{eqnarray}
S^0_q&=& n\frac{\epsilon_{q}}{E_{qi}}\coth\left(\oh\beta E_{qi}\right),
\end{eqnarray}
at temperature $ T = 1/\beta$. 

In dimensionless units $\hat{q}=q/\sqrt{2mng_f}$, $\hat{t}=ng_f t$ and $\hat{\beta}=ng_f\beta$, it is given by
\begin{widetext}
\begin{eqnarray}
\hspace{-0.2cm}
S_{\hat{q}}^0(\hat t)
=\frac{\hat{q}\coth(\hat{\beta}\hat{q}\sqrt{\hat{q}^2 + 2\sigma})}{\sqrt{\hat{q}^2+2\sigma}}
\left[1-\frac{2(1-\sigma)}{\hat{q}^2+2}
\sin^2\left(\hat{t}\hat{q}\sqrt{\hat{q}^2+2}\right)\right]\ \ \ \ \ \
\label{equationstrucutrebdg}
\end{eqnarray}
\end{widetext}
and plotted in Fig.\rf{Braggspectrobdg}.
\begin{figure}[htb]
\centering
\includegraphics[width=0.45\textwidth]{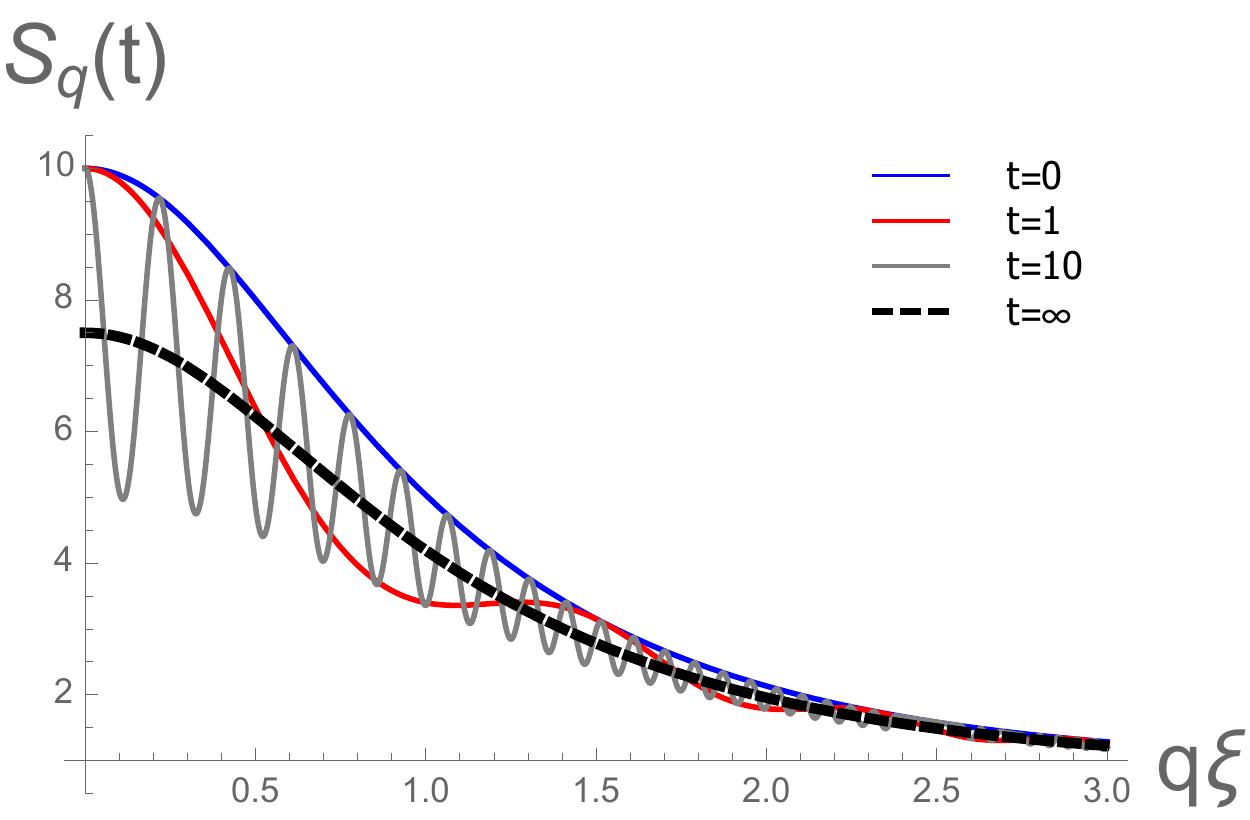}
\caption{(Color online) Time evolution of the structure function
  $S_\qv(t)$ defined in the text following a scattering length quench
  from $0.5a_f\to a_f$, referring to Eq.~\eqref{equationstrucutrebdg}
  using Bogoluibov approximation. It illustrates, following the
  quench, that the initial ground state structure function (highest
  curve) develops oscillations and becomes lower, and after some
  pre-thermalization timescale approaches the steady-state
  distribution (lowest dashed black curve). Here momentum and time are
  rescaled by $\xi\equiv 1/\sqrt{2 m n g_f}$ and $t_0\equiv 1/(n
  g_f)$, respectively. The temperature $T=10ng_f$, for a typical $^{85}$Rb
  experiment with $n=5\times10^{12} cm^{-3}$, $a_f=1100a_0$, corresponds to 16 $nK$.}
\label{Braggspectrobdg}
\end{figure}

Utilizing above Bogoluibov analysis, we have further shown that the
higher-order correction, $\delta S_\qv^B(t)$ in 3d at $T=0$ is given
by
\begin{eqnarray}
\hspace{-0.2cm}
\delta S_\qv^B(0)
&=&\int_\kv\left[u_{ki}v_{ki}u_{-\kv+\qv,i}v_{-\kv+\qv,i}
+u_{ki}^2v_{-\kv+\qv,i}^2\right],
\nonumber\\
&=&\int_\kv\left[\frac{g_i^2n^2 +
(\varepsilon_{ki}+E_{ki})(\varepsilon_{-\kv+\qv,i}-E_{-\kv+\qv,i})}
{4E_{ki}E_{-\kv+\qv,i}}\right],\nonumber\\
&\approx&\frac{g_i^2 n^2}{2 
c_i^2 }\int_\kv\left[\frac{1}{k^2+\xi_i^2k^4}+O(q)\right],\nonumber\\
&\propto&n(n a_i^3)^{1/2}\left[1+O(q)\right],
\end{eqnarray}
and for weak interaction ($na^3_s\ll1$) it is subdominant to $S_\qv^0(t)$. It can, however,
become important at finite temperature, lower dimensions and strong interactions.

\subsection{RF spectroscopy}
\label{sec:RFspectroscopy}
Radio frequency (RF) spectroscopy is another important probe that has
been fruitfully utilized to study spectroscopy and dynamics of
resonant Fermi \cite{JinBCS-BEC} and Bose \cite{Wild} gases. Quite closely
related to photoemission spectroscopy of solid state materials, the RF
signal is the number of atoms $N_b(\omega_{RF})$, that undergoes a
hyperfine transition from the many-body state of interest, $E_a$ to a
weakly interacting state $E_b = E_a + \omega_0$, in response to the
stimulated RF pulse at frequency $\omega_{RF}$.

For a weak RF pulse, the governing Hamiltonian
\begin{eqnarray}
  \hat H&=&\hat H(\hat a_\kv,\hat a^\dagger_\kv)+
  \sum_{\kv}(\epsilon_\kv + \omega_0)\hat b_\kv^\dagger \hat b_\kv +
  \sum_{\kv} I(t)\hat b_\kv^\dagger \hat a_\kv + h.c.,\nonumber\\
  &\equiv& \hat H_0 + \hat H_{RF}(t),
\end{eqnarray}
is a sum of the interacting Hamiltonian for the system of $\hat a_\kv$
bosons studied in previous subsections, the noninteracting vacuum
Hamiltonian for the $\hat b_\kv$ bosons, and the RF pulse coupling
operator $\hat H_{RF}(t)$ that drives the transitions between the two
hyperfine states, allowing a conversion of $\hat a_\kv$ into $\hat
b_\kv$.

The RF spectroscopy signal $N_{b}(\omega)$ measures the number of $\hat b$
atoms transferred for an RF pulse at frequency $\omega$. It can be
evaluated via $N_{b}(\omega)=\int_0^\infty dt \langle \hat J(t)\rangle$,
where $\hat J(t)$ is the $\hat a\rightarrow\hat  b$ ``current'' operator
\begin{eqnarray}
\hat J(t)\equiv \dot{\hat N}_b 
&=& -i\sum_\kv [\hat b_\kv^\dagger \hat b_\kv, \hat H],\nonumber\\
&=&-i\sum_\kv \left[I(t) \hat b_\kv^\dagger \hat a_\kv 
- I^*(t) \hat a_\kv^\dagger \hat b_\kv\right].\ \ \ \ 
\end{eqnarray}
%
%
%

Appropriate for experiments, we focus on a weak RF pulse and calculate
the response signal perturbatively in $I(t)$, working in the
interaction representation, with $\hat J_I(t) = e^{i\int_0^tdt' \hat
  H_0}\hat J e^{-i\int_0^tdt' \hat H_0}$,
\begin{eqnarray}
\langle \hat J(t)\rangle
&=&-i\int_0^t dt'\langle\psi|\left[\hat J_I(t), \hat H_{RF}^I(t')\right]|\psi\rangle,\\
&=&\int_0^t dt'\sum_{\kv}I^*(t')I(t)\langle \alpha_0|
\hat a_{\kv}^\dagger(t')\hat a_\kv(t)| \alpha_0\rangle\nonumber\\
&&\times e^{i(\epsilon_k +  \omega_0)(t-t')} + c.c..
\label{Jsignal}
\end{eqnarray}
Guided by the experimental protocol \cite{Makotyn}, above we have taken
the initial $t=0^-$ state $|\psi\rangle=|\alpha_0\rangle|0_b\rangle$
to be a product of a vacuum of $\hat b$ atoms, $|0_b\rangle$ and a SF
condensate of $\hat a$ atoms, $|\alpha_0\rangle$, a vacuum of the
Bogoluibov quasi-particles, $\hat \alpha_\kv|\alpha_0\rangle = 0$ for the
pre-quench interaction $g_i$.  The analysis can be straightforwardly
generalized to other initial conditions and finite temperature.

It is clear from \rf{Jsignal} that the RF signal is not generically
proportional to the momentum distribution function $n_k(t)=\langle
\hat a_{\kv}^\dagger(t)\hat a_\kv(t)\rangle$. The latter requires a sufficiently
narrow pulse so as to keep $t\approx t'$. Furthermore, a narrow
excitation bandwidth is required. Under these conditions indeed we
expect that at time $t$ the number of atoms $\hat b_\kv$ produced by the RF
pulse is proportional to the number of atoms $\hat a_\kv$ with momentum
$\kv$, such that the resonance condition $E_{kf} - \epsilon_k - \omega_0
= \omega_{RF}$ is satisfied.

Following the experiment \cite{Wild}, we take the RF pulse to be a
real part of
\begin{eqnarray}
I(t) =I_0 e^{-(t-t_0)^2/\tau^2}e^{-i\omega_{RF}t},
\end{eqnarray}
with a carrier frequency $\omega_{RF}$ and a Gaussian envelope of
width $\tau\gg 1/\omega_{RF}$, ensuring that the excitation is at a
well-defined frequency. At the same time, in order to probe the
evolving condensate dynamics at a specific time $t$, a short
pulse that is narrow on the time scale of the ramp time (that can be
made as short as a few microseconds) and on the characteristic
many-body time scale (experimentally on the order of few hundred
microseconds) that controls the condensate evolution, is required. In JILA experiment \cite{Wild},
the width $\tau$ ranges from $25\mu s$ to $200\mu s$ with $\omega_{RF}\approx 2\pi\times 50$kHz.

From the analysis of the previous section, the correlator inside
Eq.~\eqref{Jsignal} is given by
\begin{eqnarray}
&&\hspace{-0.85cm}
\langle \hat a^{\dagger}_k(t')\hat a_k(t)\rangle=C_k^{11}(t',t),\ \ \ \ \ \ \ \\
&=&u_k^2\sinh^2\Delta \theta_k e^{iE_{k f}(t'-t)}
+v^2_k \cosh^2\Delta \theta_ke^{-iE_{k f}(t'-t)}\nonumber\\
&&+u_kv_k\cosh\Delta \theta_k \sinh\Delta \theta_k
(e^{iE_{k f}(t'+t)}+e^{-iE_{k f}(t'+t)}).\nonumber
\label{Jcorrelator}
\end{eqnarray}
Using it inside Eq.~\eqref{Jsignal} and leaving the detailed analysis
to Appendix~\ref{appendix:RFspectroscopy}, in the limit of $t\gg t_0\gg \tau\gg\omega_{RF}^{-1}$ we
obtain
\begin{widetext}
\begin{eqnarray}
N_b(\omega_{RF})
&=&2\pi \tau^2 I^2_0
\sum_{\kv}\left[e^{-\frac{1}{2}(\epsilon_k+ \omega_0-\omega_{RF}-E_k)^2\tau^2}
u^2_k \sinh^2\Delta \theta_k
+e^{-\frac{1}{2}(\epsilon_k + \omega_0-\omega_{RF}+E_k)^2\tau^2} 
v^2_k\cosh^2\Delta \theta_k\right.\nonumber\\
&&\left.+ u_kv_k\sinh2\Delta \theta_k e^{-\frac{1}{2}((\epsilon_k +
    \omega_0-\omega_{RF})^2+E^2_k)\tau^2}\cos 2E t_0\right].
\label{RFnumberdetailT}
\end{eqnarray}
\end{widetext}

Although the general result is quite involved, it simplies
considerably in various important limits. For the case of broad pulse
$\tau\omega_{RF}\gg 1$ with a well-defined frequency, the Gaussian
factors reduce to energy-conserving $\delta$-functions.  In the
simplest equilibrium case, where the ground state's $n_k$ ($=v^2_k$ in
the Bogoluibov approximation) is probed, $\Delta\theta_k=0$, and we
find
\begin{eqnarray}
N^{gs}_b(\omega_{RF})&=& (2\pi)^{3/2}\tau I^2_0 
\sum_{\kv}\delta(\omega_{RF} - \omega_0 - \epsilon_k - E_k)n_k\nonumber\\
_{\omega_{RF}\gg\omega_0}
&=&\frac{\tau I_0^2 V}{\sqrt{2\pi m}}\frac{C_{gs}}{|\omega_{RF}-\omega_0|^{3/2}}.
\label{Nequil}
\end{eqnarray}
In the last equality we focussed on the large frequency tail probed in
the experiments \cite{Wild} and $C_{gs}$ is Tan's contact, that in the
Bogoluibov approximation is given by $C_{gs}^B = 16\pi^2 n^2 a_s^2$.

For a measurement of the large frequency tail,
$\omega_{RF}\gg\omega_0$ following a quench at $t=0$, it is clear from
Eq.~\rf{RFnumberdetailT} that only the second term contributes, giving
\begin{eqnarray}
  N_b(\omega_{RF})&=& (2\pi)^{3/2}\tau I^2_0 
  \sum_{\kv}\delta(\omega_{RF} - \omega_0 - \epsilon_k - E_k)\nonumber\\
&&\hspace{2.5cm}\times\ n^f_k\cosh^2\Delta \theta_k\nonumber\\
_{\omega_{RF}\gg\omega_0}
  &=&\frac{\tau I_0^2 V}{\sqrt{2\pi m}}\frac{C_{f}}{|\omega_{RF}-\omega_0|^{3/2}}.
\label{Nnonequil}
\end{eqnarray}
where within Bogoluibov approximation
\begin{eqnarray}
C_f = 16\pi^2n^2 a_f^2 .
\end{eqnarray}
This indicates that, while the overall momentum distribution function
$n_k(t)$ exhibits nontrivial post-quench dynamics, the large tail of RF
spectrum is {\em not} affected by the quench dynamics, and provides
information about short-scale correlations in the ground state of the
final state.

\section{Finite-rate ramp}
\label{finiteRateRamp}
\begin{figure}[htb]
 \centering
  \includegraphics[width=65mm]{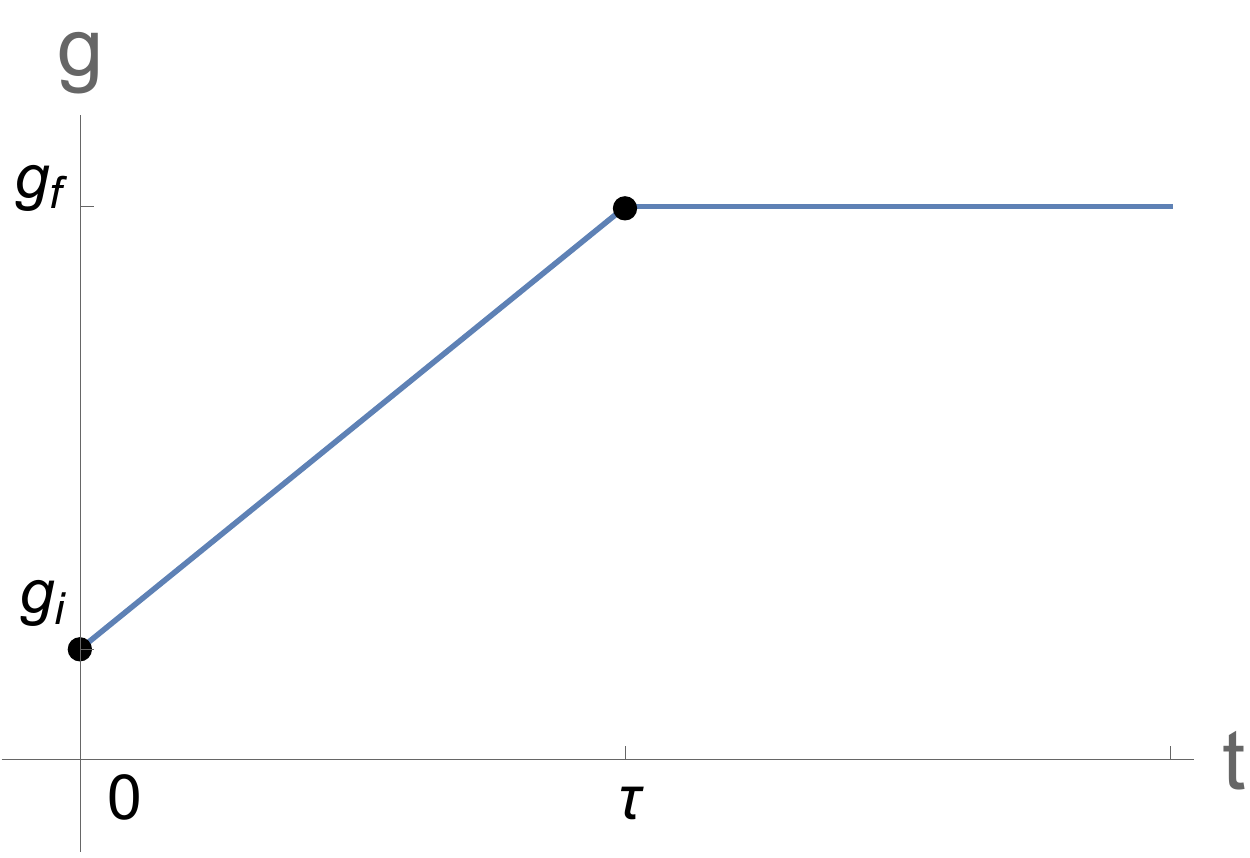}
 \caption{(Color online) Protocol of a linear ramp of coupling
   strength $g$. Starting with $g=g_i$ at ${t}=0$, the coupling
   strength is ramped to $g=g_f$ over ramp time $\tau$.}
\label{fig:rampdiagram}
\end{figure}

Having studied the idealized case of a {\em sudden} $g_i\rightarrow
g_f$ {\em quench}, we now analyze the dynamics following a more
experimentally realistic {\em finite}-rate ramp. We model it by
an idealized time-dependent coupling 
\begin{eqnarray}
g(t)&=&\left\{\begin{array}{ll}
g_i + (g_f - g_i)t/\tau,&\mbox{for $t < \tau$},\\
g_f ,&\mbox{for $t > \tau$},
\end{array}\right.
\label{gt}
\end{eqnarray}
with ramp time $\tau$, illustrated in Fig.~\ref{fig:rampdiagram}.

To this end, we solve the corresponding Heisenberg equations of motion
\begin{eqnarray}
i\begin{pmatrix}\dot{\hat a}_\kv\\ 
-\dot{\hat a}^{\dagger}_{-\kv}
\end{pmatrix}
&=&
\begin{pmatrix}
\epsilon_k + n g(t)&n g(t)\\
n g(t)& \epsilon_k(t) + n g(t)
\end{pmatrix}
\begin{pmatrix}\hat a_\kv(t)\\
\hat a^{\dagger}_{-\kv}(t)
\end{pmatrix}\;\;\;\;\;\;\;\;\;
\end{eqnarray}
by expressing the atomic operators $\hat a_\kv(t)$, $\hat a^\dagger_\kv(t)$ in
terms of the Bogoluibov quasi-particles $\hat \alpha_\kv$,
$\hat \alpha^\dagger_\kv$ of $\hat H(t=0)$ at the start of the ramp
\begin{eqnarray}
\label{eq33}
\begin{pmatrix}
\hat a_\kv\\\hat a^{\dagger}_{-\kv}
\end{pmatrix}=
\begin{pmatrix}u_k(t)&v^*_k(t)\\
v_k(t)&u^*_k(t)
\end{pmatrix}
\begin{pmatrix}
\hat \alpha_\kv\\
\hat \alpha^{\dagger}_{-\kv}
\end{pmatrix}
\equiv U_k(t)
\begin{pmatrix}
\hat \alpha_\kv\\
\hat \alpha^{\dagger}_{-\kv}
\end{pmatrix}.\;\;\;
\end{eqnarray}

The dynamics is then encoded in the time evolution of a spinor
$(u_k(t),v_k(t))$, with components satisfying
\begin{subequations}
\begin{align}
\label{finiteeoma}
i\dot{u}_\kh&=(\hat{k}^2+\hat g(t))u_\kh+\hat g(t)v_\kh,\\
-i\dot{v}_\kh&=(\hat{k}^2+\hat g(t))v_\kh+\hat g(t)u_\kh,
\label{finiteeomb}
\end{align}
\end{subequations}
where $\hat g(t)\equiv g(t)/g_f$, $\hat{t}\equiv n g_f t$ and
$\hat{k}^2\equiv k^2/(2m n g_f)$. In term of above dimensionless variables, Eq.~\eqref{gt} becomes
\begin{eqnarray}
\hat g(\hat t)&=&\left\{\begin{array}{ll}
\sigma +\gamma\hat{t},&\mbox{for $t < \tau$},\\
1 ,&\mbox{for $t > \tau$},
\end{array}\right.
\label{dimensionlessgt}
\end{eqnarray}
where we have defined a dimensionless ramp rate $\gamma\equiv (1 - \sigma) /(ng_f\tau)$. We  then solve these numerically, subject
to the initial conditions 
\bse
\begin{eqnarray}
u_k(0)
&=&\frac{1}{2^{1/2}}\left(\frac{\kh^2+\sigma}
{\sqrt{\kh^2(\kh^2+2\sigma)}}+1\right)^{1/2},\\
v_k(0)
&=&-\frac{1}{2^{1/2}}\left(\frac{\kh^2+\sigma}
{\sqrt{\kh^2(\kh^2+2\sigma)}}-1\right)^{1/2},
\label{finiteini}
\end{eqnarray}
\ese
that diagonalize the initial Hamiltonian at $t=0$. 

We focus on the momentum distribution at $T=0$
\begin{equation}
n_k(t)=\langle 0^-|a^{\dagger}_k(t)a_k(t)|0^-\rangle=|v_k(t)|^2,
\end{equation}
and condensate fraction
\begin{equation}
n_c(t)=1-n_d(t)=1-V\int \frac{d^3 k}{(2\pi)^3}n_{k}({t}).
\end{equation}
We apply this analysis to interpret experiments by Claussen, et al.,
\cite{Claussen}, where dynamics of finite-rate ramp pulse was studied as a
function of ramp time $\tau$ and heretofore remained
unexplained.

\begin{figure}[htb]
\centering
\includegraphics[width=0.45\textwidth]{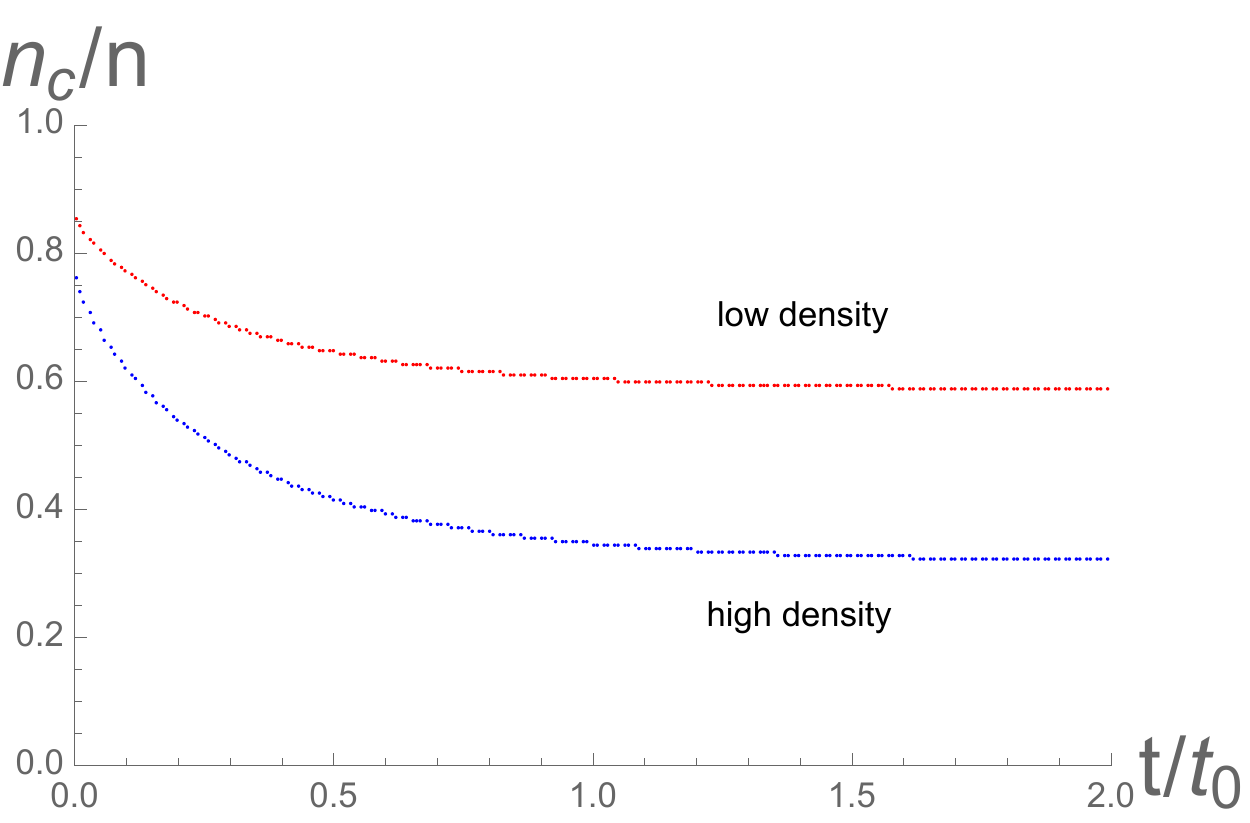}
\caption{(Color online) Dynamics of the condensate fraction $n_c(t)/n$
  for high density $n=1.9\times 10^{13}cm^{-3}$ (lower blue curve) and low density $n=0.7\times
  10^{13}cm^{-3}$ (upper red curve), after a linear interaction ramp $g(t)$ from $g_i=0.001g_f$ to $g_f$ with dimensionless ramp rate $\hat\gamma\equiv (1 - \sigma) /(\tau n g_f)
  =5$. Following the experiments in Ref.\onlinecite{Claussen} the final
  scattering length is $a_f=2700 a_0$ and the corresponding pre-thermalization timescale $t_0=150\mu s$.}
\label{nctfiniteramp}
\end{figure}

In Fig.~\ref{nctfiniteramp} we plot the time dependence of the
resulting condensate fraction for two densities and fixed ramp rate.
Using the parameters reported in \cite{Claussen}, we obtain results in
qualitative agreement with these experimental measurement. 

\begin{figure}[htb]
\centering
\includegraphics[width=0.45\textwidth]{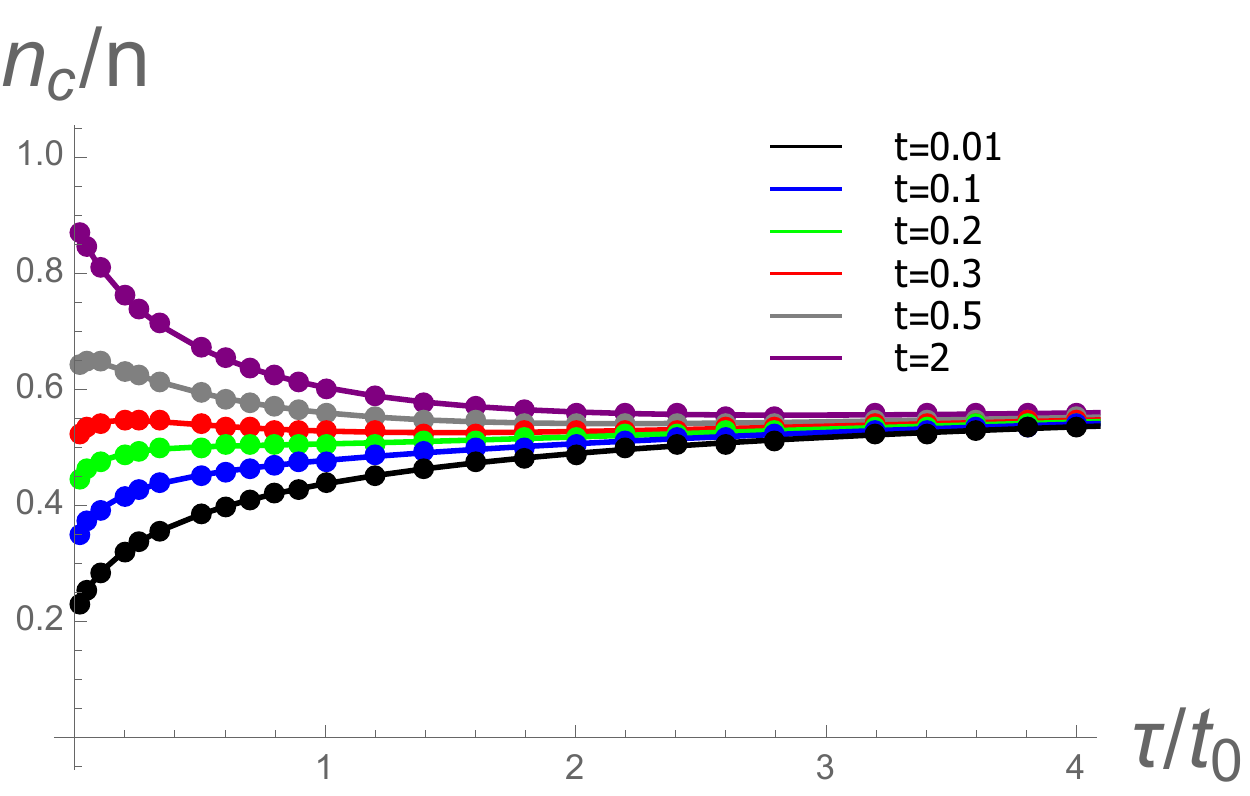}
\caption{(Color online) Dependence of condensate fraction $n_c/n$ on the
  ramp duration $\tau$ for various hold time $t$ in units of pre-thermalization timescale $t_0= \hbar/(ng_f)$ (From lowest curve to highest one, the hold times are $\hat{t}=0.01,0.1, 0.2,0.3,0.5,2$, respectively). Following the
  experiments in Ref.~\onlinecite{Claussen} we take the final scattering length
  to be $a_f=2700 a_0$ with $a_i=0.001a_f$ and $n=1.9*10^{12}cm^{-3}$.}
\label{becratedep}
\end{figure}

To explore the ramp rate dependence of the dynamics as studied by
Claussen, et al., \cite{Claussen}, in Fig.~\ref{becratedep} we plot the
condensate fraction as a function of ramp time $\tau$ (inverse ramp
rate, in units of $(1 - \sigma)/(ng_f)$). As illustrated there, we find that
the dependence on the ramp time $\tau$ is nonmonotonic and is a
function of the hold time $t$.  This can be understood by noting that
for a sudden quench (vanishing $\tau$) at long hold times, the
condensate is depleted more strongly than the ground state depletion
for $g_f$. On the other hand, at short hold times the quenched depletion is
given by the ground state for $g_i$.  In contrast, for a slow
adiabatic ramp (large $\tau$) the condensate fraction asymptotes to
the adiabatic limit corresponding to that of a ground state for $g_f$.

Thus, for short hold time the condensate fraction {\em decreases} from
$n_c^{g_i}$ to $n_c^{g_f}$ with increasing $\tau$. For long hold
times, the condensate fraction {\em increases} from pre-thermalized
value $n_c^{ss}$ to $n_c^{g_f}$ with increasing $\tau$. This behavior
is qualitatively quite similar to that found in experiments of
Ref.~\cite{Claussen}.

\section{Dynamics for deep quench}

In the present and subsequent sections we study the nonperturbative
dynamics following a deep quench, $na^3_f\gg 1$, a regime of JILA
recent experiment \cite{Makotyn} that is our main
focus \cite{YinLR14}. In contrast to the well-controlled,
perturbative dynamics of a shallow quench discussed in Sec.~\ref{sec:ShallowQuench}, for deep quenches the condensate depletion
dynamics is significant and cannot be neglected.

From the outset, we acknowledge that no rigorous solution in such a
nonperturbative regime is available even for a purely repulsive Bose
gas ground state. Nevertheless, to make progress we treat this
strongly interacting nonequilibrium dynamics utilizing a
nonperturbative but uncontrolled self-consistent Bogoluibov treatment.
This is analogous to a BCS dynamic mean-field
theory \cite{Barankov,AGR06}, with the condensate fraction $n_c(t)$
playing the role of the time-dependent order parameter.  We thus
reduce the problem to a solution of the Bogoluibov dynamics with a
time-dependent condensate fraction
that is self-consistently determined. This is a dynamical
generalization of our analysis of the strongly interacting Bose gas
ground state in Sec. \ref{sec:ndSCgs}.

Another challenge of this system is the resonant nature of the Bose
gas interaction. To handle this we employ a second
beyond-Bogoluibov approximation by replacing the scattering length
$a_f$ by the density dependent scattering amplitude $|f(k_n,a_f)|=
a_f/\sqrt{1+k_n^2a_f^2}\equiv \as_f$, and the Hartree interaction
energy $g_f n$ by $\g_fn\equiv\frac{8\pi\epsilon_F}{\sqrt{1/(k_n
    a_f)^2 + 1}}$.  This qualitatively captures the crossover from the
two-atom regime, $a_f\ll n^{-1/3}$ to a finite density limit, when
$a_f$ reaches inter-particle spacing and the scattering amplitude
saturates at $\sim k_n^{-1}$. While the detailed nature of this
crossover is ad hoc, our qualitative predictions are insensitive to
these details and only depend on the limiting values of the two
regimes.

Motivated by the experiments \cite{Makotyn}, we focus on an initial
state that is a well-established condensate. This allows us to make
progress in treating the resonant interactions by expanding in
finite-momentum quasi-particle fluctuations about a macroscopically
occupied $\kv=0$ state. Following a sudden quench, $g_i\rightarrow
g_f$, we approximate the Hamiltonian by a quadratic time-dependent
form,
\begin{eqnarray}
\label{HB}
\hspace{-0.15cm}
\hat H_f(t)&=&\oh\sum_{\kv\neq 0} 
\begin{pmatrix}
 \hat  a_\kv^\dagger & \hat a_{-\kv}\\
\end{pmatrix}
\hspace{-0.15cm}
\begin{pmatrix}
  \epsilon_k + g_f n_c(t)& g_f n_c(t)\\
  g_f n_c(t) & \epsilon_k+ g_f n_c(t)\\
\end{pmatrix}
\hspace{-0.15cm}
\begin{pmatrix}
\hat   a_\kv\\
\hat   a_{-\kv}^\dagger\\
\end{pmatrix}\nonumber\\
&\equiv&\oh\sum_{\kv\neq 0} 
\hat \Phi^\dagger_\kv(t)\cdot \hat{h}_{kf}(t)
\cdot\hat \Phi_\kv(t).
\end{eqnarray}
The key new ingredient (in contrast to Bogoluibov theory of
Sec.~\ref{sec:BdG}) is the nontrivial time-dependent condensate
density, that is self-consistently determined by the total atom
conservation,
\begin{eqnarray}
n_c(t) = n - \frac{1}{V}\sum_{\kv\neq 0} \langle 0^-|\hat a_\kv^\dagger(t)
\hat a_\kv(t)|0^-\rangle,
\label{conserveN}
\end{eqnarray}
evaluated in the pre-quench state $|0^-\rangle$ at $t=0^-$.  In a
homogeneous case, this is equivalent to a solution of the
Gross-Petaevskii equation for the condensate order parameter $\Psi_0$,
coupled to the Heisenberg equation of motion for the finite momentum
quasi-particles.  Focussing on zero temperature, we take the initial
state $|0^-\rangle$ to be the vacuum with respect to the
quasi-particles $\hat \alpha_\kv$, that diagonalize the initial
Hamiltonian, $\hat H_i=\sum_\kv E_{k i}\hat \alpha_\kv^\dagger\hat \alpha_\kv$,
characterized by a pre-quench $t = 0^-$ scattering length, $a_i$.

The corresponding Heisenberg equation of motion
\begin{eqnarray}
  i\sigma_z\partial_t\hat \Phi_\kv(t)=\hat{h}_{kf}(t)
  \cdot\hat \Phi_\kv(t)
\end{eqnarray}
for $\hat \Phi_\kv(t)=(\hat a_\kv(t), \hat a_{-\kv}^\dagger(t))$ is
conveniently encoded in terms of a time-dependent Bogoluibov
transformation $U_{k f}(t)$, 
\begin{equation}
\label{eqmquasiadiabatic}
 \hat  \Phi_\kv(t)=U_{\kv f}(t)\hat \Psi_\kv,
\end{equation}
where 
\begin{eqnarray}
\label{Ukf}
U_{kf}(t)&=&\begin{pmatrix}u_{kf}(t)&v^*_{kf}(t)\\
v_{kf}(t)&u^*_{kf}(t)
\end{pmatrix}
\end{eqnarray}
and $\hat \Psi_\kv=(\hat \beta_\kv,\hat \beta^{\dagger}_{-\kv})$ are time-independent
bosonic reference operators, that diagonalize the Hamiltonian at the
initial time $t=0^+$ after the quench, with $\hat H_f(0^+)=\sum_\kv E_{k
  f}(0^+)\hat \beta_\kv^\dagger\hat \beta_\kv$.

Equivalently, $U^\dagger_{k f}(0^+) h_{kf}(0^+) U_{k f}(0^+)=E_{k
  f}(0^+)=\sqrt{\epsilon_k^2+2g_fn_c(0^+)\epsilon_k}$, fixing the
initial condition
\bse
\begin{eqnarray}
\label{uv_inita}
  u_{kf}(0^+)&=&\sqrt{\oh\left(\frac{\epsilon_k+g_fn_c(0^+)}{E_f(0^+)} + 1\right)},\\
  v_{kf}(0^+)&=&-\sqrt{\oh\left(\frac{\epsilon_k+g_fn_c(0^+)}{E_f(0^+)}
      - 1\right)},
\label{uv_initb}
\end{eqnarray}
\ese
for spinor $\psi_{k f}(t)\equiv (u_{kf}(t),v_{kf}(t))$, that evolves
according to
\begin{eqnarray}
\label{Heom}
i\sigma_z\partial_t{\psi}_{k f}(t)=\hat{h}_{kf}(t)
\cdot{\psi}_{k f}(t).
\end{eqnarray}

As for the Bogoluibov analysis in Sec.~\ref{sec:BdG}, because the initial
state $|0^-\rangle$ is a vacuum of $\hat \alpha_k$, it is convenient to
further express $\hat \Phi_\kv(t)=(\hat a_\kv(t), \hat a_{-\kv}^\dagger(t))$ in terms
of the pre-quench quasi-particle basis
$\hat \Psi_{\kv}(0^-)=(\hat \alpha_\kv,\hat \alpha^{\dagger}_\kv)$,
\bse
\begin{eqnarray}
\hat \Phi_\kv(t)&=&U_{kf}(t)U_{kf}^{-1}(0^{+})U_{ki}(0^{-})\hat \Psi_{\kv}(0^-),\\
&\equiv& U_k(t)\hat \Psi_{\kv}(0^-).
\end{eqnarray}
\ese

The post-quench dynamics is thus fully determined by the
self-consistent solutions $\psi_{k f}(t)$ of \rfs{Heom}, together with
the atom number conservation constraint, \rf{conserveN}. This can be
obtained numerically in essentially exact way, as we will demonstrate
in Sec.~\ref{sec:exactnumerical}.

\subsection{Quasi-adiabatic self-consistent approximation}
\label{ndSCqs}
Despite availability of the numerical solution, to gain further
physical insight it is of interest to obtain an approximate analytical
solution. To this end we note that for a given slowly evolving
condensate density satisfying ${\dot{n}_c(t)}/{n}\ll{E^3_{kf}}/(\hbar ng\epsilon_k)=(\epsilon_k)^{1/2}(\epsilon_k+2gn_c)^{3/2}/(\hbar ng)$ (see Eq.~\eqref{quasigammaeqm} and \cite{YinLR14,KainLing}), \rfs{Heom} can be well-approximated
by an instantaneous, quasi-adiabatic Bogoluibov transformation of
$\hat H_f(t)$ (see Appendix~\ref{appendix:UTquasi}),
\begin{eqnarray}
\label{UTquasi}
U_{kf}(t)=
\begin{pmatrix}
u_k(t)e^{-i\int_0^t E_{kf}(t')}&v_k(t)e^{i\int_0^t E_{kf}(t')}\\
v_k(t)e^{-i\int_0^t E_{kf}(t')}&u_k(t)e^{i\int_0^t E_{kf}(t')}\\
\end{pmatrix}.\quad 
\end{eqnarray}
In above, $(u_k(t),v_k(t))$ is the instantaneous eigenstate of the
single-particle Hamiltonian $\hat h_{kf}(t)$, with time dependence
entering only through the time dependent condensate density, $n_c(t)$.
Such approximation is in the spirit of the WKB quasi-local treatment
of a smoothly varying potential \cite{Shankar}.

More specifically the solution is given by
\begin{equation}
\begin{split}
u_k(t)&=\sqrt{\frac{1}{2}\left(\frac{\epsilon_k+g_fn_c(t)}{E_{kf}(t)}+1\right)},\\
v_k(t)&=-\sqrt{\frac{1}{2}\left(\frac{\epsilon_k+g_fn_c(t)}{E_{kf}(t)}-1\right)},\\
E_{kf}(t)&=\sqrt{\epsilon_k(\epsilon_k+2g_fn_c(t))},
\end{split}
\end{equation}
with initial condition given by \rf{uv_inita}\rf{uv_initb}.

After a tedious but conceptually straightforward calculation that
utilizes above relations, we obtain the momentum distribution function
\begin{widetext}
\begin{equation}
\begin{split}
n_k(t)&=\langle 0^-|\hat a^{\dagger}_\kv(t)\hat a_\kv(t)|0^-\rangle\\
&=\frac{\epsilon^2_k+\epsilon_k(g_in+g_fn+g_fn_c(t))+2g_fg_in_c(t)n+2g_fn_c(t)n(g_f-g_i)\sin^2(\int_0^t \sqrt{\epsilon_k(\epsilon_k+2g_fn_c(t'))} dt')}{2\sqrt{\epsilon_k(\epsilon_k
+2g_fn_c(t))}\sqrt{(\epsilon_k+2g_in)}\sqrt{(\epsilon_k+2g_fn)}}-\frac{1}{2},\\
\end{split}
\label{quasin}
\end{equation} 
\end{widetext}
with the condensate density $n_c(t)$ self-consistently determined
according to $n_c(t)=n - \sum_{\boldsymbol{k}\neq0} n_k(t)$.

By construction, the above expression for $n_k(t=0)$ reduces to the 
pre-quench momentum distribution function 
\begin{equation}
n_k(t=0)=\frac{1}{2}\left[\frac{\epsilon_k + g_i
    n_c(0)}{E_{ki}}-1\right]=n_{ki},
\end{equation}
as required by continuity. Furthermore for $g_f=g_i$, i.e., in the
absence of a quench, the time-dependent part of $n_k$ drops out and
again reduces to $n_{ki}$.

Using \rf{quasin} the self-consistency condition reduces to a
dimensionless form
\begin{equation}
\begin{split}
1-\hat{n}_c=n^0_d F_d(\hat{n}_c,\sigma),
\end{split}
\label{quasiSC}
\end{equation}
where $n^0_d=8/(3\sqrt{\pi})({n a_f^3})^{1/2}$ is the depletion
corresponding to the ground state of quenched Hamiltonian,
$q\equiv\sqrt{k^2/{(2mng_f)}}$ and $\hat{n}_c(t)\equiv n_c(t)/n$ are
dimensionless variables, and
\begin{widetext}
\begin{equation}
\begin{split}
F(\hat{n}_c,\sigma,t)=3\sqrt{2}\int dq q^2\left[\frac{(q^4+{q}^2(\sigma+1+\hat{n}_c)+2\sigma\hat{n}_c
+2\hat{n}_c(1-\sigma)\sin^2(\int_0^t ngt\sqrt{q^2(q^2+2\hat{n}_c)}))}{2\sqrt{{q}^2({q}^2+2\hat{n}_c)}\sqrt{({q}^2+2\sigma)}\sqrt{({q}^2+2)}}
-\frac{1}{2}\right],
\end{split}
\label{quasiscf}
\end{equation}
\end{widetext}
is the quench-induced depletion-enhancement factor.

\begin{figure}[htb]
\centering
\includegraphics[width=0.45\textwidth]{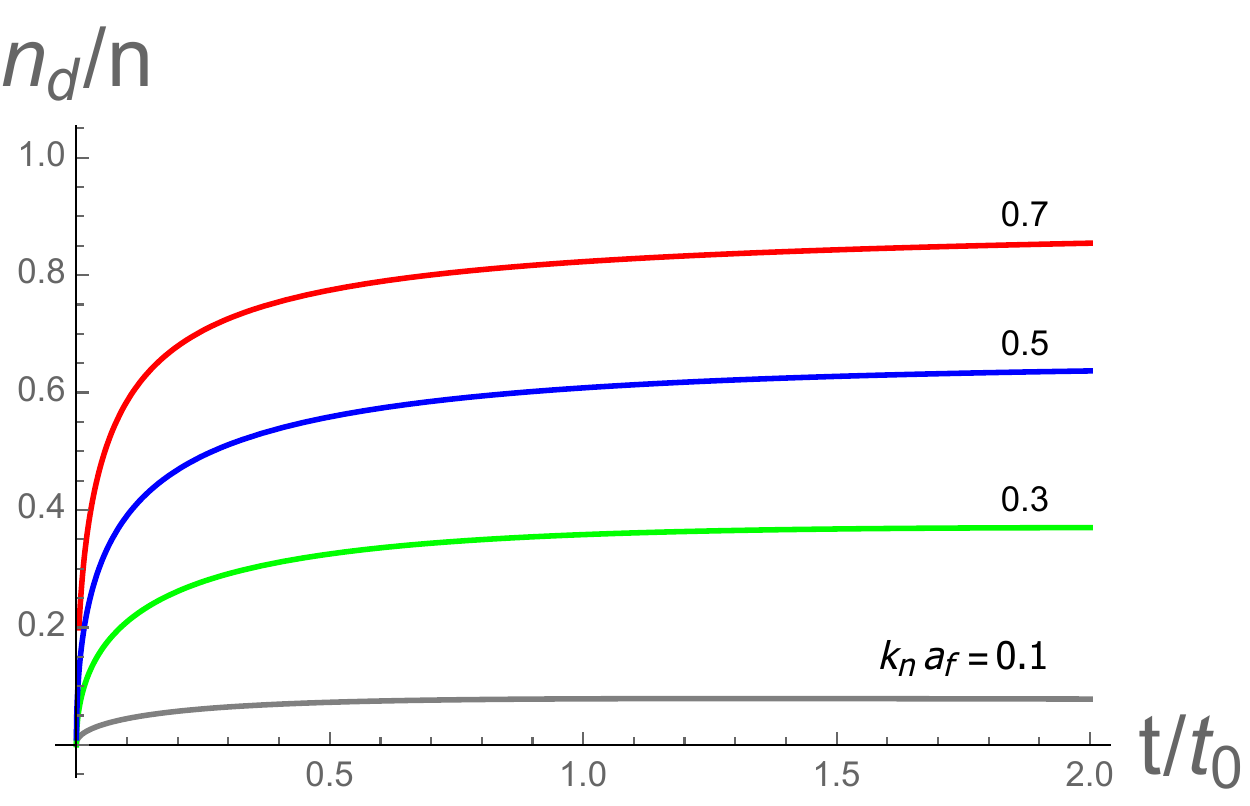}
\caption{(Color online) Time evolution of the condensate depletion fraction
  $n_d(t)/n$ (treated within a quasi-adiabatic self-consistent dynamic field
  analysis, referring to Eq.~\eqref{quasiSC}), following a
  scattering length quench from $k_na_i=0.01$ to various $k_na_f$ in a resonant Bose gas. Here we normalize the time with the pre-thermalization timescale $t_0= 1/ng_f=m/(4\pi a_f n)$ associated with $k_na_f=1$ (where $k_n\equiv n^{1/3}$).}
\label{fig:quasind}
\end{figure}

We solve Eqs.\eqref{quasiSC},\rf{quasiscf} numerically and plot the
depletion fraction $\hat{n}_d(t)=1-\hat{n}_c(t)$ as a function of time
in Fig.~\ref{fig:quasind}.

We observe that the depletion fraction increases smoothly with time on
the scale $t_0=m/(4\pi a_f n )$, reaching a stationary steady-state $n_d^{ss}$, that is an increasing
function of the quench depth $k_na_f$. Even for a deep quench to a
unitary point, the self-consistent treatment ensures that the
depletion, always remains below the total atom density.  The slow time
dependence of $n_d(t)$ justifies the quasi-static approximation for
the high momenta ($k \gtrsim 1/\xi$) quasi-particles, but fails for
the low-momenta ($k \lesssim 1/\xi$) Goldstone modes. We further note
that the asymptotic depletion $n_d^{ss}$ always significantly exceeds the depletion
for the ground state corresponding to the quenched scattering length
$a_f$. Thus not surprisingly the thermal equilibrium is never reached
in our effectively integrable harmonic model.

\begin{figure}[htb]
\centering
\includegraphics[width=0.45\textwidth]{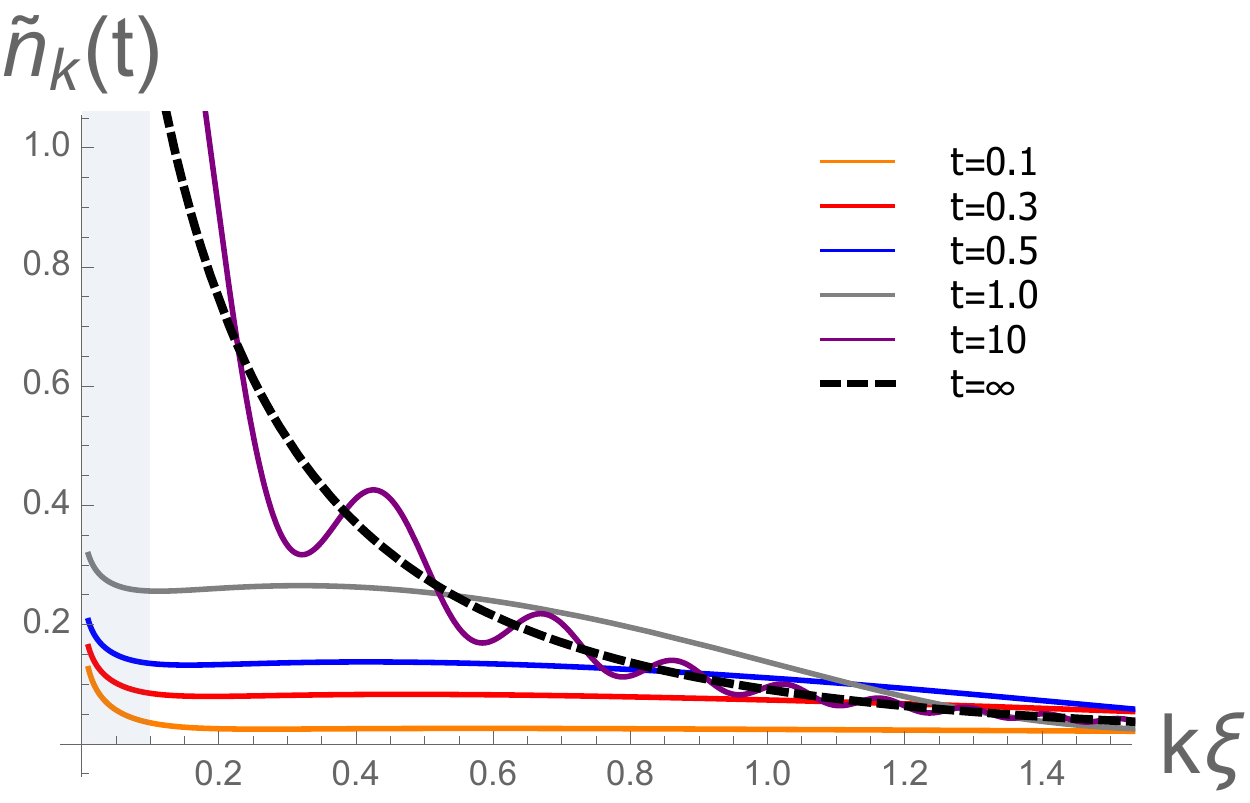}
\caption{(Color online) Time evolution of the (column-density)
  momentum distribution function, $\tilde{n}_{\kv_\perp}(t)\equiv\int dk_z n_\kv(t)$ following a scattering length
  quench $k_na_i=0.01\rightarrow k_na_f=0.5$ (where $k_n\equiv n^{1/3}$) in a resonant Bose gas, computed with quasi-adiabatic self-consistent approximation. Here we normalize the time with the pre-thermalization timescale $t_0= 1/ng_f=m/(4\pi a_f n)$ associated with $k_na_f=1$. Here momentum and time are rescaled with $\xi\equiv 1/\sqrt{2 m n g_f}$ and $t_0\equiv 1/(n g_f)$, respectively. The grey region indicates a range of momenta not resolved
in JILA experiments, due to initial inhomogeneous real space density
profile and finite trap size.}
\label{nktquasi}
\end{figure}

Having computed the condensate depletion and the associated condensate
density, $n_c(t)$, Eq.~\eqref{quasin} immediately gives us the
momentum distribution function $n_k(t)$, that we illustrate in
Fig.~\ref{nktquasi}.
Following a quench, the initially narrow (for $g_i\ll g_f$) momentum
distribution function (corresponding to pre-quench BEC state) displays
rich dynamics. Within 2-body interaction scale it quickly develops a
large momentum tail corresponding to the strong atom-atom interaction
$g_f$. With time, the suddenly turned on interaction promotes an
increasing number of atom pairs from the condensate to finite momentum
excitations.  The momentum distribution tail fills in from high to low
momenta as pair-excitation dynamics at momentum $k$ dephases with
frequency $2E_{kf}$.  Thus, at time $t$, $n_k(t)$ establishes a
pre-thermalized power-law steady-state for momenta $k > k_{pth}(t)$,
latter set by $E_{k_{pth},f} t\approx 1$. Equivalently, it takes time
\bse
\begin{eqnarray}
t_{pth}&\approx&\frac{1}{E_{k_{pth},f}},\\
&\sim&\left\{\begin{array}{ll}
1/k^2,&\mbox{for $k\gg 1/\xi$},\\
1/k,&\mbox{for $k\ll 1/\xi$},
\end{array}\right.
\label{t_pth}
\end{eqnarray}
\ese 
for the pre-thermalization to reach a stationary state down to
momentum $k$, a distinctive feature that is consistent with JILA
experiments \cite{Makotyn}.

\begin{figure}[htb]
\centering
\includegraphics[width=0.45\textwidth]{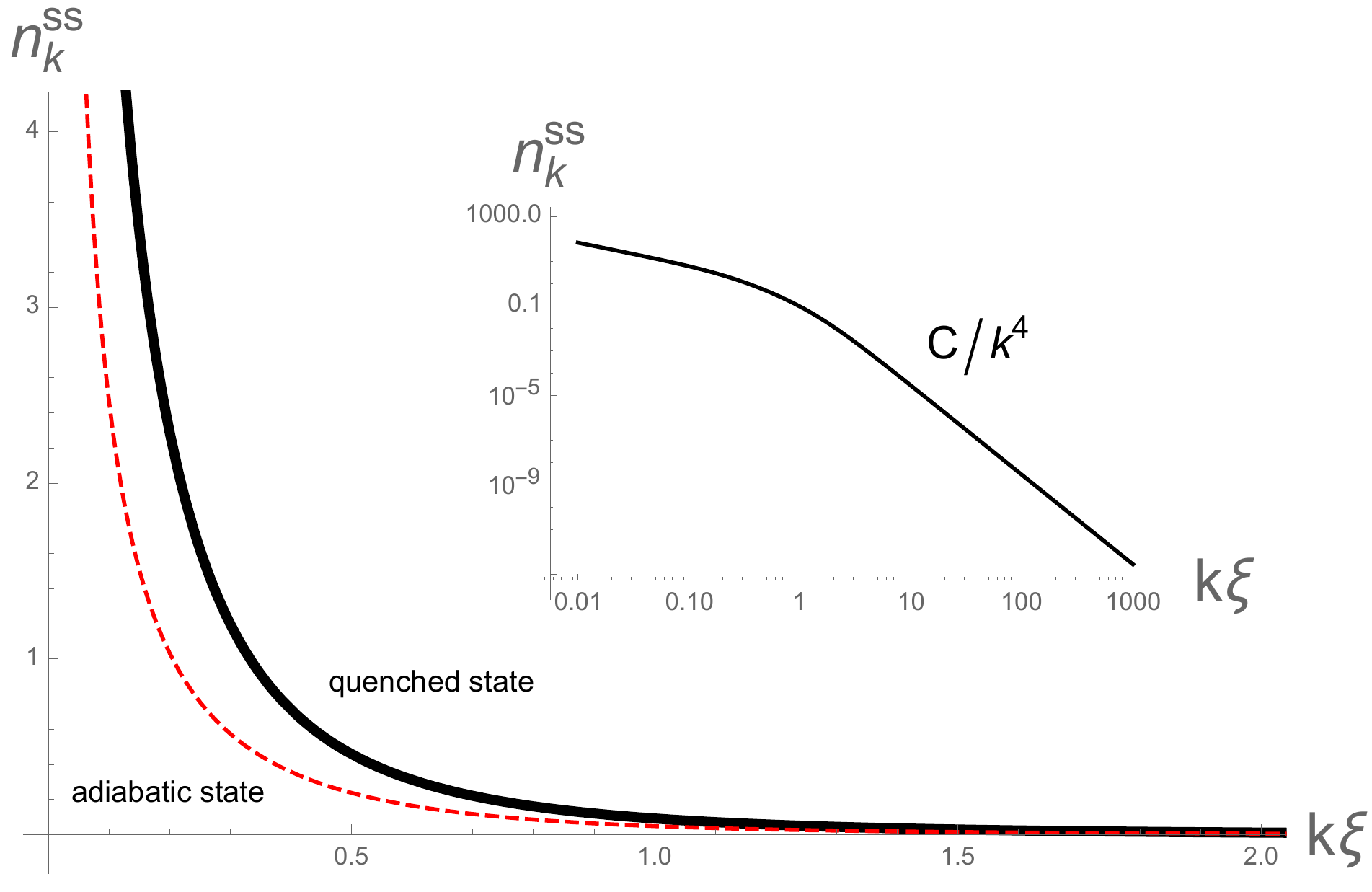}
\caption{(Color online) A long-time nonequilibrium steady-state
  momentum distribution function $n_k^{ss}$ of a resonant Bose gas
  following a scattering length quench $k_na_i=0.01\rightarrow k_na_f=0.5$ (solid black curve), as compared to ground state momentum distribution at $k_na_f$ (red dashed curve). The inset illustrates the emergence of a $1/k^4$ large momentum tail, corresponding to a steady-state ``contact''.}
\label{nkssquasi}
\end{figure}

As illustrated in Fig.~\ref{nkssquasi}, in the long time limit (around
$170 \mu$-sec in $^{85}$Rb experiments \cite{Makotyn}) a quenched Bose gas
approaches a pre-thermalized stationary state, as reflected by a
time-independent power-law momentum distribution
\begin{eqnarray}
n_k(t)&=&\frac{\hat{k}^4+\hat{k}^2(\sigma+1+\hat{n}^{ss}_c)
+\hat{n}^{ss}_c(1+\sigma)}{2\sqrt{\hat{k}^2(\hat{k}^2
+2\hat{n}^{ss}_c)}\sqrt{(\hat{k}^2+2\sigma)}\sqrt{(\hat{k}^2+2)}}-\frac{1}{2},\nonumber\\
 &\sim&C^{ss}/k^4, \qquad\mbox{for $k\xi\gg 1$},
\label{nkssBogoliubovResults}
\end{eqnarray}
where $C^{ss}=(4\pi a_f n)^2\left[n^{ss}_c/{n}+(1-\sigma)^2\right]$ is the nonequilibrium analog of Tan's
contact \cite{Tan}.  Within the above self-consistent Bogoluibov
approximation the quasi-particles do not scatter, precluding full
thermalization. The resulting final state remains nonequilibrium,
completely determined by the depth-quench parameter $\sigma$,
characterized by a diagonal density matrix ensemble.

With the above solution of the self-consistent post-quench dynamics,
we can now also calculate other physical observables, such as, for
example the structure function measured in Bragg spectroscopy. Using
above analysis for $S_\qv(t)$ in Eq.~\eqref{Sqt} we find
\begin{equation}
\begin{split}
S_{\hat{q}}(t)&=\coth(\hat{\beta}\hat{q}\sqrt{\hat{q}^2 + 2\sigma})\frac{\hat{q}}{\sqrt{\hat{q}^2+2\sigma}}\\
&\quad\times \bigg(\frac{\sqrt{\hat{q}^2+2}}{\sqrt{\hat{q}^2+2\hat{n}_c(t)}}\\
&\qquad-\frac{2(1-\sigma)\sin^2 (\hat{q}t \sqrt{\hat{q}^2+2\hat{n}_c(t)})}{\sqrt{\hat{q}^2+2}\sqrt{\hat{q}^2+2\hat{n}_c(t)}}\bigg),
\end{split}
\label{quasiscf2}
\end{equation}
where $\hat{q}=q/\sqrt{2mng_f}$, $\hat{t}=ng_f t$ and $\hat{\beta}=ng_f\beta$.

\begin{figure}[htb]
 \centering
  \includegraphics[width=80mm]{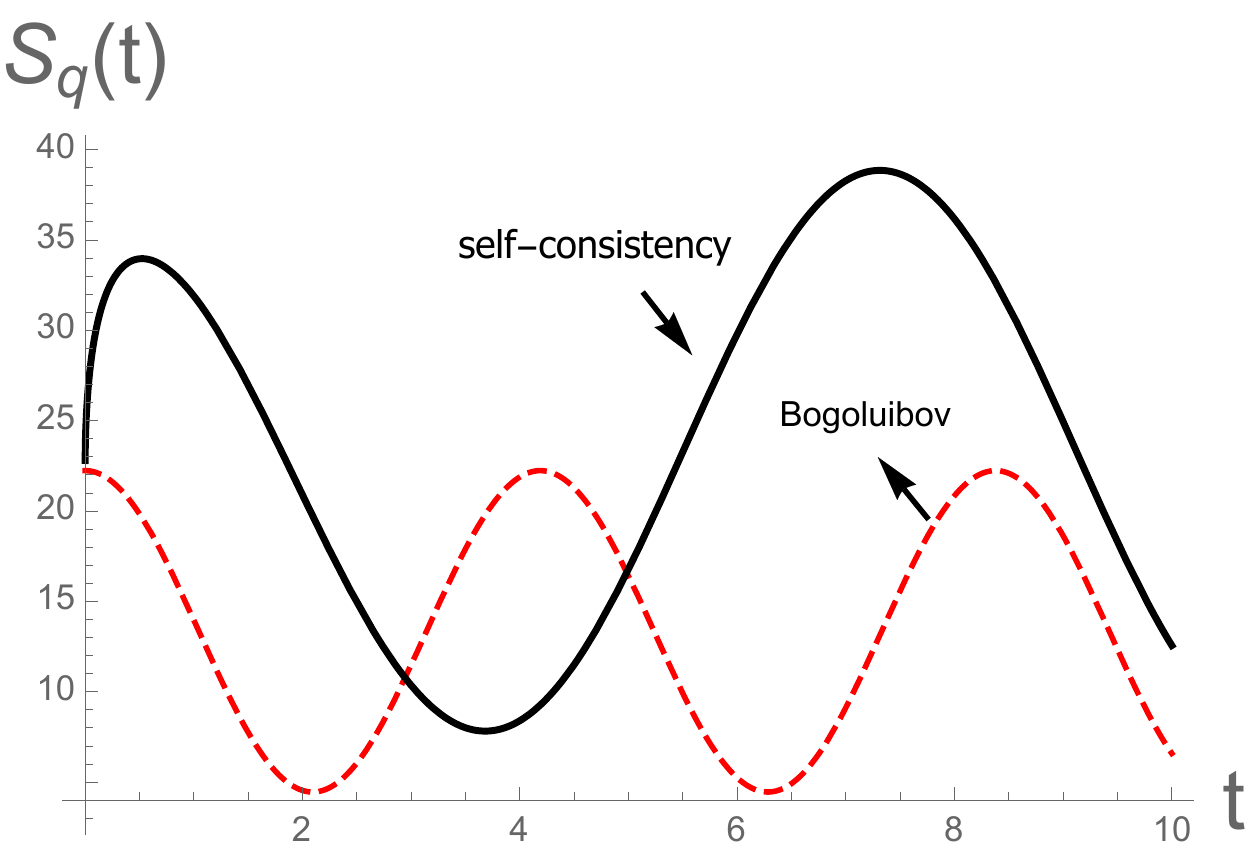}
 \caption{(Color online) Oscillation of structure function $S_\qv(t)$
   (treated within a quasi-adiabatic self-consistent dynamic field
   analysis, thick black curve, referring to Eq.~\eqref{quasiscf2}) as
   a function of time, following a scattering length quench from
   $0.1a_f\to a_f$ with $k_na_f=0.7$ (where $k_n\equiv n^{1/3}$) at momentum $k\xi=0.5$, as
   compared to Bogoluibov approximation (dashed red curve).}
\label{fig:Braggsktcomparison}
\end{figure}

The results are then illustrated in Fig.~\ref{SqDMFT} and \ref{fig:Braggsktcomparison}.
The role of self-consistency is clear: in Fig.~\ref{SqDMFT}, as compared with Fig.~\ref{Braggspectrobdg}, self-consistency  exchanges the relative position of initial and final asymptotic steady-state curve; while in Fig.~\ref{fig:Braggsktcomparison} it shifts the phase as well as modifies the frequency of the structure function oscillation. We expect these features to be experimentally testable by going to a deep quench regimes,  $k_na_f \gg 1$.

\begin{figure}[htb]
 \centering
  \includegraphics[width=80mm]{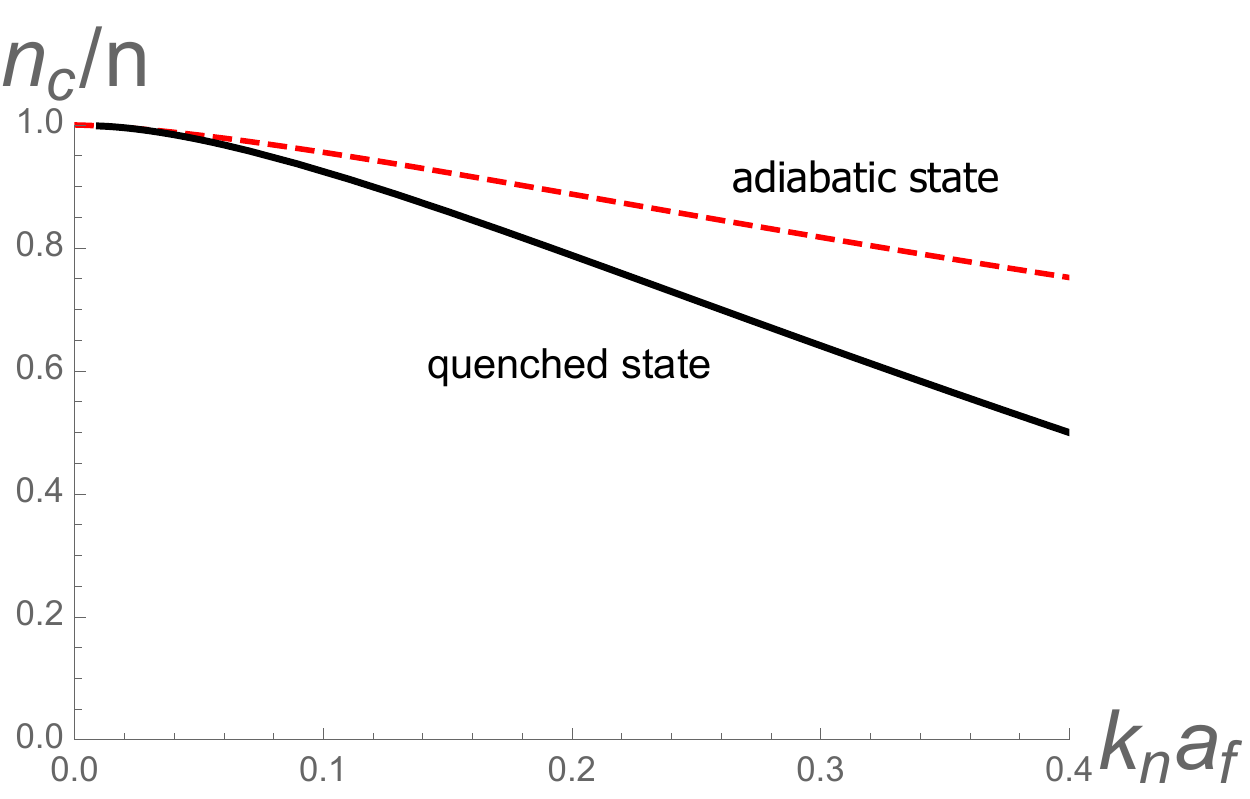}
 \caption{(Color online) Quenched steady state condensate fraction
   $n_c/n$ as a function of $k_na_f$ (solid black curve, treated with
   quasi-adiabatic self-consistent dynamic field, referring to
   Eq.\eqref{quasiSC}), as compared with the ground state condensate
   fraction at $k_na_f$ (dashed red curve), both calculated with
   self-consistency on $n_c$.}
\label{fig:ssnc}
\end{figure}

We emphasize that above analysis utilizes a quasi-adiabatic
approximation, valid for ${\dot{n}_c(t)}/{n}\ll{E^3_{kf}}/(\hbar ng\epsilon_k)$. As
mentioned above we expect it to break down for sufficiently small
momenta for slow Goldstone modes as well as large $k_na_f$ value, where $\dot{n}_c(t)/n$ is large.

\subsection{Exact numerical solution to post quench dynamics}
\label{sec:exactnumerical}

In this subsection we test the validity of above quasi-adiabatic
approximation by analyzing the post-quench dynamics through an
essentially exact numerical solution of the Heisenberg equation of
motion \rf{Heom}. Consistent with our expectations we find that while
the former provides an accurate description for a shallow quench and
high momenta, it fails quantitatively (though not qualitatively) for
$k_na_f\gg 1$ and low momenta, $k\ll 1/\xi$.

As derived in previous subsection, the dynamics is governed by
Eq.~\rf{Heom} for $\psi_\kv(t)=(u_{kf}(t),v_{kf}(t))$, that relate atomic
excitations, $\hat a_\kv$ to Bogoluibov quasi-particles $\hat\beta_\kv$. Here we
solve Eq.~\rf{Heom} numerically together with the number conservation
condition on the condensate fraction. In dimensionless form, the
equations of motion are given by 
\begin{equation}
\label{exacteom}
\begin{split}
i\dot{u}_k&=(\hat{k}^2+\bar{n}(t))u_k+\bar{n}(t)v_k,\\
-i\dot{v}_k&=(\hat{k}^2+\bar{n}(t))v_k+\bar{n}(t)u_k,\\
\end{split}
\end{equation}
with the initial conditions fixed by a requirement that at $t=0$, $\psi(0^+)$
diagonalizes $\hat H_f(0^+)$,
\bse
\begin{eqnarray}
\label{exactinicona}
u_k(t=0)&=&\sqrt{\frac{1}{2}[E^{-1}_{kf}\left(\epsilon_k+n_c(0)
    g_f\right)+1]},\nonumber\\
&=&\sqrt{\frac{1}{2}
\left(\frac{\hat{k}^2+1}{\sqrt{\hat{k}^2(\hat{k}^2+2)}}+1\right)},\\
v_k(t=0)&=&-\sqrt{\frac{1}{2}[E^{-1}_{kf}
\left(\epsilon_k+n_c(0)g_f\right)-1]},\nonumber\\
&=&-\sqrt{\frac{1}{2}
\left(\frac{\hat{k}^2+1}{\sqrt{\hat{k}^2(\hat{k}^2+2)}}-1\right)},
\label{exactiniconb}
\end{eqnarray}
\ese
where $\hat{t}\equiv n g_f t$, $\hat{k}^2\equiv k^2/(2m n g_f)$ and
$\bar{n}(t)\equiv n_c(t)/n$.

Decoupling the $u_k(t)$ and $v_k(t)$ components
\begin{equation}
\label{eqmndot}
\begin{split}
\ddot{u}&=[-k^2(k^2+2\bar{n}(t))+i\frac{\dot{\bar{n}}(t)}{\bar{n}(t)}(k^2)]u+\frac{\dot{\bar{n}}(t)}{\bar{n}(t)}\dot{u},\\
\ddot{v}&=[-k^2(k^2+2\bar{n}(t))-i\frac{\dot{\bar{n}}(t)}{\bar{n}(t)}(k^2)]v+\frac{\dot{\bar{n}}(t)}{\bar{n}(t)}\dot{v},
\end{split}
\end{equation}
more clearly reveals the relation of these exact equations to the
quasi-adiabatic approximation of previous subsection. Indeed the
latter is obtained by neglecting $\dot{\bar{n}}(t)/{\bar{n}(t)}$
relative to the instantaneous Bogoluibov dispersion $E_{kf}(t)$,
clearly only possible for sufficiently large momenta.

To fully account for the self-consistent dynamics of $n_c(t)$, here we
solve iteratively the full set of equations \eqref{eqmndot} (or
equivalently Eqs.\eqref{exacteom}, \eqref{exactinicona}\eqref{exactiniconb}) and
\eqref{exactsc}. With this solution in hand we can compute an
arbitrary physical quantity.

Focussing on experimentally accessible momentum distribution, we
compute
\begin{equation}
\begin{split}
n_k(t)&=\langle 0^-|\hat a^{\dagger}_\kv(t)\hat a_\kv(t)|0^- \rangle\\
&=|(u_k(t)\sinh\Delta\theta_k-v^*_k(t)\cosh\Delta\theta_k)|^2,
\end{split}
\end{equation}
together with the atom number self-consistency condition
\begin{equation}
\begin{split}
\bar{n}(t)&=1-\frac{8}{\sqrt{\pi}}(2na^3_f)^{1/2}\int d\hat{k}{\hat{k}}^2|(u_k(t)\sinh\Delta\theta_k\\
&\quad-v^*_k(t)\cosh\Delta\theta_k)|^2.
\end{split}
\label{exactsc}
\end{equation}
We illustrate the results in Figs.~\ref{nktplot},\ref{ndtplot}, from
which we observe that the numerically computed $n_k(t)$ and $n_d(t)$
quite closely qualitatively resemble the approximate quasi-adiabatic
counterparts. Yet, they differ quantitively, particularly in the case
of deep quench and for small momenta. 
The asymptotic time-averaged value of $n_d(t)$ always considerably
exceeds the corresponding ground state depletion and thus the
pre-thermalized system remains out of equilibrium.

\begin{figure}[htb]
 \centering
  \includegraphics[width=80mm]{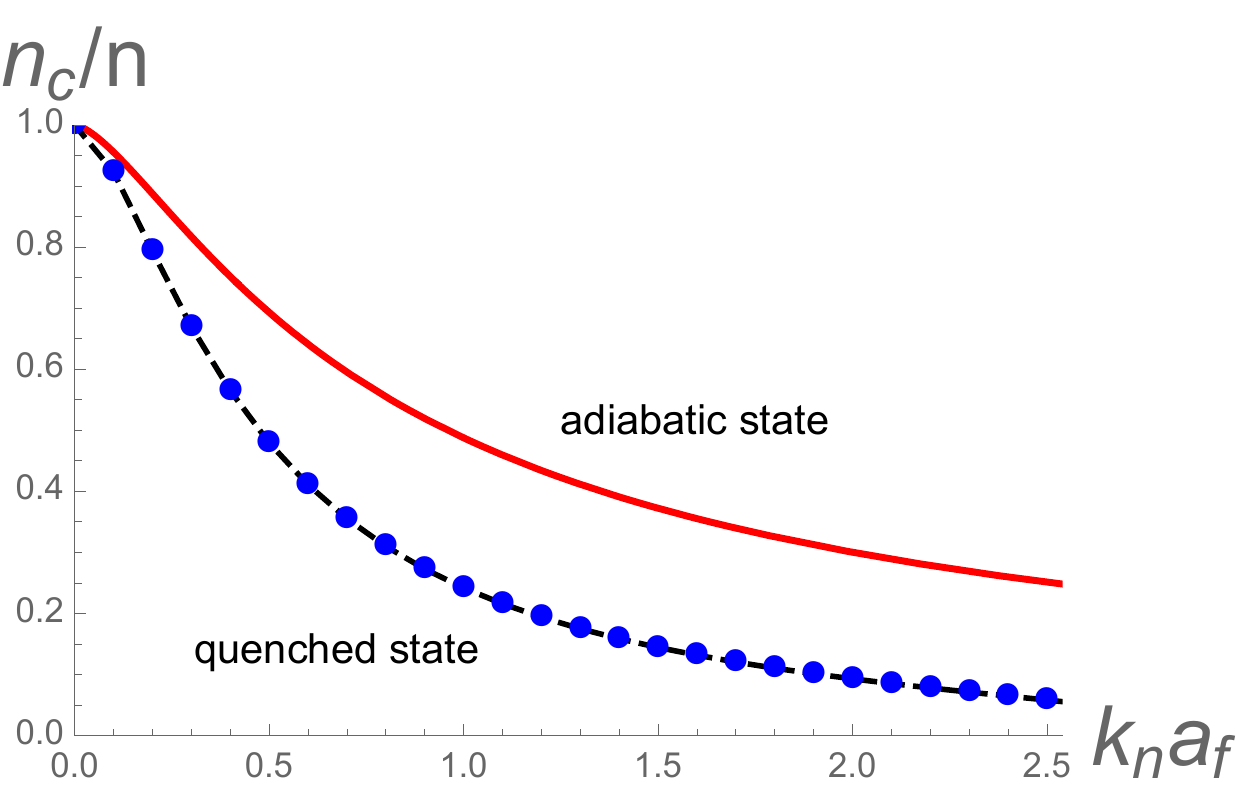}
 \caption{(Color online) Quenched steady-state condensate fraction
   (dash-dotted blue curve) as a function of $k_na_f$, following a
   quench from $k_na_i=0.01\to k_na_f$ (where $k_n\equiv n^{1/3}$), as compared to the ground
   state condensate fraction at $k_n a_f$ (solid red curve, same as in
   Fig. 2), both calculated within self-consistent dynamic field
   approximation. }
\label{ndtnumerics}
\end{figure}

\begin{figure}[htb]
 \centering
  \includegraphics[width=80mm]{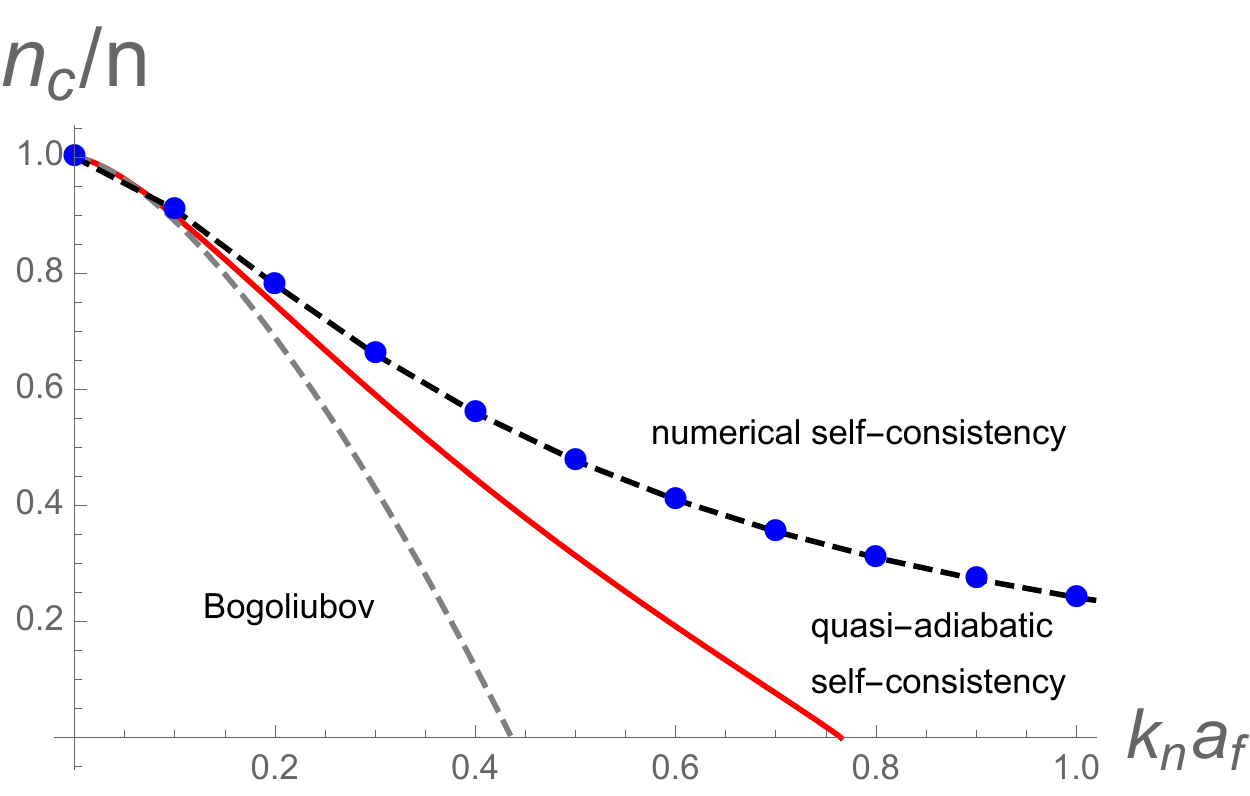}
 \caption{(Color online) Comparison of results from three different
   approaches to computation of the post-quench steady state
   condensate fraction $n_c/n$: numerical self-consistency
   (dash-dotted blue), quasi-adiabatic self-consistency (solid red)
   and Bogoluibov (dashed grey).}
\label{comparison3ways}
\end{figure}

In Fig.~\ref{comparison3ways} we compare the numerical solution with
corresponding quantities obtained via various approximate approaches
of previous sections. We find that for $k_na_f\ll 1$, both the
quasi-adiabatic self-consistent solution and numerical self-consistent
solution, reduce to that of a straight Bogoluibov approximation, but
deviate with increasing depth of the quench, $k_na_f$.
We observe that in contrast to the adiabatic approximation, the full
numerical solution predicts that the condensate fraction remains finite 
for arbitrary large $k_n a_f$, arguing that our earlier conjecture of 
a nonequilibrium phase transition to a ``normal'' state is likely
incorrect \cite{YinLR14}. 

\subsection{Generalized Gibbs Ensemble}
In the analysis above we found that following a scattering length
quench a nonequilibrium state, characterized by a stationary momentum
distribution function of atoms emerges in the long time limit. It is
thus natural to explore whether this state can be captured by a
Generalized Gibbs Ensemble (GGE) \cite{Rigola,RigolOlshanniNaturePRL07}.

At the simplest level of harmonic Bogoluibov description, the final
stationary state is completely determined by the initial post-quench
momentum distribution function of the quasi-particles $\hat \beta_k$. The
latter is in turn specified by the initial, $a_{i}$ and final
$a_{f}$ scattering lengths, i.e., by the initial ground state
$|0^-\rangle$ (vacuum of $\hat \alpha_k$) and the post-quench Hamiltonian
$\hat H(0^+)$, through the relation \rf{betaalpha} derived in
Sec.~\ref{sec:ShallowQuench}.

Since at this harmonic level the energy eigenvalues $E_{kf}$ for each
momentum are separately conserved, the distribution of $\hat\beta_\kv$
occupations can clearly be captured with GGE
\begin{equation}
\begin{split}
\hat\rho_{GGE}=Z^{-1}_{GGE}e^{-\sum_\kv\lambda_k E_{kf}\hat \beta^\dagger_\kv\hat \beta_\kv},
\end{split}
\label{rhogge}
\end{equation}
where $Z_{GGE}=Tr\left[e^{-\sum_\kv\lambda_\kv
    E_{kf}\hat \beta^\dagger_\kv\hat \beta_\kv}\right]$ and $\lambda_k$ are the
Lagrange multipliers (inverse of effective temperatures) for each
conserved mode $\kv$. These are fixed by requiring 
\begin{eqnarray}
n_k^\beta&\equiv&\langle 0^-|\hat \beta^\dagger_\kv\hat \beta_\kv|0^-\rangle
=\langle\hat \beta^\dagger_\kv\hat \beta_\kv\rangle_{GGE} \equiv Tr(
   \hat  \beta^\dagger_\kv\hat \beta_\kv\hat\rho_{GGE}).\nonumber\\
\label{rhogge}
\end{eqnarray}
The analysis from Sec.~\ref{sec:ShallowQuench} gives the left hand side
\begin{equation}
n_k^{ \beta} = \frac{1}{4}\left(\frac{E_{kf}}{E_{ki}}
+\frac{E_{ki}}{E_{kf}}\right)-\oh, 
\end{equation}
determing 
\begin{equation}
\lambda_k=\frac{1}{E_{kf}}\ln\left(\frac{n_k^{ \beta} + 1}{n_k^{ \beta}}\right).
\end{equation}

We now want to see if the long-time atomic momentum distribution
function $n_k(t\rightarrow\infty)$ can be characterized by the GGE.

\subsubsection{shallow quench}
For a shallow quench, captured by purely harmonic Bogoluibov 
approximation we have
\begin{equation}
\begin{split}
n_k(t)&=\langle 0^-| \hat a^\dagger_\kv(t)\hat a_\kv(t)|0^-\rangle,\\
&=v^2_k + (u^2_k+v^2_k)\langle\hat \beta^\dagger_\kv(t)\hat \beta_\kv(t)\rangle\\
&\quad-u_kv_k\langle\hat \beta_\kv(t)\hat \beta_{-\kv}(t)
+\hat \beta^\dagger_\kv(t)\hat \beta^\dagger_{-\kv}(t)\rangle.
\end{split}
\label{rhogge}
\end{equation}
In the long time limit, the time-dependence of the off-diagonal last
terms dephases away, and only first two terms survive. The steady-state momentum distribution $n^{ss}_k$ then becomes
\begin{equation}
\begin{split}
n^{ss}_k= v^2_k + (u^2_k+v^2_k)\langle\hat \beta^\dagger_\kv\hat \beta_\kv\rangle
\end{split}
\label{rhogge}
\end{equation}
Since $\langle \hat \beta^\dagger_\kv\hat \beta_\kv\rangle=\langle 
\hat \beta^\dagger_\kv\hat \beta_\kv\rangle_{GGE}$, it is clear that in this
purely harmonic approximation the GGE does describe the steady-state distribuition. 

\subsubsection{deep quench}

As we demonstrated in previous subsections, for a deep quench, a
self-consistency of condensate density must be implemented. This
results to an effective time dependent Hamiltonian. In the simplest
quasi-adiabatic approximation, we find
\begin{equation}
\begin{split}
n_k(t)&=v^2_k(t) + (u^2_k(t)+v^2_k(t))\langle\hat \beta^\dagger_\kv\hat \beta_\kv\rangle\\
&\quad-u_k(t)v_k(t)\langle\hat \beta_\kv(t)\hat \beta_{-\kv}(t)+\hat \beta^\dagger_\kv(t)\hat \beta^\dagger_{-\kv}(t)\rangle.
\end{split}
\label{rhogge}
\end{equation}
This leads to a steady-state distribution
\begin{equation}
\begin{split}
n^{ss}_k=(v^{ss}_k)^2 + ((u^{ss}_k)^2+(v^{ss}_k)^2)\langle\hat \beta^\dagger_\kv\hat \beta_\kv\rangle,
\end{split}
\label{rhogge}
\end{equation}
where
\bse
\begin{align}
u^{ss}_k&=\sqrt{\frac{1}{2}\big(\frac{\epsilon_k+n^{ss}_cg_f}{E_{kf}}+1\big)},\\
v^{ss}_k&=-\sqrt{\frac{1}{2}\big(\frac{\epsilon_k+n^{ss}_cg_f}{E_{kf}}-1\big)},\\
E_{kf}&=\sqrt{{\epsilon_k}^2+2n^{ss}_cg_f\epsilon_k},
\end{align}
\ese
and $n^{ss}_c$ the steady-state condensate density determined by the
self-consistency condition. The latter spoils the GGE description of
the long-time distribution even in this approximation. 

Indeed beyond the quasi-adiabatic approximation the inability of GGE to capture the long-time distribution is clear from the avoided sharp phase transition from superfluid phase to normal phase, illustrated in Fig.~\ref{comparison3ways}.

\section{Excitation energy}
\label{appendix:excitationenergy}

We now turn to a study of the excitation energy $E_{exc}$ following a quench,
defined by
\begin{equation}
\begin{split}
E_{exc}=\langle 0^{-}|\hat H^f|0^{-}\rangle-\langle 0_f|\hat H^f|0_f\rangle,
\end{split}
\label{excenergy}
\end{equation}
as the difference between the expectation value of the post-quench
Hamiltonian in the initial state and the ground state energy of the
same Hamiltonian. For a closed system and unitary energy conserving
dynamics, this quantity is an important measure of the long time
nonequilibrium stationary state, and in particular the resulting
temperature for the equilibrated state.

Below, we first study $E_{exc}$ within perturbative Bogoluibov
approximation valid for a shallow sudden quench and a dilute gas
characterized by $n a_s^3\ll 1$.  Within this approximation the ground
state energy with repulsive interactions (i.e., here for a resonant
problem ignoring the bound molecular state \cite{JasonHoPaperOnUpperBranch,ZhouPRA08}) is given by the LHY result
\begin{equation}
\begin{split}
\label{LHYenergy}
E_{gs}=\langle 0_f|\hat H^f|0_f\rangle=\frac{2\pi n a_f}{m}\left[1+\frac{128}{15\sqrt{\pi}}({na^3_f})^{1/2}\right].
\end{split}
\end{equation}
Our focus is then on the calculation of $\langle
0^{-}|\hat H^f|0^{-}\rangle$.

We will then generalize this analysis to arbitrary strength
interactions, relating the excitation energy to Tan's
contact \cite{Tan}. We then conclude by studying the excitation energy
for a finite-rate ramp.

\subsection{Sudden quench}

\subsubsection{Bogoluibov approximation}

Within a sudden quench Bogoluibov approximation a straightforward
analytical treatment is possible. To this end, leaving details to Appendix~\ref{appendix:energyafterquench},
we expand the Hamiltonian about the condensed state,
\begin{eqnarray}
\hspace{-0.5cm}
\hat{H}_f&\approx&\frac{g_f}{2V}N^2 +
\oh\sum_{\kv\neq 0}\left[(\epsilon_k + g_f n) \hat a_\kv^\dagger \hat a_\kv 
 + g_f n \hat a_{-\kv} \hat a_\kv + h.c. \right],\nonumber\\
\end{eqnarray}
that to quadratic order can be diagonalized as analyzed in
Sec.~\ref{sec:BdG}, giving
\begin{eqnarray}
\hat H_f&=&\oh g_f n^2 V - \sum_{\kv\neq 0}\left[\epsilon_k+g_f n_c(0^+) -
  E_{kf}(0^+)\right]\nonumber\\
&+&\sum_{\kv\neq 0} E_{kf}(0^+)\hat \beta_\kv^\dagger\hat \beta_\kv.
\end{eqnarray}

The first two constant terms give the LHY ground-state energy (with UV cutoffs in the second term cancelled by the cutoff dependent terms coming from $g_f$ in the first term after it is expressed in terms of scattering length, $a_f$ as detailed in Appendix~\ref{appendix:energyafterquench}.). They clearly cancel in the subtraction in Eq.~\eqref{excenergy}, giving excitation energy
density ${\cal E}_{exc}\equiv E_{exc}/V$
\begin{eqnarray}
  {\cal E}_{exc}&=&\frac{1}{V}\sum_{\kv\neq 0} E_{kf}(0^+)\left[\langle
    0^-|\hat\beta_\kv^\dagger\hat\beta_\kv|0^-\rangle
    -\langle 0_f|\hat \beta_\kv^\dagger\hat \beta_\kv|0_f\rangle\right].\nonumber\\
  \label{eps}
\end{eqnarray}
The last term vanishes at $T=0$, since by definition $|0_f\rangle$ is a
vacuum of $\hat \beta_\kv$. Given that $|0^-\rangle$ is a vacuum of the
Bogoluibov quasi-particles $\hat \alpha_\kv$ associated with the pre-quench
Hamiltonian, $\hat H_i$, it is convenient to express $\hat \beta_\kv$ in terms
of $\hat \alpha_\kv$, using the relations \rf{betaalpha}, \rf{eq65} worked out in
Sec. ~\ref{sec:ShallowQuench}. Evaluating the expectation value
\begin{eqnarray}
\langle 0^-|\hat \beta_\kv^\dagger\hat \beta_\kv|0^-\rangle
&=&\sinh^2\Delta\theta,\\
\nonumber\\
&=&\oh\left[\frac{\epsilon_k+(g_f+g_i)n}
{\sqrt{(\epsilon_k+2g_in)(\epsilon_k+2g_fn)}}-1\right],\nonumber
\end{eqnarray}
gives
\begin{eqnarray}
  {\cal E}_{exc}&=&\oh\int\frac{d^3k}{(2\pi)^3}
\sqrt{\epsilon_k^2+2g_fn\epsilon_k}\nonumber\\
&&\times\left[\frac{\epsilon_k+(g_f+g_i)n}
{\sqrt{(\epsilon_k+2g_in)(\epsilon_k+2g_fn)}}-1\right].
\end{eqnarray}
Simple analysis shows that ${\cal E}_{exc}$ exhibits a (UV divergent)
contribution 
\begin{eqnarray}
  {\cal E}_{exc}^\Lambda&=&\oh\int^\Lambda\frac{d^3k}{(2\pi)^3}
\frac{(g_f-g_i)^2n^2}{2\epsilon_k},\nonumber\\
&=&\frac{m n^2}{4\pi^2 }
(g_f-g_i)^2\Lambda.
\label{eps_excLambda}
\end{eqnarray}
set by the microscopic range $r_0 \sim 1/\Lambda$ of the
two-body potential. This remains the case even when the couplings
$g_{i,f}$ are eliminated in favor of the physical scattering lengths
$a_{i,f}$, using
\begin{subequations}
\begin{align}
g &= \frac{\tilde{g}}{1 - \frac{m}{2\pi^2 }\tilde{g}\Lambda}
=\frac{4\pi }{m}\frac{a_s}{1 - \frac{2}{\pi}a_s\Lambda},\\
&\approx\frac{4\pi }{m} a_s(1 + \frac{2}{\pi}a_s\Lambda),
\label{gTOa}
\end{align}
\end{subequations}
and to first order of $a_s\Lambda$ (assuming $a_s\Lambda\ll 1$)
\begin{eqnarray}
  {\cal E}_{exc}^\Lambda&=&4(1-\sigma)^2 \frac{  n^2a_f}{m}a_f\Lambda.
\label{eps_excLambdanew}
\end{eqnarray}

The remaining finite part of ${\cal E}_{exc}$ is then given by
$\tilde{\cal E}_{exc} = {\cal E}_{exc} - {\cal E}_{exc}^\Lambda$, 

\begin{equation}
\tilde{\cal E}_{exc}=
-\frac{128\pi^{1/2}  a_f n^2}{15m}(n a_f^3)^{1/2}
\left[\sigma^{3/2}(3\sigma-5)+2\right].
\label{eps_exc}
\end{equation}

It is negative for all $\sigma=g_i/g_f$ and leads to

\begin{equation}
\label{eps_excR}
\begin{split}
{\cal E}_{exc}&=4(1-\sigma)^2 \frac{  n^2a_f}{m}a_f\Lambda\\
&\quad-\frac{128\pi^{1/2}}{15}\frac{  n^2 a_f }{m}(n a_f^3)^{1/2}
\left[\sigma^{3/2}(3\sigma-5)+2\right],
\end{split}
\end{equation}

This expression vanishes as $(\sigma-1)^2$ in no quench $\sigma = 1$ limit.  Although a negative finite correction
$\tilde{\cal E}_{exc}$ is disconcerting, the total excitation energy
density ${\cal E}_{exc}$ is indeed positive in the dilute regime $(n
a_f^3)^{1/2}\ll 1 \ll a_f\Lambda$, required for the validity of the
Bogoluibov approximation \cite{commentEexc_positive}.


The potential-range (UV cutoff) dependence of ${\cal E}_{exc}$ may at
first sight appear surprising (even when expressed in terms of the
physical scattering lengths, that renders all equilibrium properties
finite). However, as we will see below, this result arises from an
unphysical feature of the model protocol, namely an infinitely fast
quench. We reexamine this UV dependence below by studying a more
physical model with a finite-rate ramp.

\subsubsection{beyond Bogoluibov approximation and relation to Tan's contact}

Below we present a more general analysis of the excitation energy,
without relying on the expansion about the condensed state, by relating
it to other physical quantities like the ground state energy and Tan's
contact \cite{Tan}.

We begin with the basic model Hamiltonian of resonant bosons
\begin{widetext}
\begin{eqnarray}
\hat{H}_f&=&\sum_{\kv\neq 0}\epsilon_k \hat a_\kv^\dagger \hat a_\kv+\frac{g_f}{2V}\sum_{\kv_1,\kv_2,\qv}\hat a_{-\kv_1+\qv/2}^\dagger
\hat a_{\kv_1+\qv/2}^\dagger \hat a_{-\kv_2+\qv/2} \hat a_{\kv_2+\qv/2},\nonumber\\
&=&\hat H_i+\frac{g_f-g_i}{2V}\sum_{\kv_1,\kv_2,\qv}\hat a_{-\kv_1+\qv/2}^\dagger
\hat a_{\kv_1+\qv/2}^\dagger \hat a_{-\kv_2+\qv/2} \hat a_{\kv_2+\qv/2},\nonumber\\
\end{eqnarray}
\end{widetext}
where the bare interaction coupling $g$ is expressible in terms of
the renormalized coupling $\tilde{g}^{-1} = g^{-1}+ m\Lambda/(2\pi^2 )$,
related to the scattering length $a_s(g)$,
\begin{eqnarray}
\tilde{g}=\frac{4\pi  a_s}{m} = \frac{g}
{1 + m\Lambda g/(2\pi^2 )}. 
\label{gRas}
\end{eqnarray}

With the initial (pre-quench) state $|0^-\rangle\equiv|0_i\rangle$
the vacuum of the pre-quench Hamiltonian, $\hat H_i$, the excitation
energy density is then given by
\begin{widetext}
\begin{eqnarray}
E_{exc}&=&\langle 0_i|\hat H_i|0_i\rangle-\langle 0_f|\hat H_f|0_f\rangle
+\frac{g_f-g_i}{2V}\sum_{\kv_1,\kv_2,\qv}\langle 0_i|\hat a_{-\kv_1+\qv/2}^\dagger
\hat a_{\kv_1+\qv/2}^\dagger \hat a_{-\kv_2+\qv/2} \hat a_{\kv_2+\qv/2}|0_i\rangle,\nonumber\\
&=&E^i_{gs}-E^f_{gs}
+\frac{g_f-g_i}{2V}\sum_{\kv_1,\kv_2,\qv}\langle 0_i|\hat a_{-\kv_1+\qv/2}^\dagger
\hat a_{\kv_1+\qv/2}^\dagger \hat a_{-\kv_2+\qv/2} \hat a_{\kv_2+\qv/2}|0_i\rangle.
\label{Eexc_formal}
\end{eqnarray}
\end{widetext}
For a dilute weakly interacting gas, $n a_s^3\ll 1$, we can evaluate the
first two (ground state energy) terms within Bogoluibov approximation
for the initial and final Hamiltonians, using the LHY result,
Eq.~\rf{eps_gsLHY} for $g_i$, $g_f$. The last term can be related to
Tan's contact.

To this end, we first note that the expectation value of the quartic
interaction is related to Tan's contact \cite{Tan,Braaten},
\begin{eqnarray}
  C&=&(mg)^2
  \langle\hat \psi^\dagger\hat \psi\hat \psi^\dagger\hat \psi\rangle,
\label{Cpsi4}
\end{eqnarray}
that in Bogoluibov approximation is given by
\begin{eqnarray}
  C \approx (4\pi n
  a_s)^2\left(1+\frac{64}{3}(n a^3_s/\pi)^{1/2}\right),
  \label{Contact}
\end{eqnarray} 
and is UV cutoff $\Lambda = 1/r_0$ independent. The ground state
energy density is also expressible in terms of the contact
\bse
\begin{eqnarray}
  {\cal E}_{gs} &=& \frac{1}{V}\sum_{\kv}\epsilon_k\left(n_k - \frac{C}{k^4}\right) 
+ \frac{  C}{8\pi m a_s},\\
&\approx& \frac{2\pi 
  n^2a_s}{m}\left(1 + \frac{128}{15}(n a^3_s/\pi)^{1/2}\right),
\label{EgsBogoluibov}
\end{eqnarray}
\ese
with the last equality computed within the Bogoluibov limit.

Using Eq.~\rf{Eexc_formal} and \rf{Cpsi4}, the excitation energy
density is thus given by:
\begin{eqnarray}
{\cal E}_{exc}&=&{\cal E}^i_{gs}-{\cal E}^f_{gs}
+\oh(g_f-g_i)\langle 0_i|\hat \psi^\dagger
\hat \psi\hat \psi^\dagger\hat \psi|0_i\rangle,\nonumber\\
&=&{\cal E}_{gs}^i-{\cal E}_{gs}^f
+\frac{(g_f-g_i)}{2m^2g_i^2}C_i.
\end{eqnarray}

Recalling from scattering analysis, that the microscopic UV cuttoff-dependent
interaction $g$ is given by
\begin{eqnarray}
  g&=&\frac{4\pi  a}{m}\left(1-\frac{2}{\pi} a_s/r_0\right)^{-1},
\label{g_as}
\end{eqnarray}
allows us to express ${\cal E}_{exc}$ in terms of the more physical
scattering lengths
\begin{eqnarray}
  {\cal E}_{exc}
  &=&{\cal E}^i_{gs}-{\cal E}^f_{gs}
  +\frac{ C_i}{8\pi m a_i^2}(a_f-a_i)
  \frac{1-\frac{2}{\pi}a_i/r_0}{1-\frac{2}{\pi}a_f/r_0}.\;\;\;\;\;\;\;\;\;\;
  \label{Eexc_formalContact}
\end{eqnarray}
As is clear from $a_s(g)$ in \eqref{gRas} plotted in
Fig.~\ref{fig:frdiagram}, the scattering length falls into two distint
ranges $0 < |a_s| < \oh\pi r_0$ and $|a_s| > \oh\pi r_0$, where from
\rf{gRas} the latter is only accessible for attractive interactions,
$g < 0$.  Analyzing above expression in the first range and within the
Bogoluibov approximation (using \rf{Contact},\rf{EgsBogoluibov}), to
lowest order we recover the UV cutoff dependent result
\rf{eps_excLambdanew} of the previous subsection,
\begin{eqnarray}
  {\cal E}_{exc}
  &\approx&\frac{2\pi n^2}{m}(a_i - a_f)\left[
1-\frac{1-\frac{2}{\pi}a_i/r_0}{1-\frac{2}{\pi}a_f/r_0}\right],\;\;\;\;\;\;\\
  &\approx&\frac{4 n^2}{m}(a_i - a_f)^2/r_0
\label{Eexc_Bogoluibov}
\end{eqnarray}

In the complementary more physically interesting regime $|a_s| > \oh\pi
r_0$, we instead have
\begin{eqnarray}
{\cal E}_{exc}&=&{\cal E}^i_{gs}-{\cal E}^f_{gs}
+\frac{ C_i}{8\pi m}(a_i^{-1}-a_f^{-1}),\nonumber\\
\label{Eexc_formalContact2}
\end{eqnarray}
that in the Bogoluibov limit $n a^3_s\ll 1$ (i.e., $r_0\ll |a_s| \ll
n^{-1/3}$) reduces to 
\begin{eqnarray}
{\cal E}_{exc}&\approx&\frac{4\pi n^2a_i}{m}\left[1-\oh\left(\frac{a_f}{a_i}
+\frac{a_i}{a_f}\right)\right].
\label{Eexc_formalContact3}
\end{eqnarray}
For weak (no bound state) attractive interactions $a_i < 0$ this
expression is positive and as required vanishes for the case of
no-quench, $\sigma = a_f/a_i=1$.

For a strong resonant interactions, beyond Bogoluibov regime,
excitation energy reduces to
\begin{eqnarray}
{\cal E}_{exc} &=& \frac{1}{V}\sum_{\kv}\epsilon_k
\left(\delta n^i_k - \delta n^f_k\right) 
+ \frac{1 }{4\pi m}\left[\frac{C_i}{a_i}
-\oh \frac{C_i + C_f}{a_f}\right],\nonumber\\
\label{EgsStrong}
\end{eqnarray}
where $\delta n_k\equiv n_k - C/k^4$ is the momentum distribution with
large momentum tail subtracted off.

We observe, that for $a_s > 0$ the excitation energy appears to be
negative. However, in this regime, $a_s > \oh\pi r_0$, for $a_s > 0$
the interaction $g$ is necessarily attractive (see \rf{gRas} showing
that for $g > 0$, $a_s$ is limited below $\oh\pi r_0$) and exhibits a
molecular bound state that lies below atomic BEC continuum. Thus the
initial purely atomic condensate state with $a_s > 0$ is therefore not
a ground state (the molecular bound state is) and thus there is no a
priori reason to expect for the change in energy to be positive under
a quench. We thus conjecture that the negative excitation energy
$\eps_{exc} < 0$ is a reflection of such resonant interaction.

Finally, as we will show next, the UV cutoff dependent excitation
energy, \rf{Eexc_Bogoluibov} is a reflection of the unphysical
infinitely fast quench, a divergence that in a more physical situation
of a finite-rate ramp is cut off by the ramp rate.

\subsection{Finite-rate ramp}

In this subsection we analyze the excitation energy following a finite-rate ramp of the coupling strength, for simplicity focussing on a
linear ramp, defined by Eq.~\eqref{gt}, \eqref{dimensionlessgt} in
Sec.~\ref{finiteRateRamp}, characterized by a dimensionless rate
$\gamma$ and related ramp time $\tau\equiv (1-\sigma)
/(ng_f\gamma)$. Below we will show that above short-scale divergence
for a sudden quench is regularized by a finite ramp rate $\gamma$.

\subsubsection{scaling analysis}

To this end we first conjecture that for a finite-rate ramp (nonzero
ramp time) the dominant singular part of excitation energy is
generalized to
\begin{eqnarray}
{\cal E}_{exc}^\Lambda(\gamma)
&=&\frac{4(\sigma - 1)^2  n^2 a_f}{m}a_f\Lambda f(E_\Lambda\tau ),\\
&=&\frac{4(\sigma - 1)^2  n^2 a_f}{m}a_f\Lambda f\left(E_\Lambda(1-\sigma)/(ng_f\gamma)\right),\nonumber
\label{excenergyramp}
\end{eqnarray}
where $E_\Lambda={\Lambda^2}/{2m}\approx 1/(2m r_0^2)$ is the UV
cutoff energy scale (corresponding to range of the potental
$r_0\sim\Lambda^{-1}$), that sets the ramp rate scale.

We can deduce the asymptotic form of the scaling function $f(x)$ from
the knowledge of the behavior of ${\cal E}_{exc}^\Lambda(\gamma)$ in
sudden quench and adiabatic limits. In the former case of
$\gamma\rightarrow \infty$, cleary $f(x)=1$ so that \rf{eps_excLambdanew} is
recovered. In the latter case of $\gamma\rightarrow 0$, we expect the
system to track the ground state and thus
${\cal E}_{exc}^\Lambda(\tau\rightarrow\infty)\rightarrow 0$, and require
for the UV cutoff $\Lambda$ to drop out.

The latter condition thus requires that $f(x\rightarrow\infty)=\kappa/\sqrt{x}$ 
($\kappa$ is a dimensionless constant), so that
\bse
\begin{eqnarray}
{\cal E}_{exc}^\Lambda(\gamma)
&\underset{\gamma\rightarrow 0}{=}
&\kappa \frac{8\sqrt{2\pi}n^2 a_f}{m}(n a_f^3)^{1/2}(1-\sigma)^{3/2}
\sqrt{\gamma},\;\;\;\;\;\;\;\;\;\quad\\
&\equiv&\frac{1}{4}\kappa{\cal E}_0(1-\sigma)^{3/2}
\sqrt{\gamma},
\label{scalingFunction}
\end{eqnarray}
\ese
scaling as the square-root of the ramp rate, with ${\cal E}_0\equiv
32\sqrt{2\pi}\frac{n^2 a_f}{m}(n a_f^3)^{1/2}$. 
This is consistent with the general predictions \cite{Polkovnikovnaturephysics}, with the
specific exponent of $1/2$ appearing here.

Before turning to a more microscopic analysis, we note that an
estimate of experimental ramp rate is $\gamma\approx 10^{-10}$ eV and
of UV energy cutoff $E_\Lambda\approx 10^{-7}$ eV
\cite{Makotyn}. Thus, in JILA experiments $E_\Lambda/\gamma\gg 1$,
with the finite ramp rate expecting to cutoff the dependence on the
microscopic cutoff $\Lambda$, and the excitation energy scaling
proportional to ${\cal E}_{exc}\sim \sqrt{\gamma}$.

\subsubsection{microscopic and numerical analysis}

As a complementary approach, we can use a microscopic model of a
finite-rate ramp protocol, Sec.~\ref{finiteRateRamp}, together with a
numerical analysis to compute the resulting excitation energy. 

Leaving the detailed calculations to Appendix~\ref{appendix:energyafterquench} we
find that the energy right after the finite-rate ramp is given by
\begin{widetext}
\begin{equation} 
\begin{split}
  {\cal E}_{total}
&=2\pi\frac{  n^2a_f}{m}
+32\sqrt{2\pi}\frac{ n^2a_f}{m}({na^3_f})^{1/2} \int dk k^2\bigg[(k^2+1)|v_k(\tau)|^2-\frac{1}{2}(u_k(\tau)v^*_k(\tau)+v_k(\tau)u^*_k(\tau))+\frac{1}{4k^2}\bigg],
\end{split}
\end{equation}
\end{widetext}
where $u_k(t),v_k(t)$ are solutions of Eqs.\rf{finiteeoma}\rf{finiteeomb} (see Eq.~\eqref{excenergyfiniteramp}). 
Subtracting the LHY ground state energy density ${\cal E}_{gs}$ \rf{LHYenergy}, the excitation
energy density is then given by 
\begin{equation}
{\cal E}_{exc}={\cal E}_{total}-{\cal E}_{gs}={\cal E}_0 f(\sigma,\Lambda,\gamma),
\end{equation}
where
\begin{equation}
\begin{split}
\label{excfactorF}
f(\sigma,\Lambda,\gamma)&=\int dk k^2\bigg[(k^2+1)|v_k(\tau)|^2\\
&\quad-\frac{1}{2}\left(u_k(\tau)v^*_k(\tau)+v_k(\tau)u^*_k(\tau)\right)+\frac{1}{4k^2}\bigg]
-\frac{4\sqrt{2}}{15}\\
\end{split}
\end{equation}
is a dimensionless function that can be evaluated using numerical
solutions for $u_k(\tau)$ and $v_k(\tau)$.

\begin{figure}[htb]
 \centering
  \includegraphics[width=80mm]{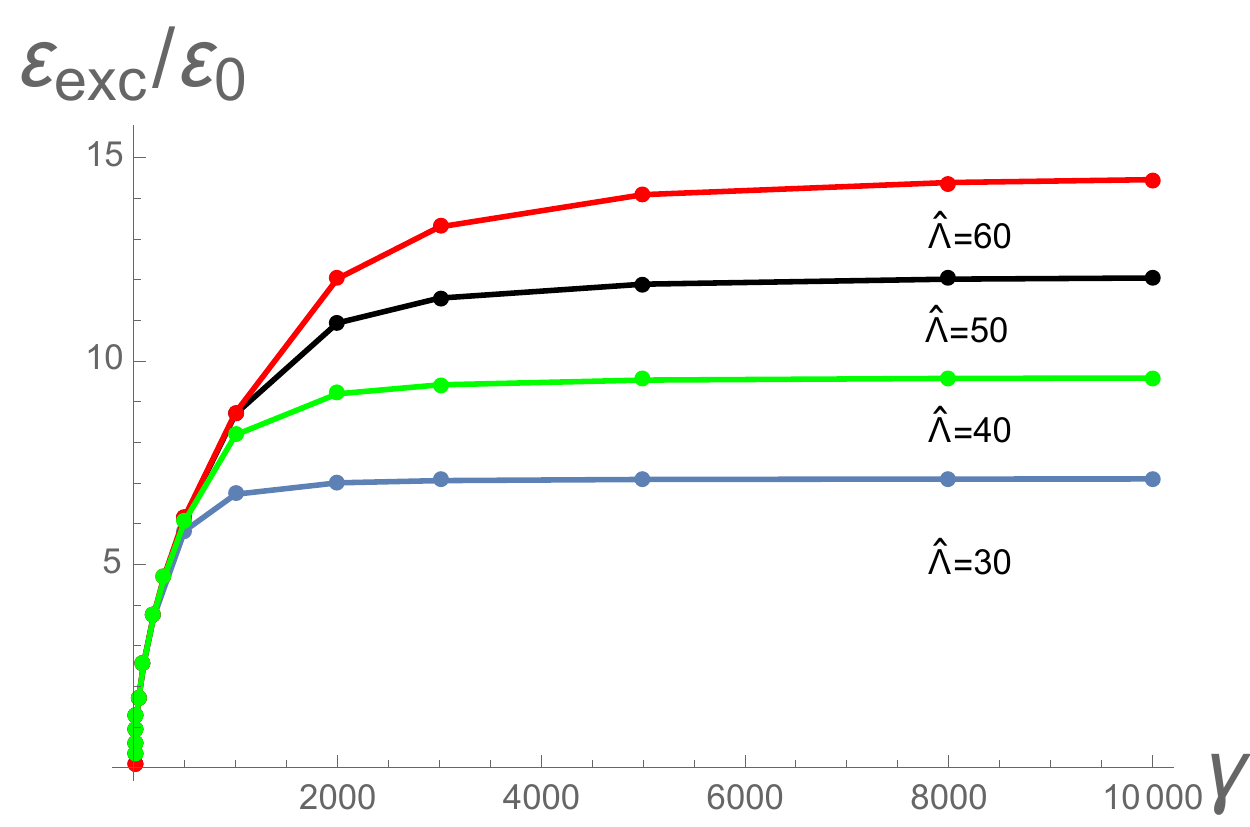}
 \caption{(Color online) Excitation energy (scaled by LHY correction
   to the ground state energy) following a scattering length ramp
   $0.5a_f\to a_f$ as a function of ramp rate $\gamma$ for different
   scaled momentum cutoff $\hat{\Lambda}=\Lambda\xi$ (here $\xi\equiv1/\sqrt{2 m n g_f}$ is the coherence length). For large ramp
   rate $\gamma$ (fast ramp), the excitation energy ${\cal E}_{exc}$
   (defined in the text) grows linearly with the UV cutoff $\Lambda$,
   while for small rate (slow ramp) the cutoff dependence drops out
   and is replaced by a square-root of the ramp rate $\gamma$.}
\label{excratedep}
\end{figure}

Displaying the results in Fig.~\ref{excratedep}, we observe that for a
small ramp rate $\gamma$, the cutoff dependence drops out of the
excitation energy, as curves with different cutoffs $\Lambda$
collapse. In the opposite limit of $\gamma\to\infty$, the excitation
energy recovers the linear cutoff-depedence displayed for the
sudden quench case, in Eq.~\eqref{eps_excLambdanew}.

\begin{figure}[htb]
 \centering
  \includegraphics[width=90mm]{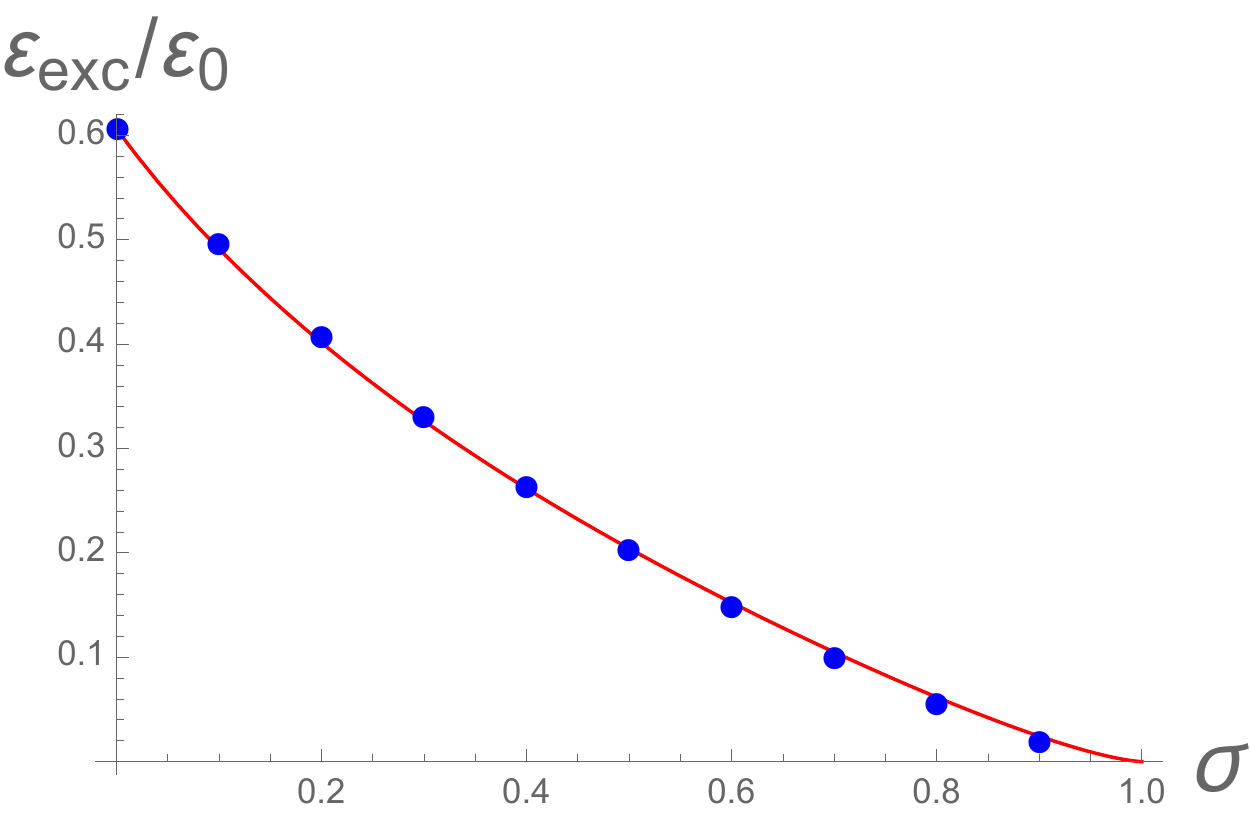}
 \caption{(Color online) Excitation energy (scaled by LHY correction to the ground state energy)  following a scattering length ramp from $a_i\to a_f$ as a function of quench depth $\sigma={a_i}/{a_f}$. The blue data points are obtained for each chosen $\sigma$ at dimensionless ramp $\hat\gamma\equiv (1 - \sigma) /(\tau n g_f)
  =10$ and scaled momentum cutoff $\hat{\Lambda}=\Lambda\xi=60$; the red curve represents fitting function
$y=0.58(1-x)^{3/2}$. }
\label{energysigma}
\end{figure}

We also verify the square-root prediction of the scaling theory for slow ramp rate, Eq.\eqref{scalingFunction}, in Fig.~\ref{energyrampplot}, by ``zooming-in'' Fig.~\ref{excratedep}. The $\sigma$ (quench depth) dependence in Eq.~\eqref{scalingFunction} is also confirmed by inspecting Fig.~\ref{energysigma}.

With this we conclude our analysis of the excitation energy and turn
to the study of the dynamic analog of Tan's contact \cite{Tan}.

\section{Contact} \label{sec:contact} 

\subsection{Ground state contact}

Contact, $C$ is a remarkable physical parameter introduced by
Tan \cite{Tan}, that enters in a large variety of physical
observables. Most notably, it appears as a coefficient of the universal
large momentum tail of the ground-state momentum distribution function 
\begin{eqnarray}
C&=&\lim_{k\to\infty}k^4 n_k
\label{Ck4}
\end{eqnarray}
and as a response of the ground-state energy density ${\cal E}_{gs}\equiv E_{gs}/V$ 
to the tuning of the scattering length, the so-called adiabatic theorem,
\bse
\begin{eqnarray}
C&=&-{8\pi m}{ }\frac{d{\cal E}_{gs}}{d a^{-1}_s},
\label{dEda}\\
&=&\left({m  g}{ }\right)^2
\langle\hat\psi^{\dagger}\hat\psi^{\dagger}\hat\psi\hat\psi\rangle,
\label{Cpsi4new}
\end{eqnarray}
\ese
with the second relation to the interaction energy (already noted in
the previous section) obtained via the Hellmann-Feynman
theorem \cite{Shankar}. As we show in Appendix~\ref{appendix:groundstatecontact},
above can be straightforwardly evaluated in the ground state within
the dilute Bogoluibov approximation \cite{refsBogoluibovContact}.
Though contact is quite different for fermions and bosons, in
equilibrium, these relations are expected to hold independent of
statistics.

The contact was first successfully measured in the ground state of
stable fermionic gases, with relations experimentally
verified \cite{JinRF}. More recently, the contact was studied
in a resonant bosonic gas via Bragg spectroscopy, utilizing the
adiabatic theorem, \rf{dEda}) \cite{Wild} and more directly
from the large frequency $1/\omega^{3/2}$ tail (frequency analog of
$1/k^4$ momentum tail, \rf{Ck4}; see \rf{Nequil}) of the RF
spectroscopy signal \cite{Wild}. However, because a
resonant Bose gas is fundamentally unstable and evaporates through
the three-body decay, these measurements are intrinsically
nonequilibrium, requiring a dynamical analysis of the contact. 

\subsection{Dynamical contact}
\label{sec:dynamicalcontact}
We thus examine the contact and its associated relations for a
resonant Bose gas dynamics following a quench. Immediately after the
quench the states remain unchanged $|0^-\rangle = |0^+\rangle$ and
only the coupling changes, $g_i\rightarrow g_f$. Thus, the relation
between two forms of $C$ defined in \rf{dEda} and \rf{Cpsi4new}
remains valid,
\begin{eqnarray}
  C_{dE}(0^+)&=&-{8\pi m}{ }\frac{d{\cal E}_f}{d a_f^{-1}}
\label{dEaf}\nonumber\\ 
  &=&\left({m  g_f}{ }\right)^2
  \langle 0^+|\hat\psi^{\dagger}\hat\psi^{\dagger}\hat\psi\hat\psi
|0^+\rangle\nonumber\\ 
&=&\left(\frac{g_f}{g_i}\right)^2 C_E(0^-),
\end{eqnarray}
despite the fact that $|0^+\rangle$ is not an eigenstate of
$\hat H(0^+)\equiv \hat H_f$ and thus Hellmann-Feynman theorem no longer
applies. 

However, the contact $C_{dE}$ is then clearly not continuous across the
quench, and using \rf{gRas},\rf{g_as} acquires a UV cutoff dependence
$\Lambda=1/r_0$, that drops out only in the $a_{i,f}\gg r_0$ limit
\bse
\begin{eqnarray}
\hspace{-0.5cm} 
C_{dE}(0^+)&=&\left(\frac{g_f}{g_i}\right)^2 C_E(0^-),\\
&=& C_E(0^-),\;\;\; \mbox{$a_{i,f}\gg r_0$},\\
&=&\left(\frac{a_f}{a_i}\right)^2
\left(1 + \frac{4}{\pi r_0}(a_f-a_i)\right) C_E(0^-),\;\;\; 
\mbox{$a_{i,f}\ll r_0$}\nonumber\\.
\label{contactt0}
\end{eqnarray}
\ese
This is consistent with cutoff dependence found in the excitation
energy, \rf{Eexc_Bogoluibov}. On the other hand the momentum
distribution function only depends on the state and is thus continuous
across the quench. Thus, the contact $C_n$, defined by the large
momentum tail of the distribution function, \rf{Ck4} is continuous
across the quench and is therefore distinct from $C_E$.

\begin{figure}[ht]
 \centering
  \includegraphics[width=80mm]{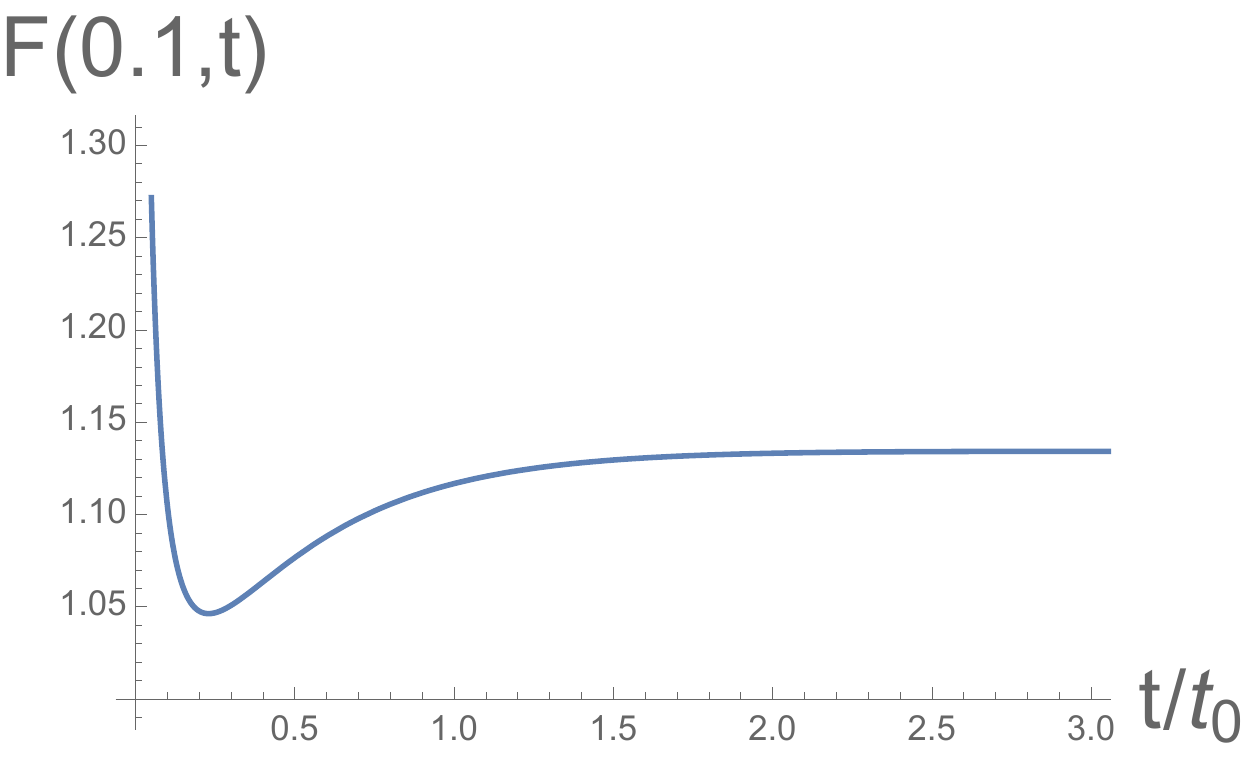}
 \caption{(Color online) Contact enhancement factor $F_C(\sigma,t)$
   above the corresponding ground state value as a function of time
   following a scattering length quench $0.1a_f\to a_f$, in units of
   pre-thermalization time scale $t_0=\hbar /ng_f=m/(4\pi a_f n
   \hbar)$. For a typical $^{85}$Rb experiment with $n=5\times10^{12}
   cm^{-3}$, $a_s=1100a_0$, $t_0\approx360\mu s$.}
\label{fig:contactsuddenquench}
\end{figure}

Utilizing the analysis of Sec.~\ref{sec:ShallowQuench}, we next
compute these contact quantities at time $t$ after the quench. We
first study the contact $C_E(t)$ defined by the quartic interaction,
\rf{Cpsi4new}. Relegating the calculation details to
Appendix~\ref{app:dynamicalcontact}, within the Bogoluibov
approximation we find
\begin{eqnarray}
\label{dynamicalcontactLHY}
C_E(t)=(4\pi n a_f)^2+F_C(\sigma,t)C^f_{LHY},
\end{eqnarray}
where the $C^f_{LHY}$ is the LHY correction to the ground state contact for quenched Hamiltonian with $a_f$ 
\begin{eqnarray}
C^f_{LHY}=(4\pi n a_f)^2\frac{64}{3\sqrt{\pi}}(na^3_f)^{1/2}
\end{eqnarray}
and the time-dependent enhancement factor due to the quench is given by
\begin{equation}
\label{contact_h}
\begin{split}
F_C(\sigma,t)&=\frac{\sigma^{3/2}+3\sqrt{\sigma}
+3\sqrt{1-\sigma}\mathrm{arccos}{\sqrt{\sigma}}}{4}\\
&\quad+\frac{3\sqrt{2}}{8}\int
dyy^2\frac{y(1-\sigma)}{(y^2+2)\sqrt{y^2+2\sigma}}\\
&\qquad\times\cos[2\hat{t}\sqrt{y^2(y^2+2)}]\\
\end{split}
\end{equation}
and illustrated in Fig.~\ref{fig:contactsuddenquench}. 

Immediately after the quench, at $t=0^+$, the quantity $F_C(\sigma,t)$ can be evaluated analytically, giving the contact 
\begin{equation}
\begin{split}
  C_E(0^+)
&=(4\pi n a_f)^2\left[1+\frac{64}{3\sqrt{\pi}}({na_i^3})^{1/2}\right]\\
&\quad+64\pi a^2_fn^2\Lambda(a_f-a_i),
\end{split}
\label{contacttzero}
\end{equation}
which is the Bogoluibov limit of the general result in
Eq.~\eqref{contactt0}. This UV cutoff-dependence is reflected in the large
value of the numerically evaluated contact near $t=0^+$, in
Fig.~\ref{fig:contactsuddenquench}. As time evolves after a quench, the
contact decreases dramatically within a short window of time, with the
cutoff-dependence quickly vanishing. After reaching a minimum it then
slowly grows to a finite steady-state value, $C_{ss}$.

\begin{figure}[ht]
 \centering
  \includegraphics[width=80mm]{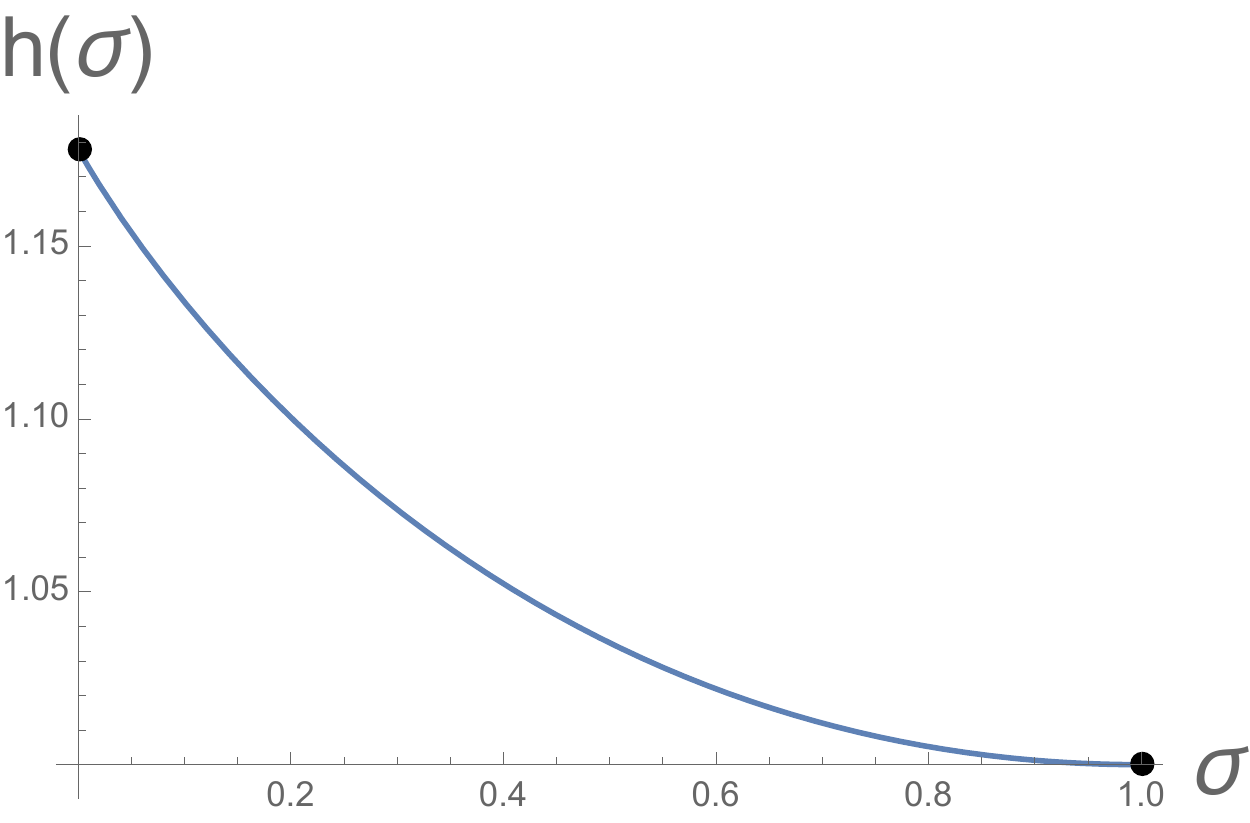}
 \caption{(Color online) Asymptotic Contact enhancement factor
   $h(\sigma)\equiv F_C(\sigma,t\to\infty)$ following a scattering
   length quench $a_i\to a_f$ as a function of quench depth
   $\sigma={a_i}/{a_f}$. Two dots correspond to maximum enhancement
   $h(0)=3\pi/8$ (non-interacting initial state or unitarity final
   state) and minimum enhancement $h(1)=1$ (no quench), respectively.}
\label{fig:contactsuddenquenchasym} 
\end{figure}

At long times, the sinusoid in \rf{contact_h} averages out and
contact reaches a steady-state value 
\begin{equation}
\label{contactFasym}
\begin{split}
h(\sigma)\equiv F_C(\sigma,t\to\infty)=\frac{\sigma^{3/2}+3\sqrt{\sigma}+3\sqrt{1-\sigma}\mathrm{arccos}{\sqrt{\sigma}}}{4}\\
\end{split}
\end{equation}
plotted in Fig.~\ref{fig:contactsuddenquenchasym}. This steady-state
contact is greater than the contact in the ground state for the same
scattering length $a_f$.

Finally, we examine the contact associated with the tail of the
momentum distribution function after the quench, which is given by
Eq.~\eqref{quasin} following a deep quench and
Eq.~\eqref{nkbogoliubov} for a shallow quench, respectively. From
these we straightforwardly obtain
 \begin{equation}
\begin{split}
C_{n}=\lim_{k\to\infty}k^4 n_k=(4\pi a_f n)^2\left[\frac{n^{ss}_c}{n}+(1-\sigma)^2\right]
\end{split}
\end{equation}
and the shallow quench result is obtained by setting $n^{ss}=n$ using
Bogoluibov approximation.  Clearly, this is also independent of
time. Thus, out of equilibrium, the three forms of the contact, $C_n$,
$C_E$, $C_{dE}$ no longer coincide, like they do in the ground state.

Above analysis of various forms of contact in the nonequilibrium state
thus shows that no direct relation of the coefficient of the
$1/\omega^{3/2}$ tail in RF spectroscopy \cite{Wild} to the equilibrium
contact and its other ground state relations can be made.

\section{Summary and open directions}

In this manuscript we studied the dynamics of a resonant Bose gas
following shallow and deep scattering length quenches and ramps,
confining to a metastable regime of a positive scattering length.
Utilizing a dynamic field theory extension of the Bogoluibov theory,
which self-consistently accounts for a large depletion and a
time-dependent condensate density, we approximately solved for the
full post-quench evolution of the system. From this we then computed a
variety of physical observables, such as the evolution of the momentum
distribution function, the associated condensate depletion, the
time-dependent structure function, the RF spectroscopy signal, the
excitation energy and various forms of a ``nonequilibrium
contact''. We found, that following initial transient dynamics, the
Bose gas exhibits a pre-thermalization to a stationary state
(characterized e.g., by a stationary momentum distribution function)
that differs qualitatively from the corresponding ground
state. Because of integrability of the approximate model, that does
not include quasi-particles scattering, the system never exhibits full
thermalization to a ground state. Despite the simplicity of our model
and approximate analysis, our results are in reasonable qualitative agreement
with recent JILA experiments \cite{Makotyn}.

Although we made significant progress in understanding the post-quench
dynamics of a resonant Bose system, our work leaves a number questions
for a future investigation. Our present study utilized a
single-channel model and focussed on the upper-branch physics with a
tunable positive scattering length, thereby neglecting the closed
molecular channel. The latter may in fact be quite significant,
enriching the dynamics by allowing coherent condensate oscillations
not only into pairs of atomic quasi-particles in the upper branch, but
also into molecular condensate and molecular quasi-particles. This
extension can be quite naturally treated within a two-channel model,
where the closed molecular channel is explicitly included. It would
allow one to address the dynamics not only within the superfluid phase
but across quantum and classical phase transitions, most notably
across the quantum Ising transition between atomic and molecular
superfluids and throughout the atomic-molecular phase diagram
\cite{RPWmolecularBECprl04,StoofMolecularBEC04,RPWmolecularBECpra08}.

Another crucial ingredient missing in our model is the quasi-particle
scattering. This is responsible for a time-independent quasi-particle
momentum distribution function, that is completely fixed by the
initial state, characterized by $a_i$ and the final scattering length
$a_f$. This feature is responsible for the absence of thermalization
of the system. It is thus desirable to extend the present model to
include quasi-particle scattering, that can be handled through the
Boltzmann equation for the quasi-particle distribution function. In
such a generalized model, the dynamics of the atomic observables
(e.g., atomic momentum distribution and structure functions) will
consist of two contributions, Heisenberg evolution of atoms due to
quasi-particle unitary dynamics, coupled to the evolution of the
quasi-particle momentum distribution function governed by the
Boltzmann equation with collision integrals. We expect that such
dynamics will exhibit a second, longer time scale, set by the
quasi-particle scattering that will lead to true long-time
thermalization.

Finally, to treat the effects of interactions more systematically, it
is desirable to have a full nonequilibrium Schwinger-Keldysh field
theoretic formulation. We leave these and a number of other open
question for future research \cite{LRunpublished}.

\acknowledgements

We thank P. Makotyn, D. Jin, and E. Cornell for sharing their data
with us before publication, and acknowledge them, A. Andreev, D. Huse,
V. Gurarie, and A. Kamenev for stimulating discussions. This research
was supported by the NSF through DMR-1001240, and by the Simons
Investigator award from the Simons Foundation.

\appendix



\section{Energy conservation}
\label{app:energyconserve}
In this appendix we study the time evolution of the total energy
following a deep quench.  Although energy is conserved under exact
unitary evolution of a closed system, it is less clear whether it
remains so for the time-dependent self-consistent Bogoluibov
approximation employed in deep quenches. We demonstrate below that
within this approximation, that neglects anomalous averages of finite
momentum excitations, the total energy is indeed conserved.

To this end we study the time derivative of the full time-dependent
Hamiltonian, including the constant mean-field parts derived in
Sec.~\ref{model}, \rf{Htotal}. It is given by
\begin{equation}
\begin{split}
\hat{H}_{total}&=\sum_{\kv\neq 0} \bigg[(\epsilon_k+gn_c(t))\hat{a}^{\dagger}_\kv\hat{a}_\kv+\frac{g}{2}n_c(t)(\hat{a}_\kv\hat{a}_{-\kv}+\hat{a}^{\dagger}_\kv\hat{a}^{\dagger}_{-\kv})\bigg]\\
&\quad+ g[n_c(t){n}_d(t)+\frac{1}{2}n^2_c(t)+{n}^2_d(t)],
\end{split}
\label{energychangetotal}
\end{equation}
where $g$ is the final interaction $g_f$ to which the system is
quenched, 
and the energy is evaluated as 
\begin{equation}
E_{total}(t)=\langle
0^-|\hat H_{total}(t)|0^-\rangle = E_1(t) + E_2(t) + E_3(t).
\end{equation}

The time derivative of last mean-field term, $E_3$ is given by
\begin{equation}
\begin{split}
\frac{dE_3}{dt}
&=g[\dot{n}_c{n}_d+n_c\dot{{n}}_d+n_c\dot{n}_c+2{n}_d\dot{{n}}_d],\\
&=g{n}_d\dot{{n}}_d,
\end{split}
\end{equation}
where we used the atom conservation constraint $n=n_c(t)+n_d(t)$,
giving $\dot{n}_c(t)+\dot{n}_d(t)=0$.

A time derivative of the first term $E_1(t)$ is
\begin{equation}
\begin{split}
\frac{dE_1}{dt}&\equiv\frac{d}{dt}
\sum_{\kv\neq 0}\left(\epsilon_k+gn_c(t)\right)
\langle 0^-| \hat{a}^{\dagger}_\kv(t)\hat{a}_\kv(t)|0^-\rangle,\\
&=\sum_{\kv\neq 0} g\dot{n}_c \langle \hat{a}^{\dagger}_\kv\hat{a}_\kv\rangle+(\epsilon_k+gn_c(t))
\langle\dot{\hat{a}}^{\dagger}_\kv \hat a_\kv+\hat{a}^{\dagger}_\kv\dot{\hat{a}}_\kv\rangle,
\end{split}
\label{dE1}
\end{equation}
and of the second term $E_2(t)$
\begin{equation}
\begin{split}
\frac{dE_2}{dt}&\equiv\frac{g}{2}\frac{d}{dt}\sum_{\kv\neq 0} n_c(t)
\langle 0^-|\hat{a}_\kv(t)\hat{a}_{-\kv}(t)+\hat{a}^{\dagger}_\kv(t) \hat{a}^{\dagger}_{-\kv}(t)|0^-\rangle,\\
&=\sum_{\kv\neq 0} \frac{g}{2}\dot{n}_c\langle \hat{a}_\kv\hat{a}_{-\kv}+\hat{a}^{\dagger}_\kv\hat{a}^{\dagger}_{-\kv}\rangle+\frac{g}{2}n_c\langle\dot{\hat{a}}_\kv\hat{a}_{-\kv}+\hat{a}_\kv\dot{\hat{a}}_{-\kv}\\
&\quad
+\dot{\hat{a}}^{\dagger}_\kv\hat{a}^{\dagger}_{-\kv}+\hat{a}^{\dagger}_\kv\dot{\hat{a}}^{\dagger}_{-\kv}\rangle.
\end{split}
\label{dE2}
\end{equation}
Using the Heisenberg equation of motion to eliminate time derivatives
of atom operators we find
\begin{equation}
\begin{split}
\dot{\hat{a}}_\kv&=\frac{1}{i }[(\epsilon_k+gn_c)\hat{a}_\kv+gn_c \hat{a}^{\dagger}_{-\kv}],\\
\dot{\hat{a}}_{-\kv}&=\frac{1}{i }[(\epsilon_k+gn_c)\hat{a}_{-\kv}+gn_c \hat{a}^{\dagger}_{\kv}].\\
\end{split}
\end{equation}
With this \rf{dE1} and \rf{dE2} reduce to
\begin{equation}
  \begin{split}
    \frac{dE_1}{dt}=g\dot{n}_c n_d+\frac{1}{i }\sum_{\kv\neq 0}
g n_c(\epsilon_k+gn_c)\langle 
\hat{a}^{\dagger}_{\kv}\hat{a}^{\dagger}_{-\kv}-\hat{a}_{\kv}\hat{a}_{-\kv}\rangle
  \end{split}
\end{equation}
and
\begin{equation}
\begin{split}
\frac{dE_2}{dt}&=\sum_{\kv\neq 0} \frac{g}{2}\dot{n}_c \langle \hat{a}_\kv\hat{a}_{-\kv}+\hat{a}^{\dagger}_\kv\hat{a}^{\dagger}_{-\kv}\rangle\\
&\quad+\frac{gn_c}{i }((\epsilon_k+g n_c)\langle \hat{a}_\kv\hat{a}_{-\kv}-\hat{a}^{\dagger}_\kv\hat{a}^{\dagger}_{-\kv}\rangle.
\end{split}
\end{equation}
For the total energy we then obtain,
\begin{equation}
\begin{split}
\frac{dE_{total}}{dt}
&=\sum_{\kv\neq 0} \frac{g}{2}\dot{n}_c[\langle \hat{a}_\kv\hat{a}_{-\kv}\rangle+\langle
\hat{a}^{\dagger}_\kv\hat{a}^{\dagger}_{-\kv}\rangle]\approx 0,\\
\end{split}
\end{equation}
where in the last approximation we neglected anomalous correlator of
excited atoms. More precisely, following Sotiriadis and Cardy \cite{SotiriadisCardy10}, we
observe that while the conventional definition of the energy is not
conserved, the shifted one $E_{shifted}\equiv E_{total} -\frac{g}{2}\int dt
\dot{n}_c[\langle \hat{a}_\kv\hat{a}_{-\kv}\rangle+\langle
\hat{a}^{\dagger}_\kv\hat{a}^{\dagger}_{-\kv}\rangle]$ approximately is.

\section{$U(t)$ for quasi-adiabatic approximation}
\label{appendix:UTquasi}

In this seciton, we fill in the technical details leading to $U(t)$ for quasi-adiabatic approximation in Eq.~\eqref{UTquasi}. The operator part of time-dependent Hamiltonian is
\begin{eqnarray}
\begin{split}
\hat H(t)&=\frac{1}{2}\sum_{\kv\neq 0}((\epsilon_k+n_c(t)g_f)(\hat{a}^{\dagger}_k\hat{a}_\kv+
\hat{a}^{\dagger}_{-\kv}\hat{a}_{-\kv})\\
&\quad+n_c(t)g_f(\hat{a}^{\dagger}_\kv\hat{a}^{\dagger}_{-\kv}+\hat{a}_\kv\hat{a}_{-\kv})).
\end{split}
\end{eqnarray} 
It can be instantaneously diagonalized by
\begin{equation}
\begin{pmatrix}\hat{a}_\kv(t)\\\hat{a}^{\dagger}_{-\kv}(t)\end{pmatrix}=\begin{pmatrix}u_k(t)&v_k(t)\\
v_k(t)&u_k(t)\end{pmatrix}\begin{pmatrix}\hat\gamma_\kv(t)\\ \hat\gamma^{\dagger}_{-\kv}(t)\end{pmatrix},
\end{equation}
and rewritten as
\begin{equation}
\begin{split}
\label{E2_2}
\hat H(t)=-\frac{1}{2}\sum_{\kv\neq 0}(\epsilon_k+n_c(t)g-E_{kf}(t))\\
+\frac{1}{2}\sum_{\kv\neq 0} E_{kf}(t)(\hat\gamma^{\dagger}_\kv \hat\gamma_\kv+\hat\gamma^{\dagger}_{-\kv} \hat\gamma_{-\kv}),
\end{split}
\end{equation}
where 
\begin{equation}
\begin{split}
u_k(t)&=\sqrt{\frac{1}{2}(\frac{\epsilon_k+g_fn_c(t)}{E_k(t)}+1)},\\
v_k(t)&=-\sqrt{\frac{1}{2}(\frac{\epsilon_k+g_fn_c(t)}{E_k(t)}-1)},\\
E_{kf}(t)&=\sqrt{\epsilon_k(\epsilon_k+2g_fn_c(t))}.
\end{split}
\end{equation}
The time-dependence of $\hat\gamma^{\dagger}_\kv(t)$ and $\hat\gamma_\kv(t)$ is
obtained from the Heisenberg equation of motion,
\begin{equation}
\frac{d\hat\gamma_\kv}{dt}=i[\hat\gamma_\kv,\hat H]+\frac{\partial \hat\gamma_\kv}{\partial t},
\end{equation}
where the last term accounts for the explicit time-dependence in
Hamiltonian. To compute it we first express $\hat\gamma^{\dagger}_\kv(t)$
and $\hat\gamma_\kv(t)$ in terms of $\hat a^{\dagger}_\kv(t)$ and $\hat a_\kv(t)$.
\begin{equation}
\begin{pmatrix}\hat\gamma_\kv(t)\\ \hat \gamma^{\dagger}_{-\kv}(t)\end{pmatrix}=\begin{pmatrix}u_k(t)&-v_k(t)\\
-v_k(t)&u_k(t)\end{pmatrix}\begin{pmatrix}\hat a_\kv(t)\\\hat a^{\dagger}_{-\kv}(t)\end{pmatrix},
\end{equation}
\begin{widetext}
Then
\begin{equation}
\begin{split}
\frac{\partial}{\partial t}\begin{pmatrix}\hat \gamma_\kv(t)\\ \hat \gamma^{\dagger}_{-\kv}(t)\end{pmatrix}&=\begin{pmatrix}\dot{u}_k(t)&-\dot{v}_k(t)\\
-\dot{v}_k(t)&\dot{u}_k(t)\end{pmatrix}\begin{pmatrix}\hat a_\kv(t)\\\hat a^{\dagger}_{-\kv}(t)\end{pmatrix},\\
&=\begin{pmatrix}\dot{u}_k(t)&-\dot{v}_k(t)\\
-\dot{v}_k(t)&\dot{u}_k(t)\end{pmatrix}\begin{pmatrix}u_k(t)&v_k(t)\\
v_k(t)&u_k(t)\end{pmatrix}\begin{pmatrix}\hat \gamma_\kv(t)\\ \hat \gamma^{\dagger}_{-\kv}(t)\end{pmatrix},\\
&=\begin{pmatrix}0&\frac{g\dot{n}_c(t)\epsilon_k}{2E^2_{kf}}\\
\frac{g\dot{n}_c(t)\epsilon_k}{2E^2_{kf}}&0\end{pmatrix}\begin{pmatrix}\hat \gamma_\kv(t)\\ \hat \gamma^{\dagger}_{-\kv}(t)\end{pmatrix}.
\end{split}
\end{equation}
Now the equation of motions become
\begin{equation}
\label{quasigammaeqm}
\frac{d}{dt}\begin{pmatrix}\hat \gamma_\kv(t)\\ \hat \gamma^{\dagger}_{-\kv}(t)\end{pmatrix}=\begin{pmatrix}-i E_{kf}(t)&\frac{g\dot{n}_c(t)\epsilon_k}{2E^2_{kf}}\\
\frac{g\dot{n}_c(t)\epsilon_k}{2E^2_{kf}}&i E_{kf}(t)\end{pmatrix}\begin{pmatrix}\hat \gamma_\kv(t)\\ \hat \gamma^{\dagger}_{-\kv}(t)\end{pmatrix}.
\end{equation}
Assuming $\dot{n}_c(t)$ changes slowly compared to other timescales (or more explicitly ${\dot{n}_c(t)}/{n}\ll{E^3_{kf}}/(\hbar ng\epsilon_k)$), we can ignore the off-diagonal terms in \eqref{quasigammaeqm} and have
\begin{equation}
\frac{d}{dt}\begin{pmatrix}\hat \gamma_\kv(t)\\ \hat \gamma^{\dagger}_{-\kv}(t)\end{pmatrix}\approx\begin{pmatrix}-i E_{kf}(t)&0\\
0&i E_{kf}(t)\end{pmatrix}\begin{pmatrix}\hat \gamma_\kv(t)\\ \hat \gamma^{\dagger}_{-\kv}(t)\end{pmatrix},
\end{equation}
from which we can solve $\hat \gamma^{\dagger}_\kv(t)$ and $\hat \gamma_\kv(t)$ as
\begin{equation}
\label{E3}
\hat \gamma_\kv(t)=\hat \gamma_\kv e^{-i\int_0^t dt' E_{kf}(t')},\;\;\;\;\;\;\;
\hat \gamma^{\dagger}_{-\kv}(t)=\hat \gamma^{\dagger}_{-\kv}e^{i\int_0^t dt'
  E_{kf}(t')},
\end{equation}
thus
\begin{equation}
\begin{split}
\begin{pmatrix}\hat a_\kv(t)\\\hat a^{\dagger}_{-\kv}(t)\end{pmatrix}
&=\begin{pmatrix}u_k(t)&v_k(t)\\
v_k(t)&u_k(t)\end{pmatrix}\begin{pmatrix}e^{-i\int_0^t dt E_{kf}(t')}&0\\0&e^{i\int_0^t dt E_{kf}(t')}\end{pmatrix}\begin{pmatrix}\hat \gamma_\kv\\\hat \gamma^{\dagger}_{-\kv}\end{pmatrix},\\
&=\begin{pmatrix}u_k(t)e^{-i \int_0^t dt E_{kf}(t')}&v_k(t) e^{i \int_0^t dt E_{kf}(t')}\\
v_k(t) e^{-i \int_0^t dt E_{kf}(t')}&u_k(t) e^{i \int_0^t dt E_{kf}(t')}\end{pmatrix}\begin{pmatrix}\hat \gamma_\kv\\\hat \gamma^{\dagger}_{-\kv}\end{pmatrix}.
\end{split}
\end{equation}
Comparing this with Eq.~\eqref{eqmquasiadiabatic} , we find
\begin{equation}
\begin{split}
U(t)=\begin{pmatrix}u_k(t)e^{-i \int_0^t dt E_{kf}(t')}&v_k(t) e^{i \int_0^t dt E_{kf}(t')}\\
v_k(t) e^{-i \int_0^t dt E_{kf}(t')}&u_k(t) e^{i \int_0^t dt E_{kf}(t')}\end{pmatrix}.
\label{eq6}
\end{split}
\end{equation}

\section{Energy after quench}
\label{appendix:energyafterquench}
In this section we evaluate the total energy of the system after the sudden quench. 
 Separating the energy into kinectic part and interaciton part
\begin{equation}
\begin{split}
\label{eq42}
\langle 0^{-}|\hat H^f|0^{-}\rangle=\langle 0^{-}|\hat H^f_{KE}|0^{-}\rangle+\langle 0^{-}|\hat H^f_{int}|0^{-}\rangle,
\end{split}
\end{equation}
with
\begin{equation}
\begin{split}
\label{eq43}
\hat H^f_{KE}=\frac{1}{2}\sum_{\kv\neq 0}\epsilon^0_k(\hat a^{\dagger}_\kv\hat a_\kv+
\hat a^{\dagger}_{-\kv}\hat a_{-\kv}),\\
\end{split}
\end{equation}
\begin{equation}
\begin{split}
\hat H^f_{int}=\frac{1}{2}Vg_fn^2+\frac{1}{2}\sum_{\kv\neq 0} [ng_f(\hat a^{\dagger}_\kv\hat a_\kv+
\hat a^{\dagger}_{-\kv}\hat a_{-\kv})+ng_f(\hat a^{\dagger}_\kv\hat a^{\dagger}_{-\kv}+\hat a_\kv\hat a_{-\kv})],
\end{split}
\label{energyinta}
\end{equation}
we then use Bogoluibov transformation to evaluate them respectively by expressing $\hat a_\kv$ in terms of pre-quench basis $\hat \alpha_\kv$, obtaining
\begin{equation}
\begin{split}
\label{eq43}
\hat H_{KE}=\sum_{\kv\neq0}\epsilon_k\hat a^{\dagger}_\kv\hat a_\kv=
\sum_{\kv\neq0}\epsilon_k\left(|u_k|^2\hat \alpha^{\dagger}_\kv\hat \alpha_\kv
-u^*_kv_k\hat \alpha^{\dagger}_\kv\hat \alpha^{\dagger}_{-\kv}-u_kv^*_k\hat \alpha_{-\kv}\hat \alpha_\kv+|v_k|^2\hat \alpha_{-\kv}\hat \alpha^{\dagger}_{-\kv}\right),
\end{split}
\end{equation}
and
\begin{equation}
\label{eq31}
\begin{split}
\hat H_{int}&=\frac{1}{2}Vg_fn^2+ng_f\sum_{\kv\neq0} [|v_k|^2-\frac{1}{2}(u^*_kv_k+u_kv_k^*)]\\
&\quad+\frac{1}{2}ng_f\sum_{\kv\neq0} \left[\left(|u^2_k|+|v^2_k|-(u^*_kv_k+u_kv^*_k)\right)(\hat \alpha^{\dagger}_\kv\hat \alpha_{\kv}+\hat \alpha^{\dagger}_{-\kv}\hat \alpha_{-\kv})
+(u^2_k+v^2_k-2u_kv_k)(\hat \alpha^{\dagger}_\kv\hat \alpha^{\dagger}_{-\kv}+\hat \alpha_{\kv}\hat \alpha_{-\kv})\right].
\end{split}
\end{equation}

Since $\hat\alpha_\kv|0^-\rangle = 0$, we have
\begin{equation}
\langle 0^{-}|\hat H_{KE}|0^{-}\rangle=\sum_{\kv\neq0}\epsilon_k|v_k|^2,
\label{energykin}
\end{equation}
and
\begin{equation}
\label{energyint}
\begin{split}
&\langle 0^{-}|\hat H^f_{int}|0^{-}\rangle=\frac{1}{2}V\tilde{g}_fn^2+n\tilde{g}_f\sum_{\kv\neq0} \left[|v_k|^2-\frac{1}{2}(u^*_kv_k+u_kv_k^*)+\frac{n\tilde{g}_f}{4\epsilon_k}\right],
\end{split}
\end{equation}
during which coupling $g$ has been expanded to second order
\begin{equation}
\label{eq20}
\begin{split}
g=\frac{4\pi a}{m}+\frac{(4\pi a)^2}{m^2V}\sum_{\kv\neq0}\frac{1}{2\epsilon_k}\equiv\tilde{g}+\frac{\tilde{g}^2}{V}\sum_{\kv\neq0}\frac{1}{2\epsilon_k}.
\end{split}
\end{equation}
Therefore, the total energy is
\begin{equation}
\begin{split}
E_{tot}(t=0^+)&=\langle 0^{-}|\hat H_{KE}+\hat H_{int}|0^{-}\rangle,\\
&=\frac{1}{2}V\tilde{g}_fn^2+n\tilde{g}_f\sum_{\kv\neq0}\left [\left(\frac{\epsilon_k}{n\tilde{g}}+1\right)|v_k|^2
-\frac{1}{2}(u^*_kv_k+u_kv_k^*)+\frac{n\tilde{g}_f}{4\epsilon_k}\right],\\
&=\frac{2\pi na_f}{m}
+\frac{32\sqrt{2\pi} na_f}{m}({na^3_f})^{1/2} \int dk k^2\left[(k^2+1)|v_k|^2-\frac{1}{2}(u_kv^*_k+v_ku^*_k)+\frac{1}{4k^2}\right].
\label{excenergyfiniteramp}
\end{split}
\end{equation}

For a sudden quench, the expressions for $u_k$ and $v_k$ are simple
\begin{equation}
\begin{split}
\label{uvsudden}
u_k&=\sqrt{\frac{1}{2}(\frac{\epsilon_k+n\tilde{g}_i}{\sqrt{\epsilon_k(\epsilon_k+2n\tilde{g}_i)}}+1)},\\
v_k&=-\sqrt{\frac{1}{2}(\frac{\epsilon_k+n\tilde{g}_i}{\sqrt{\epsilon_k(\epsilon_k+2n\tilde{g}_i)}}-1)}.\\
\end{split}
\end{equation}
Plugging Eq.~\eqref{uvsudden} into \eqref{energykin} and \eqref{energyint}, we obtain the kinetic energy as
\begin{equation}
\begin{split}
\langle 0^{-}|\hat H_{KE}|0^{-}\rangle&=\sum_{\kv\neq0}\frac{1}{2}\epsilon_k(\frac{\epsilon_k+n\tilde{g}_i}{\sqrt{\epsilon_k(\epsilon_k+2n\tilde{g}_i)}}-1),\\
&=\frac{1}{2}n\tilde{g}_i(2m\tilde{g}_i n)^{3/2}\frac{4\pi V}{(2\pi)^3}\int dy y^2\left[\frac{y^2+1}{\sqrt{y^2(y^2+2)}}-1\right],\\
&=N\frac{4a_i n}{m}\Lambda a_i-\frac{128\sqrt{\pi} a_i n}{5m}N({na^3_i})^{1/2},
\end{split}
\end{equation}
the interaction energy as
\begin{equation}
\label{quenchenergyint}
\begin{split}
\langle 0^{-}|\hat H^f_{int}|0^{-}\rangle&=\frac{1}{2}V\tilde{g}_fn^2+\frac{1}{2}n\tilde{g}_f\sum_{\kv\neq0}(\epsilon_k/\sqrt{\epsilon_k(\epsilon_k+n\tilde{g}_f)}-1+n\tilde{g}_f/2\epsilon_k),\\
&=\frac{1}{2}Vn^2\tilde{g}_f+\frac{1}{2}n\tilde{g}_f(2m\tilde{g}_fn)^{\frac{3}{2}}\frac{4\pi V}{(2\pi)^3}\int dy y^2\left(\frac{y}{\sqrt{y^2+2\sigma}}-1+\frac{1}{2y^2}\right),\\
&=\frac{2\pi a_fn}{m}N\left[1+\frac{64}{3}(\frac{na^3_i}{\pi})^{\frac{1}{2}}\right]
+(1-2\sigma)N\frac{4a_f n}{m}\Lambda a_f,
\end{split}
\end{equation}
and the total energy as
\begin{equation}
\begin{split}
E_{tot}&=\langle 0^{-}|\hat H_{KE}+\hat H_{int}|0^{-}\rangle=\frac{4(1-\sigma)^2 na_f}{m}Na_f\Lambda-\frac{128\sqrt{\pi} na_i}{5m}N({na^3_i})^{1/2}+\frac{2\pi na_f}{m}N\left[1+\frac{64}{3\sqrt{\pi}}({na^3_i})^{1/2}\right],
\end{split}
\end{equation}
which is Eq.~\eqref{eps_excR} in the text. The ground state energy can be easily obtained by setting $\sigma=1$, and one obtains
\begin{equation}
\begin{split}
E_{tot}=\frac{2\pi a_sn}{m}N[1+\frac{128}{15}(\frac{na^3_s}{\pi})^{\frac{1}{2}}].\\
\end{split}
\end{equation}
as the cutoff dependences of kinetic energy and interaction energy
cancel each other, recovering the LHY result as expected.

\section{Contact}

\subsection{Ground state contact} 
\label{appendix:groundstatecontact}

In this paper, we follow E. Braaten et.al \cite{Braaten} and take the
working definition of contact to be
\begin{equation}
 \begin{split} 
 C=(mg)^2 \langle\int dr\hat \psi^{\dagger}(r,t)\hat \psi^{\dagger}(r,t)\hat \psi(r,t)\hat \psi(r,t))\rangle=2m^2g\langle\hat  H_{int}\rangle /V.
\end{split}
 \label{contactint}
  \end{equation} At
$T=0$ for $na^3_s\ll1$, the interaction energy of Bose gas is given in
Appendix \ref{appendix:energyafterquench}. For ground state, $\langle O|\hat H_{int}|O\rangle$ can be
evaluated by applying $\sigma=1$ to Eq.~\eqref{quenchenergyint}, which
gives
 \begin{equation} 
  \begin{split}
  \langle O|\hat H_{int}|O\rangle
&=\frac{2\pi a_sNn}{m}\left[1+\frac{64}{3\sqrt{\pi}}(na^3_s)^{1/2}\right]-\frac{4\Lambda a^2_sNn}{m}
,\\
&=2Nn/m\left[a_s\pi\left(1+\frac{64}{3\sqrt{\pi}}(na^3_s)^{1/2}\right)-2a^2_s\Lambda\right].
\end{split}
\end{equation} 
The last term contains the same divergence as the bare interaction
$g$, and we show below they exactly cancel each other to give a finite
contact.
\begin{equation}
 \label{contactlambdacancel}
\begin{split}
C&=2m^2g\langle O|\hat H_{int}|O\rangle/V,\\
&=2m^2\left(\frac{4\pi a_s}{m}+\frac{8\Lambda a^2_s}{m}\right)\langle O|\hat H_{int}|O\rangle/V,\\
&=8m(\pi a_s+2\Lambda a^2_s)\langle O|\hat H_{int}|O\rangle/V,\\
&=(4\pi a_s)^2n^2\left[1+\frac{64}{3\sqrt{\pi}}({na^3_s})^{1/2}+\frac{128}{3{\pi}^{3/2}}a_s\Lambda({na^3_s})^{1/2}+O(\Lambda^2 a^2_s)\right].
\end{split}
\end{equation}
Thus to the order of $(na^3_s)^{1/2}$, the contact value for ground state at $T=0$ is 
\begin{equation}
\label{contactgrounddef}
\begin{split}
C=(4\pi a_s)^2nN\left[1+\frac{64}{3\sqrt{\pi}}({na^3_s})^{1/2}\right].\\
\end{split}
\end{equation}

For bosons in thermal equilibrium, one central Tan's relation is the
adiabatic theorem, which relates the energy change with respect to
scattering length to the contact. The theorem states the following
thing
\begin{equation}
\begin{split}
C={8\pi m a^2_s}\frac{{d\cal E}_{gs}}{da_s}.\\
\end{split}
\end{equation}
Since the ground state energy is given by Eq.~\eqref{LHYenergy}, it is straightforward to show that 
\begin{equation}
\begin{split}
{8\pi m a^2_s}\frac{d{\cal E}_{gs}}{da_s}=(4\pi a_s n)^2\left[1+\frac{64}{3\sqrt{\pi}}({na^3_s})^{\frac{1}{2}}\right].
\end{split}
\end{equation}
Thus we have verified the adiabatic theorem in ground state.

Another important Tan's relation is the momentum theorem, which
relates contact to the high momentum tail of the momentum distribution function
\begin{equation}
\begin{split}
C=\lim_{k\to\infty}k^4 n_k.
\end{split}
\end{equation}
For ground state at $T=0$, momentum distribution $n_k$ is given by
Eq.~\eqref{nkt0}, giving
\begin{equation}
\begin{split}
\lim_{k\to\infty}k^4 n_k=C_0+O(1/k^2),
\end{split}
\end{equation}
with $C_0=(4\pi a_s n)^2$. Thus we recover the lowest order of contact
obtained in Eq.~\eqref{contactgrounddef}.

We can also generalize the contact to large $n a_s^3$ case. From
Eq.~\eqref{Htotal}, the ground state energy is modified as
\begin{equation}
\begin{split}
{\cal E}_{gs}&=\langle O|\hat H_{total}|O\rangle\\
&=\frac{2\pi a_s Vn^2}{m}\left[1+\left(\frac{{n}_d}{n}\right)^2+\frac{128({na^3_s})^{1/2}}{15\sqrt{\pi}}\left(\frac{{n}_c}{n}\right)^{5/2}\right].
\end{split}
\end{equation}
Then the adiabatic theorem gives
\begin{equation}
\begin{split}
C&=8\pi m a^2_s\frac{d{\cal E}_{gs}}{da_s}\\
&=(4\pi a_s n)^2\left[1+\left(\frac{{n}_d}{n}\right)^2+\frac{64({na^3_s})^{1/2}}{3\sqrt{\pi}}\left(\frac{n_c}{n}\right)^{5/2}\right].
\end{split}
\end{equation}
It is straightforward to verify that this also agrees with contact
obtained via Eq.~\eqref{contactint}. Here, condensate density $n_c$
and depletion density $n_d$ are determined self-consistently by
Eq.~\eqref{screduced}.

\subsection{Dynamical contact} 
\label{app:dynamicalcontact}

An important quantity to determine dynamical contact is the
interaction energy $\langle 0^{-}|\hat H^f_{int}|0^{-}\rangle$. 
In this section, still assuming a sudden quench, we further study the
dynamics of interaction energy and focus on its asymptotic long time
limit, and use it to construct the dynamical contact as in
Eq. \eqref{contactint}. Using Eq.~\eqref{pseudounitarytrans} to
decompose $\hat a_k$ into post-quench basis $\hat \beta_k(t)$, as $\hat \beta_k(t)$
evolve simply according to Eq.~\eqref{quenchevo}, combined with
Eq.~\eqref{energyinta}, we obtain
\begin{equation}
\label{energyinttime}
\begin{split}
&\langle 0^{-}|\hat H^f_{int}|0^{-}\rangle\\
&=\frac{1}{2}Vg_fn^2-\frac{1}{2}\sum_{\kv\neq0} ng\frac{\epsilon_k+2ng_f-\sqrt{\epsilon_k(\epsilon_k+2ng_f)}}{\epsilon_k+2ng_f}
+\frac{1}{2}\sum_{\kv\neq0}\frac{\epsilon_k gn}{\sqrt{\epsilon_k(\epsilon_k+2ng_f)}}[\langle\beta^\dagger_\kv\beta_\kv\rangle+\langle\beta^\dagger_{-\kv}\beta_{-\kv}\rangle\\
&\quad+\langle\beta^\dagger_{\kv}\beta^\dagger_{-\kv}\rangle e^{2iE_kt}+\langle\beta_{\kv}\beta_{-\kv}\rangle e^{-2iE_kt},]\\
&=\frac{1}{2}Vg_fn^2+\frac{(n\tilde{g}_f)^2}{4\epsilon_k}
-\frac{1}{2}\sum_{\kv\neq0} ng_f\frac{\epsilon_k+2ng_f-\sqrt{\epsilon_k(\epsilon_k+2ng_f)}}{\epsilon_k+2ng_f}
+\sum_{\kv\neq0}\frac{1}{2}\epsilon_k n g_f[\frac{\epsilon_k+ng_f+ng_i-\sqrt{(\epsilon_k+2ng_i)(\epsilon_k+2ng_f)}}{(\epsilon_k+2ng_f)\sqrt{\epsilon_k(\epsilon_k+2ng_i)}},\\
&\quad+\frac{n(g_f-g_i)}{(\epsilon_k+2ng_f)\sqrt{\epsilon_k(\epsilon_k+2ng_i)}}\cos[2t\sqrt{\epsilon_k(\epsilon_k+2ng_f)}].
\end{split}
\end{equation}
Rescaling time and momentum and taking the integral, we obtain
\begin{equation}
\begin{split}
\langle 0^{-}|\hat H^f_{int}|0^{-}\rangle&=\frac{1}{2}Ng_fn+\frac{1}{2}ng_f(2mg_fn)^{\frac{3}{2}}\frac{4\pi V}{(2\pi)^3}\int dyy^2\\
&\quad\times[y\frac{y^2+1+\sigma-\sqrt{(y^2+2\sigma)(y^2+2)}}{(y^2+2)\sqrt{y^2+2\sigma}}+\frac{y}{\sqrt{y^2+2\sigma}}-1
+\frac{1}{2y^2}+\frac{y(1-\sigma)}{(y^2+2)\sqrt{y^2+2\sigma}}\cos(2\hat{t}\sqrt{y^2(y^2+2)}]\\
&=\frac{2Nna_f\pi}{m}[1+F_C(\sigma,t)\frac{64}{3\sqrt{\pi}}(na^3_f)^{1/2})]-\frac{4Nna^2_f\Lambda}{m},\\
\end{split}
\end{equation}
where $F_C(\sigma,t)=h(\sigma)+T(t,\sigma,\Lambda)$ and
\begin{equation}
\begin{split}
h(\sigma)=\frac{{\sigma}^{3/2}+3\sqrt{\sigma}+3\sqrt{1-\sigma}\mathrm{arccos}{\sqrt{\sigma}}}{4},
\end{split}
\end{equation}
\begin{equation}
\begin{split}
T(t,\sigma,\Lambda)=\frac{3\sqrt{2}}{8}\int dyy^2\frac{y(1-\sigma)}{(y^2+2)\sqrt{y^2+2\sigma}}
\times\cos[2\hat{t}\sqrt{y^2(y^2+2)}].\\
\end{split}
\end{equation}
Following Eq.~\eqref{contactlambdacancel}, to the order of
$(na^3)^{1/2}$ we obtain the dynamical contact after a quench,
 \begin{eqnarray}
C_E(t)=(4\pi n a_f)^2+F_C(\sigma,t)C^f_{LHY},
\end{eqnarray}
given in Eq.~\eqref{dynamicalcontactLHY} of the main text. In the
asymptotically long time limit, $T(t\to\infty,\sigma,\Lambda)\to0$.


\subsection{RF spectroscopy}
\label{appendix:RFspectroscopy}
In this appendix, we dervie Eq.~\eqref{RFnumberdetailT} of Sec.~\ref{sec:RFspectroscopy}. With $\hat J_I(t) = e^{i\int_0^tdt' \hat H_0}\hat Je^{-i\int_0^tdt' \hat H_0}$,

\begin{eqnarray}
\langle \hat J(t)\rangle&=&\langle\psi| \hat J(t) |\psi\rangle,\nonumber\\
&=&\langle\psi| e^{i\int_0^tdt'(\hat H_0 + \hat H_1(t'))}J 
e^{-i\int_0^tdt'(\hat H_0+\hat H_1(t'))}|\psi\rangle,\nonumber\\
&=&\langle\psi_I(t)|\hat J_I(t)|\psi_I(t)\rangle,\nonumber\\
&=&\langle\psi| e^{i\int_0^tdt'(\hat H_0 + \hat H_{RF}(t'))}e^{-i\int_0^tdt' \hat H_0}\hat J_I(t) e^{i\int_0^tdt' \hat H_0}e^{-i\int_0^tdt'(\hat H_0+\hat H_{RF}(t'))}|\psi\rangle,\nonumber\\
&=&\langle\psi| T^*\left[e^{i\int_0^tdt'\hat H_{RF}^I(t')}\right]\hat J_I(t) 
T\left[e^{-i\int_0^tdt' \hat H_{RF}^I(t')}\right]|\psi\rangle,\nonumber\\
&=&-i\int_0^t dt'\langle\psi|\left[\hat J_I(t), \hat H_{RF}^I(t')\right]
|\psi\rangle.
\end{eqnarray}
The state $|\psi\rangle=|\alpha_0, 0_b\rangle$ a product state of a
vacuum of $b$ atoms, $|0_b\rangle$ and a SF condensate of $a$ atoms,
$|\alpha_0\rangle$, corresponding to the $t=0^-$ state (ground state
for $T=0$: $\hat \alpha_\kv|\alpha_0\rangle = 0$) {\em before} the ramp
(quench) to a new scattering length, which we will take to be a vacuum
of Bogoluibov quasi-particles for $t = 0^-$ interactions at $T=0$. 

Plugging the expressions for $\hat J_I(t)$ and $\hat H_1(t)$ into above equation, we obtain the current as
\begin{equation}
\begin{split}
\langle \hat J(t)\rangle
&=\int_0^t dt'\sum_{\kv,\kv'}\langle\psi|\bigg[I(t) \hat b_\kv^\dagger(t)
  \hat a_\kv(t) - I^*(t) \hat a_\kv^\dagger(t) \hat b_\kv(t)\\
&\quad+i g n_0\sum_\kv (\hat a_{-\kv}(t) \hat a_\kv(t) - \hat a_{\kv}^\dagger(t) \hat a_{-\kv}^\dagger(t)),\quad I(t') \hat b_{\kv'}^\dagger(t') \hat a_{\kv'}(t') 
+ I^*(t') \hat a_{\kv'}^\dagger(t')
\hat b_{\kv'}(t')\bigg]|\psi\rangle,\\
&=\int_0^t dt'\sum_{\kv,\kv'}I(t)I^*(t')\langle\psi|(
\hat a_\kv(t)\hat a_{\kv'}^\dagger(t')\hat b_\kv^\dagger\hat b_{\kv'}-
\hat a_{\kv'}^\dagger(t')\hat a_\kv(t)  \hat b_{\kv'}\hat b_\kv^\dagger)|\psi\rangle e^{i\epsilon_k t - i\epsilon_{k'} t' + i\omega_0(t-t')} + c.c.,\\
&=-\int_0^t dt'\sum_{\kv}I^*(t')I(t)\langle\alpha_0|
\hat a_{\kv}^\dagger(t')\hat a_\kv(t)|\alpha_0\rangle e^{i(\epsilon_k +  \omega_0)(t-t')} 
+ c.c.,
\end{split}
\end{equation}

Now the RF spectroscopy signal can be evaluated as
\begin{equation}
\begin{split}
  N_b(\omega_{RF})&=-\int_{0}^\infty dt \langle \hat J(t)\rangle\\
 & =\int_{0}^\infty dt \int_{0}^t dt'\sum_{\kv}
  I^*(t')I(t)\langle\alpha_0|\hat a_{\kv}^\dagger(t')\hat a_\kv(t)|\alpha_0\rangle e^{i(\epsilon_k +  \omega_0)(t-t')} + c.c.,\\
&  =\oh\int_{0}^\infty dt \int_{0}^\infty dt'\sum_{\kv}
  I^*(t')I(t)\langle\alpha_0|\hat a_{\kv}^\dagger(t')\hat a_\kv(t)|\alpha_0\rangle  e^{i(\epsilon_k +  \omega_0)(t-t')} + c.c..
\label{RFnumber}
\end{split}
\end{equation}
where we utilized the $t\rightarrow t'$ symmetry to simplify the
integral.

Plugging the correlator in Eq.~\eqref{Cij} into Eq.~\eqref{RFnumber},
we obtain
\begin{equation}
\begin{split}
 N_b(\omega_{RF})&=\oh\int_{0}^\infty dt \int_{0}^\infty dt'\sum_{\kv}
  I^*(t')I(t)\langle\alpha_0|\hat a_{\kv}^\dagger(t')\hat a_\kv(t)|\alpha_0\rangle
\times  e^{i(\epsilon_k + \omega_0)(t-t')} + c.c.,\\
&=\oh\int_{0}^\infty dt \int_{0}^\infty dt'\sum_{\kv} I^2_0 e^{-(t'-t_0)^2/\tau^2}e^{-(t-t_0)^2/\tau^2}
\times [e^{i(\epsilon_k + \omega_0-\omega_{RF}-E_k)(t-t')}u^2_k (\sinh\Delta \theta_k)^2
+e^{i(\epsilon_k + \omega_0-\omega_{RF}+E_k)(t-t')}\\
&\quad\times v^2_k (\cosh\Delta \theta_k)^2+\frac{1}{2}u_kv_k\sinh2\Delta \theta_k(e^{i(\epsilon_k + \omega_0-\omega_{RF}+E_k)t}e^{-i(\epsilon_k + \omega_0-\omega_{RF}-E_k)t'}
+e^{i(\epsilon_k + \omega_0-\omega_{RF}-E_k)t}e^{-i(\epsilon_k + \omega_0-\omega_{RF}+E_k)t'})]+c.c.,\\
&=\pi \tau^2 I^2_0 \sum_{\kv}[e^{-\frac{1}{2}(\epsilon_k + \omega_0-\omega_{RF}-E_k)^2\tau^2} u^2_k (\sinh\Delta \theta_k)^2
+e^{-\frac{1}{2}(\epsilon_k + \omega_0-\omega_{RF}+E_k)^2\tau^2} v^2_k (\cosh\Delta \theta_k)^2
+\frac{1}{2}u_kv_k\sinh2\Delta \theta_k\\
&\quad\times(e^{-\frac{1}{4}(\epsilon_k + \omega_0-\omega_{RF}-E_k)^2\tau^2-\frac{1}{4}(\epsilon_k + \omega_0-\omega_{RF}+E_k)^2\tau^2} )\cos (2E t_0)],\\
&=\pi \tau^2 I^2_0\sum_{\kv} [e^{-\frac{1}{2}(\epsilon_k + \omega_0-\omega_{RF}-E_k)^2\tau^2} u^2_k (\sinh\Delta \theta_k)^2
+e^{-\frac{1}{2}(\epsilon_k + \omega_0-\omega_{RF}+E_k)^2\tau^2} v^2_k (\cosh\Delta \theta_k)^2\\
&\quad+\frac{1}{2}u_kv_k\sinh2\Delta \theta_k
\times(e^{-\frac{1}{2}((\epsilon_k + \omega_0-\omega_{RF})^2+E^2_k)\tau^2}\cos (2E t_0))],\\
\end{split}
\label{RFnumberdetail}
\end{equation}
\end{widetext}
which gives Eq.~\eqref{RFnumberdetailT} of the main text. In above
derivation we have used $\hat b_\kv(t)=\hat b_\kv e^{−i\epsilon_k t}$
for atoms in the noninteracting hyperfine state, dropped the number
non-conserving $\hat b^{\dagger}\hat b^{\dagger}$,$\hat b \hat b$ terms, neglecting a weak
condensation that is always in principle induced by the linear
coupling to the a-Bose condensate during the time that the RF coupling
pulse is on.


\begin{thebibliography}{99}
\bibitem{footnote}
Although we refer to the initial state as the ground, for a bosonic system on the BEC side, it is not, as lower-lying molecular (and higher number cluster) states exist. However, on short time scales relevant to experiments, we confine our analysis to the upper-branch \cite{JasonHoPaperOnUpperBranch, ZhouPRA08}, valid for the initial preparation far detuned from the Feshbach resonance \cite{Makotyn}.
\end{thebibliography}
\end{document}